\documentclass[]{aa}

\usepackage{graphicx}
\usepackage{txfonts}
\usepackage{natbib}
\usepackage{url}
\usepackage{multirow,bigdelim}
\usepackage{rotating}
\usepackage{placeins}
\usepackage{color}
\usepackage{soul}
\begin{document}

%USER DEFINED ShORTCUTS
\newcommand{\zabs}{\ensuremath{z_{\rm abs}}}
\newcommand{\zem}{\ensuremath{z_{\rm em}}}
\newcommand{\zqso}{\ensuremath{z_{\rm QSO}}}
\newcommand{\zgal}{\ensuremath{z_{\rm gal}}}
\newcommand{\HH}{\mbox{H$_2$}}
\newcommand{\HD}{\mbox{HD}}
\newcommand{\DD}{\mbox{D$_2$}}
\newcommand{\CO}{\mbox{CO}}
\newcommand{\dla}{damped Lyman\,$\alpha$}
\newcommand{\Dla}{damped Lyman\,$\alpha$}
\newcommand{\lya}{\ensuremath{{\rm Ly}\,\alpha}}
\newcommand{\lyb}{Ly\,$\beta$}
\newcommand{\Ha}{H\,$\alpha$}
\newcommand{\Hb}{H\,$\beta$}
\newcommand{\lyg}{Ly\,$\gamma$}
\newcommand{\lyd}{Ly\,$\delta$}

%ions A\&A style can be used both in math or normal mode
\newcommand{\ArI}{\ion{Ar}{i}}
\newcommand{\CaII}{\ion{Ca}{ii}}
\newcommand{\CI}{\ion{C}{i}}
\newcommand{\CII}{\ion{C}{ii}}
\newcommand{\CIV}{\ion{C}{iv}}
\newcommand{\ClI}{\ion{Cl}{i}}
\newcommand{\ClII}{\ion{Cl}{ii}}
\newcommand{\CoII}{\ion{Co}{ii}}
\newcommand{\CrII}{\ion{Cr}{ii}}
\newcommand{\CuII}{\ion{Cu}{ii}}
\newcommand{\DI}{\ion{D}{i}}
\newcommand{\FeI}{\ion{Fe}{i}}
\newcommand{\FeII}{\ion{Fe}{ii}}
\newcommand{\GeII}{\ion{Ge}{ii}}
\newcommand{\HI}{\ion{H}{i}}
\newcommand{\MgI}{\ion{Mg}{i}}
\newcommand{\MgII}{\ion{Mg}{ii}}
\newcommand{\MnII}{\ion{Mn}{ii}}
\newcommand{\NI}{\ion{N}{i}}
\newcommand{\NII}{\ion{N}{ii}}
\newcommand{\NV}{\ion{N}{v}}
\newcommand{\NiII}{\ion{Ni}{ii}}
\newcommand{\OI}{\ion{O}{i}}
\newcommand{\OII}{\ion{O}{ii}}
\newcommand{\OIII}{\ion{O}{iii}}
\newcommand{\OVI}{\ion{O}{vi}}
\newcommand{\PII}{\ion{P}{ii}}
\newcommand{\PbII}{\ion{Pb}{ii}}
\newcommand{\SI}{\ion{S}{i}}
\newcommand{\SII}{\ion{S}{ii}}
\newcommand{\SiII}{\ion{Si}{ii}}
\newcommand{\SiIV}{\ion{Si}{iv}}
\newcommand{\TiII}{\ion{Ti}{ii}}
\newcommand{\ZnII}{\ion{Zn}{ii}}
\newcommand{\AlII}{\ion{Al}{ii}}
\newcommand{\AlIII}{\ion{Al}{iii}}
\def\h2{$\rm H_2$}
%Other shortcuts.
\newcommand{\Ho}{\mbox{$H_0$}}
\newcommand{\angstrom}{\mbox{{\rm \AA}}}
\newcommand{\abs}[1]{\left| #1 \right|} % for absolute value
\newcommand{\avg}[1]{\left< #1 \right>} % for average
\newcommand{\kms}{\ensuremath{{\rm km\,s^{-1}}}}
\newcommand{\cmsq}{\ensuremath{{\rm cm}^{-2}}}
\newcommand{\ergs}{\ensuremath{{\rm erg\,s^{-1}}}}
\newcommand{\ergsa}{\ensuremath{{\rm erg\,s^{-1}\,{\AA}^{-1}}}}
\newcommand{\ergscm}{\ensuremath{{\rm erg\,s^{-1}\,cm^{-2}}}}
\newcommand{\ergscma}{\ensuremath{{\rm erg\,s^{-1}\,cm^{-2}\,{\AA}^{-1}}}}
\newcommand{\msyr}{\ensuremath{{\rm M_{\rm \odot}\,yr^{-1}}}}
\newcommand{\nhi}{n_{\rm HI}}
\newcommand{\fhi}{\ensuremath{f_{\rm HI}(N,\chi)}}
\newcommand{\refs}{{\bf (refs!)}}
\newcommand{\jt}{J2140$-$0321}
\newcommand{\jtl}{SDSS\,J214043.02$-$032139.2}
\newcommand{\jf}{J1456$+$1609}
\newcommand{\jfl}{SDSS\,J145646.48$+$160939.3}
\newcommand{\jz}{J0154$+$1935}
\newcommand{\jzl}{SDSS\,J015445.22$+$193515.8}
\newcommand{\jonze}{J1135$-$0010}
\newcommand{\jonzelong}{SDSS\,J113520.39$-$001053.5}
\newcommand{\PN}{\color{red} PN:~}
\newcommand{\HR}{\color{red} HR:~}
\newcommand{\RS}{\color{red} RS:~}

%Institutes
\newcommand{\iap}{Institut d'Astrophysique de Paris, CNRS-UPMC, UMR7095, 98bis bd Arago, 75014 Paris, France\label{iap}}
\newcommand{\iran}{Institute for Research in Fundamental Sciences (IPM), PO Box 19395-5531, Tehran, Iran\label{iran}}
\newcommand{\lam}{Laboratoire d’Astrophysique de Marseille, CNRS/Aix Marseille Université, UMR 7326, 13388, Marseille, France\label{lam}}
\newcommand{\uchile}{Departamento de Astronom\'ia, Universidad de Chile, Casilla 36-D, Santiago, Chile\label{uchile}}
\newcommand{\trieste}{Osservatorio Astronomico di Trieste, via G. B. Tiepolo 11, 34131 Trieste, Italy\label{trieste}}
\newcommand{\iucaa}{Inter-University Centre for Astronomy and Astrophysics, Post Bag 4, Ganeshkhind, 411\,007, Pune, India\label{iucaa}} 
\newcommand{\eso}{European Southern Observatory, Alonso de C\'ordova 3107, Vitacura, Casilla 19001, Santiago 19, Chile\label{eso}}

\hyphenation{ESDLA}
\hyphenation{ESDLAs}
%================================== TITLE =============================================

   \title{VLT/UVES observations of extremely strong intervening damped Lyman-$\alpha$ systems 
\thanks{Based on observations collected 
with the Ultraviolet and Visual Echelle Spectrograph on the Very Large Telescope 
at the European Organisation for Astronomical Research in the Southern Hemisphere, Chile, 
under Programme ID 091.A-0370(A).}}
\subtitle{Molecular hydrogen and excited carbon, oxygen and silicon at $\log N(\HI)=22.4$}
   \author{
        P.~Noterdaeme       \inst{\ref{iap}}
\and    R.~Srianand         \inst{\ref{iucaa}}
\and    H.~Rahmani          \inst{\ref{iran},\ref{lam}}
\and    P.~Petitjean        \inst{\ref{iap}}
\and    I.~P{\^a}ris        \inst{\ref{uchile},\ref{trieste}}
\and    C.~Ledoux           \inst{\ref{eso}}
\and    N. Gupta            \inst{\ref{iucaa}}
\and    S.~L{\'o}pez        \inst{\ref{uchile}}
          }

   \institute{    
\iap\  -- \email{noterdaeme@iap.fr}
\and \iucaa
\and \iran
\and \lam
\and \uchile
\and \trieste
\and \eso
             }

   \date{}

 \abstract{
We present a detailed analysis of three extremely strong intervening damped Lyman-$\alpha$ systems 
(ESDLAs, with $\log N(\HI) \ge 21.7$) observed towards quasars with the Ultraviolet and Visual Echelle 
Spectrograph on the Very Large Telescope. We measure overall metallicities of $[$Zn/H$]\,\sim$\,-1.2, 
-1.3 and -0.7 at respectively $\zabs$=2.34 towards SDSS\,J214043.02$-$032139.2
($\log N(\HI)\,=22.4\,\pm\,0.1$), 
$\zabs$=3.35 towards SDSS\,J145646.48$+$160939.3 ($\log N(\HI)\,=\,21.7\,\pm\,0.1$) and $\zabs$=2.25 towards 
SDSS\,J015445.22$+$193515.8 ($\log N(\HI)\,=\,21.75\,\pm\,0.15$).
Iron depletion of about a factor 15 compared to volatile elements is seen in the DLA towards \jt, while the 
other two show deletion typical of known DLAs. 
We detect H$_2$ towards \jt\ 
($\log N($H$_2)\,=\,20.13\,\pm\,0.07$) and \jf\ ($\log N($H$_2)\,=\,17.10\,\pm\,0.09$) and argue for 
a tentative detection towards \jz. 

Absorption from the excited fine-structure levels of \OI, \CI\ and \SiII\ are detected in 
the system towards \jt, that has the largest \HI\ column density detected so far in an intervening DLA. 
This is the first detection of \OI\ fine-structure lines in a QSO-DLA, that also provides us a 
rare possibility to study the chemical abundances of less abundant atoms like Co and Ge.
Simple single phase photo-ionisation models fail to reproduce all the observed quantities. Instead, we suggest that 
the cloud has a stratified structure: H$_2$ and \CI\ 
likely originate from both a dense ($\log n_{\rm H} \sim 2.5-3$) cold (80\,K) and warm (250\,K) phase 
containing a fraction of the total \HI\ while a warmer ($T>1000$~K) phase probably contributes significantly to the high 
excitation of \OI\ fine-structure levels. The observed \CI/H$_2$ column density ratio is surprisingly low compared 
to model predictions and we do not detect CO molecules: this suggests a possible underabundance of C by 
0.7~dex compared to other alpha elements. 
The absorber could be a photo-dissociation region close to a bright star (or a star cluster) 
where higher temperature occurs in the illuminated region. Direct detection of on-going star formation 
through e.g. NIR emission lines in the surrounding of the gas would enable a detailed physical modelling 
of the system.
}
 
   \maketitle
%________________________________________________________________

\section{Introduction}
%QSO ALS
Diffuse gaseous clouds in the Universe are primarily described by their neutral hydrogen 
column density, which is a quantity directly and accurately measurable in absorption against 
background sources. The study of \lya\ absorption lines in quasar spectra 
currently allows us to detect \HI\ 
over ten orders of magnitude in column density,  
from $N(\HI)\,\sim\,10^{12}$ to $10^{22}$~\cmsq, and over a wide range of redshifts. 

Several observations at low redshift show that clouds with $\log N(\HI)>15$  tend to cluster
around galaxies \citep[e.g.][]{Tripp98, Bowen02, Prochaska11}. 
In particular, Lyman limit systems (LLS), with $\log N(\HI)\sim 17-18$, 
likely trace cool circumgalactic environments, with a possible bimodality in metallicity 
that would represent outflows of metal-rich gas and accretion of metal-poor gas by galaxies \citep{Lehner13}. 
An inflexion in the \HI\ column density distribution function is 
seen at $\log N(\HI)\sim 19$, due to the onset of self-shielding \citep{Petitjean92}. When the column 
density reaches $\log N(\HI)\sim 20$, the gas is mostly neutral and produces strong 
\lya\ absorption with prominent Lorentzian wings. The corresponding absorbers are 
called 
Damped Lyman-$\alpha$ systems (DLAs) when $\log N(\HI)\ge 20.3$ \citep[][]{Wolfe86}. 
These systems are thought to arise when the line of sight passes at small impact parameter to a 
foreground galaxy. 
Early evidence for this came 
mostly from the similarity with \HI\ column densities in nearby galactic discs and the 
presence of metals at different levels of chemical enrichment: the smoking gun of star formation activity. 
Interestingly, this $N(\HI)$-threshold also corresponds to the value above which Galactic  
gas clouds have high fraction of cold neutral medium \citep{Kanekar11}, a prerequisite for star formation. 
Naively, one can assume that  
the higher the \HI\ column density, the closer the association between the absorbing gas and a galaxy.

High redshift DLAs have been intensively studied \citep[see][for a review]{Wolfe05}, 
with the main scope to put constraints on the formation and evolution of galaxies 
in the early Universe as well as understanding the links between star-formation 
activity and the distribution, physical and chemical state of the gas.  
Systematic surveys for DLAs yielded a measurement the cosmological density of 
neutral gas \citep[e.g.][]{Prochaska09a, Noterdaeme09dla}, which is a key ingredient 
to constrain hydrodynamical simulations of galaxy formation \citep[][]{Pontzen08, Altay13} 
and the understanding of gas consumption by star formation as well as feedback effects. 
Because metals are released in the ISM by the stars with different time-scales, studies of 
both the overall metal abundances \citep[e.g.][]{Rao05,Rafelski12} and the abundance ratios 
\citep[e.g.][]{Petitjean08, Pettini08,Zafar14} constrains the star-formation history. 
It is also possible to 
derive the physical state of the gas from the excitation of atomic and molecular 
species. In particular, the surrounding UV field can be estimated, thereby constraining 
the instantaneous star-formation rates \citep[e.g.][]{Noterdaeme07,Srianand05, Wolfe04}. 
Overall, DLAs are a very important probe of various physical processes that are at play in 
the formation of galaxies \citep[e.g.][]{Bird14}.

A steepening of the $N(\HI)$-frequency distribution at the very high column 
density end has long been discussed. It is possible that at least a fraction of 
high column density systems (in particular those with high-metallicity)  
are missed in magnitude-limited QSO surveys \citep[e.g.][]{Boisse98}. However, radio-selected DLA 
samples appeared not to differ significantly from optically-selected ones \citep[e.g.][]{Ellison05,Jorgenson06}, 
at least in terms of \HI-content. 
Another 
explanation, proposed by \citet[][]{Schaye01}, is that the high column density gas collapses 
into molecular hydrogen eventually leading to star formation. 
Somehow paradoxically, this important mechanism at the heart of galaxy formation is observationally 
supported by the absence of high column densities, molecular gas and in-situ star-formation in 
DLA samples \citep[see ][]{Krumholz09}. It is therefore important to directly search for these three 
ingredients and study the relation between them.

%molecular gas
Molecular gas has a very small covering fraction, making it unlikely to be found along a given random 
line of sight \citep{Zwaan06}. 
Nevertheless, efficient searches based on the presence of strong neutral carbon lines 
-- which is an indicator of cold gas -- have allowed us to detect for the 
first time translucent clouds (the regime in-between diffuse and dense molecular clouds, see e.g. 
\citealt{Snow06}) in high redshift DLAs 
\citep{Srianand08, Noterdaeme09co,Noterdaeme10co,Noterdaeme11}. While 
the $N(\HI)$-distribution of \CI-selected absorbers is found to be flatter than that of 
regular DLAs (i.e. \HI-selected), the former population is still dominated by 
low-$N(\HI)$ and the current sample (66 \CI\ systems, \citealt{Ledoux14}) does not 
contain very large \HI\ column densities. The strong 
reddening by these systems also reinstates the possibility of a dust bias against the cold phase.
Finally, 
\CI\ systems frequently present the 2175\,{\AA} absorption feature, although with a reduced strength 
compared to 
Milky Way lines of sight, possibly because of strong UV field originating from in-situ star-formation.

%high-N gas.
The detection of very high column density DLAs is possible through searching large datasets 
of quasar spectra. 
Using the Sloan Digital Sky Survey \citep[SDSS,][]{York00}, \citet{Noterdaeme09dla} 
showed that the slope of the $N(\HI)$ distribution at the high end is actually shallower 
than previously thought, 
with the first detection of a QSO-DLA reaching $\log N(\HI)=22$ \citep[in which H$_2$ is detected, ][]{Guimaraes12}. 
The tremendous increase of available quasar spectra \citep[][]{Paris12,Paris14} in the SDSS-III 
Baryon Oscillation Spectroscopic Survey \citep[BOSS,][]{Dawson13} allowed 
to extend the study to even higher column densities \citep{Noterdaeme12c},  
with a slope similar to what is seen from opacity-corrected 
21-cm maps in the local Universe \citep{Braun12}.  
The analysis of stacked spectra showed that extremely strong DLAs (ESDLAs, with $\log N(\HI) \ge 21.7$) are 
likely to arise from galaxies at small impact parameters ($b<$ a few kpc) with average star 
formation rates of the order of 2\,\msyr\ \citep{Noterdaeme14}. 
X-shooter observations of a system with $\log N(\HI)=22.10$ towards the quasar \jonzelong\ resulted in the 
detection of a star-forming galaxy at $b<$~1kpc from the quasar line of sight \citep[][see also \citealt{Kulkarni12}]{Noterdaeme12a}, 
making it possible the first mass determination of a DLA galaxy by weak lensing \citep{Grillo14}.

Here, we search for molecular hydrogen in extremely strong DLAs, using high-resolution spectroscopic 
observations. 
We present the observations and data reduction in Sect.~\ref{s:obs} and the abundance measurements 
in Sect.~\ref{s:ab}. 
We then investigate the chemical and physical conditions in the highest $N(\HI)$ QSO-DLA detected so far 
in Sect.~\ref{s:phys}. We summarise and conclude in Sect.~\ref{s:conclusion}.

\section{Observations and data reduction \label{s:obs}}
The three quasars were observed in service mode 
with the Ultraviolet and Visual Echelle Spectrograph 
\citep[UVES,][]{Dekker00} mounted on the Very Large Telescope UT2 of 
the European Southern Observatory under Programme ID 091.A-0370(A). 
The total 15\,600~s of science exposure of each object was divided into four 
exposures of 3900\,s each. 
All observations of SDSS\,J214043.02$-$032139.2 ($z_{\rm em}=2.48$, hereafter \jt) 
and SDSS\,J015445.22$+$193515.8 ($z_{\rm em}=2.53$, hereafter \jz) were performed using the 
standard beam splitter in 390$+$580 setting that covers roughly from 330 to 450 nm 
on the blue CCD and from 465 to 578 nm and 583 to 680 nm on the two red CCDs. 
Observations of SDSS\,J145646.48$+$160939.3 ($z_{\rm em}=3.68$, hereafter \jf) included three 
exposures using the 390$+$760 setting that roughly 
covers wavelength ranges of 567 to 754~nm and 763 to 946~nm for the two red CCDs along with 
an exposure in 390$+$580 setting. 
The slit width of 1 arcsec and a CCD readout with 2$\times$2 binning used for all the  
observations resulted in a spectral resolution power $R \approx 48\,000$ dispersed on pixels of 
$\sim$~2.0 km/s.
The log of the observations is provided in Table~\ref{t:obs}.

\begin{table*}[!ht]
\centering
\caption{Log of VLT/UVES observations \label{t:obs}}
\begin{tabular}{c c c c c c}
\hline \hline
QSO  & Observing dates  & Setting    & Exposure time       & Seeing (\arcsec) & Airmass \\
\hline
\jtl\                 & 15-06-2013/02-07-2013                 & 390$+$580  &  4 $\times$ 3900\,s  & 0.7-0.8         & 1.1       \\
\hline
\multirow{2}{*}{\jfl} & 06-05-2013/12-05-2013                 & 390$+$760  &  3 $\times$ 3900\,s  & 0.9             & 1.3-1.4       \\
                     & 07-06-2013                            & 390$+$580  &  1 $\times$ 3900\,s           & 1.0             & 1.3       \\
\hline
\jzl                  & 01-09-2013/04-09-2013/05-09-2013      & 390$+$580  &  4 $\times$  3900\,s & 0.8-1.0     & 1.4       \\ 
\hline
\end{tabular} 
\end{table*}
%Seeing quoted is TEL IA FWHM

We reduced the raw frames to extract the 1D spectra using the 
UVES Common Pipeline Library (CPL) data reduction 
pipeline release 6.3\footnote{\url{http://www.eso.org/sci/facilities/paranal/instruments/uves/doc/}} 
by implementing the optimal extraction. Polynomials of $\rm 4^{th}$ order were used to find the 
wavelength solution and the rms calibration error was found to be always less than 100 m\,s$^{-1}$.  
We shifted the individual science exposures to the heliocentric-vacuum frame correcting for 
the observatory's motion towards the line of sight at the exposure mid point and using the 
air-to-vacuum relation from \citet{Ciddor96}. We then rebinned individual spectra  
to a common wavelength array, scaled and combined them altogether using weights 
estimated from the errors in each pixel while rejecting bad pixels and 
cosmic-ray impacts at the same time. In the following, abundances are expressed relative 
to solar, i.e. [X/H]~$\equiv \log N(X)/N(H) - \log (X/H)_{\rm \odot}$, where we adopted 
the solar abundances of \citet{Asplund09}.

\section{Abundance measurements\label{s:ab}}

\subsection{Neutral hydrogen}

\begin{figure}
\centering
\includegraphics[bb=190 13 394 755,clip=,angle=90,width=\hsize]{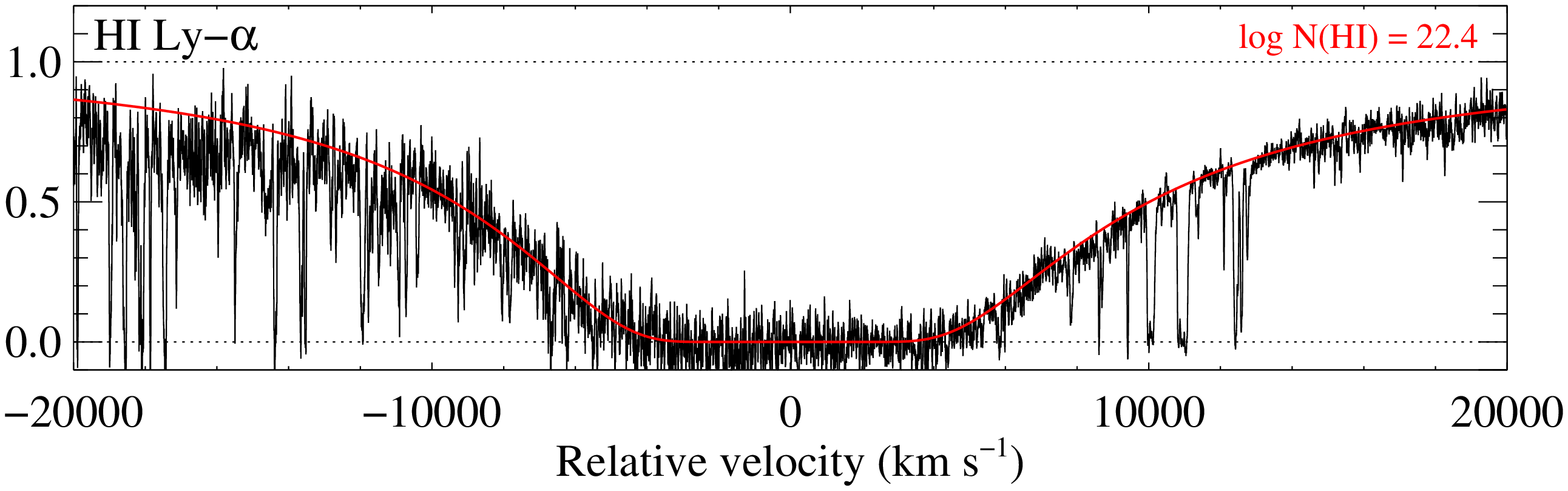}
\includegraphics[bb=164 13 394 755,clip=,angle=90,width=\hsize]{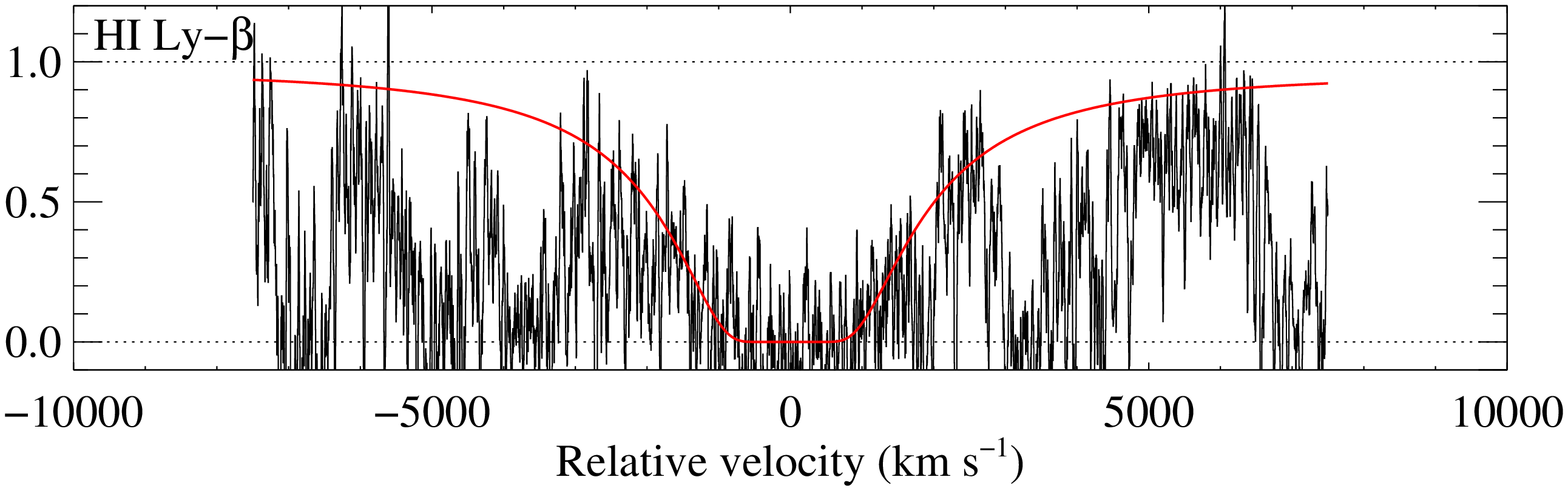}
\caption{Voigt-profile fitting to the damped \lya\ (top) and \lyb\ lines at $\zabs=2.34$ 
towards \jt. The normalised spectrum has been boxcar-smoothed by 10 pixels for presentation only. 
We note that here, and in all figures representing Voigt-profile fits to the data, 
the y-axis correspond to the normalised flux (unitless).
\label{2140:f:dla}}
\end{figure}

The DLA at $\zabs=2.34$ towards J\,2140$-$0321 presents a very 
wide dark core and extended wings that decrease the quasar continuum 
over several hundred angstroms. \citet{Noterdaeme14} used a principal 
component analysis \citep[see][]{Paris11} on the low-resolution, flux-calibrated 
BOSS spectrum to get a first estimate of the unabsorbed quasar continuum, 
including the \lya\ emission line. 
They derived $\log N(\HI)=22.35$ by fitting the damped \lya\ line. 
Fitting both the damped \lya\ and \lyb\ absorption lines in the UVES spectrum 
together with the quasar continuum, we derive $\log N(\HI)=22.40 \pm 0.10$, see 
Fig.~\ref{2140:f:dla}. 
This represents the highest column density ever 
found in a high-$z$ intervening cloud. 

Similarly, we derive the \HI\ column densities in the DLAs at 
$\zabs=3.35$ towards J\,1456$+$1609 and $\zabs=2.155$ towards 
J\,0154$+$1935.  For J\,1456$+$1609, we use the \lya, \lyb\ and 
\lyg\ to constrain the fit while only \lya\ is covered with sufficient 
signal-to-noise ratio for J\,0154$+$1935. We obtain $\log N(\HI)=21.70 \pm 0.10$ 
and $21.75 \pm 0.15$, respectively. These values are also in good agreement with 
the respective values of $\log N(\HI)=21.85$ and 21.77 derived 
from the BOSS data.

\begin{figure}
\centering
\addtolength{\tabcolsep}{-6pt}
\begin{tabular}{cc}
\multicolumn{2}{c}{\includegraphics[bb=190 13 394 755,clip=,angle=90,width=\hsize]{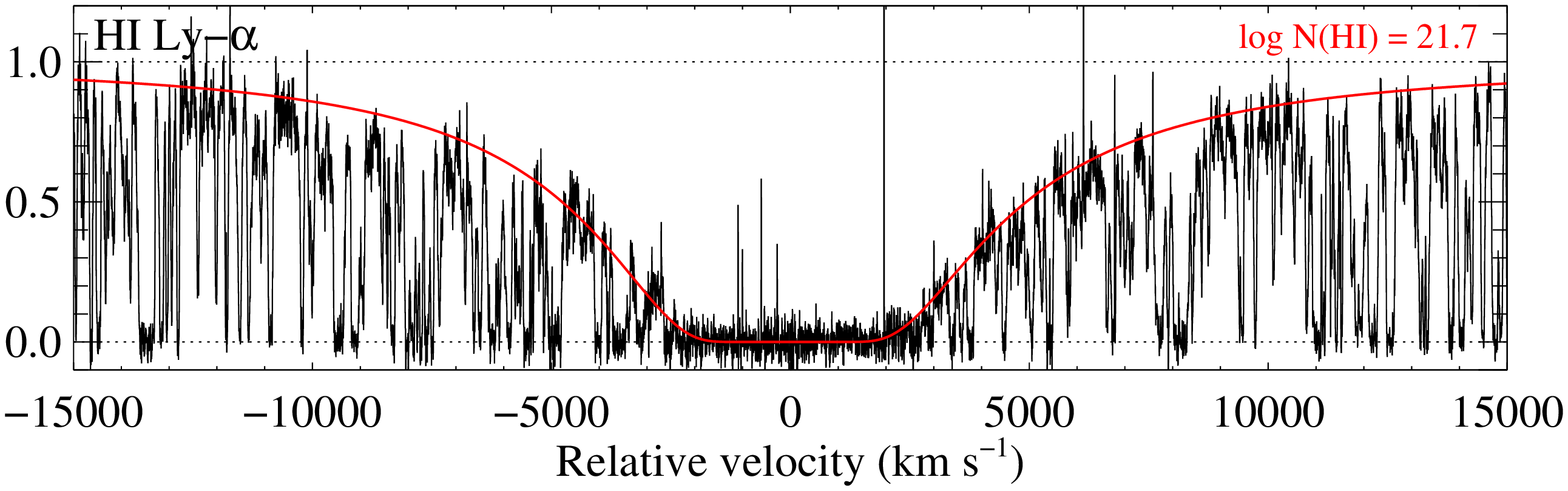}} \\
\includegraphics[bb=164 220 394 617,clip=,angle=90,width=0.5\hsize]{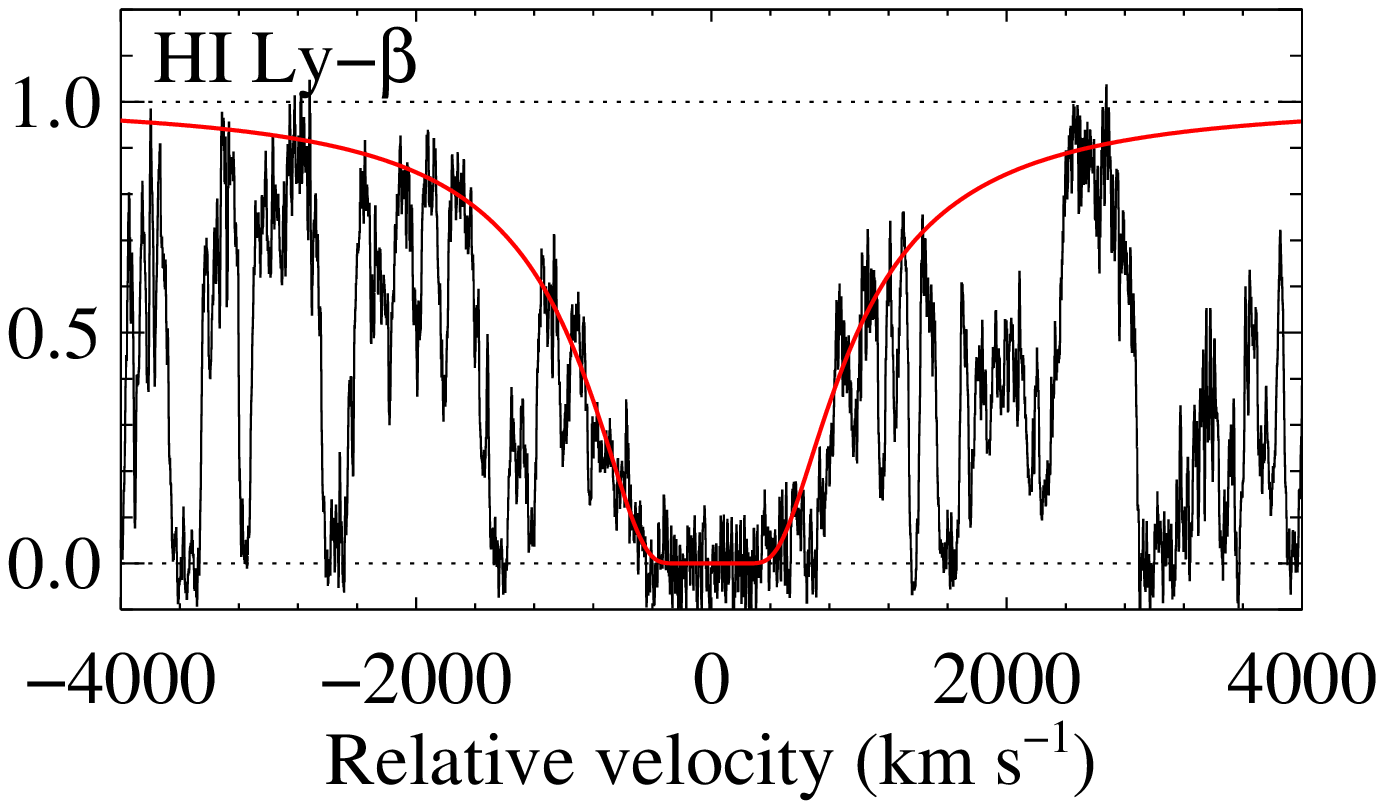} & 
\includegraphics[bb=164 220 394 617,clip=,angle=90,width=0.5\hsize]{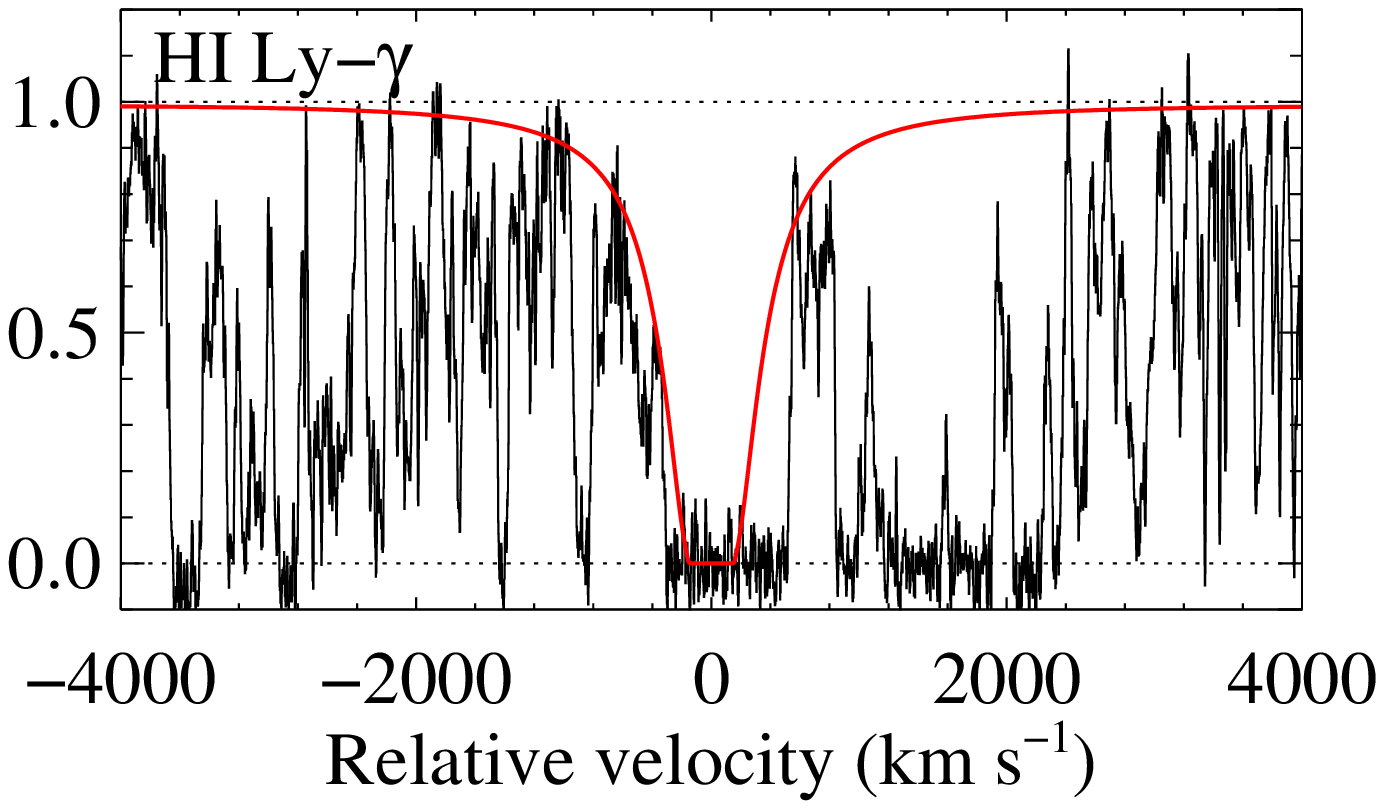} \\
\end{tabular}
\addtolength{\tabcolsep}{6pt}
\caption{Same as Fig.~\ref{2140:f:dla} for the DLA towards \jf. Boxcar-smoothed by 5 pixels for presentation only. \label{1456:f:dla}}
\end{figure}

\begin{figure}
\centering
\includegraphics[bb=164 13 394 755,clip=,angle=90,width=\hsize]{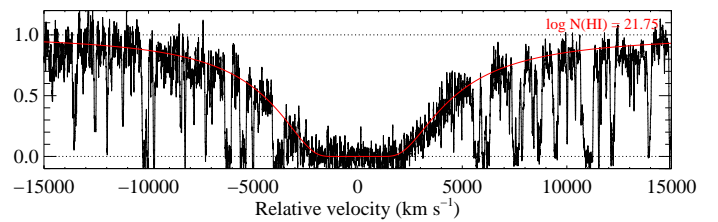}
\caption{Same as Fig.~\ref{2140:f:dla} for the DLA towards \jz. Boxcar-smoothed by 10 pixels for presentation only.\label{0154:f:dla}}
\end{figure}

\subsection{Metallicity and dust depletion}

\begin{figure*}[!ht]
\centering
 \addtolength{\tabcolsep}{-5pt}
\begin{tabular}{ccccc}
\jt, $\zabs=2.33995$ & & \jf, $\zabs=3.35183$ & & \jz, $\zabs=2.25155$ \\
 \begin{tabular}{cc}
 \includegraphics[bb=219 228 393 617,clip=,angle=90,width=0.15\hsize]{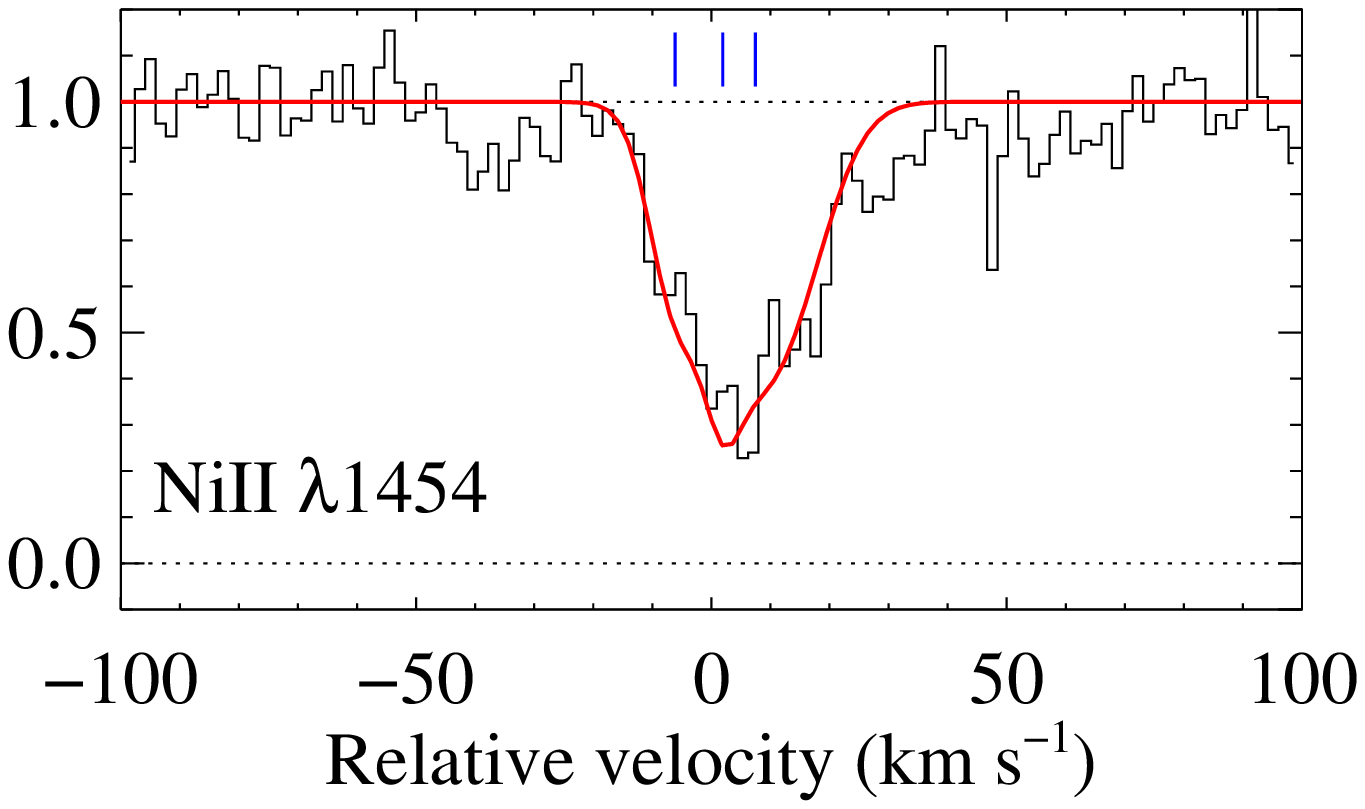}&
 \includegraphics[bb=219 228 393 617,clip=,angle=90,width=0.15\hsize]{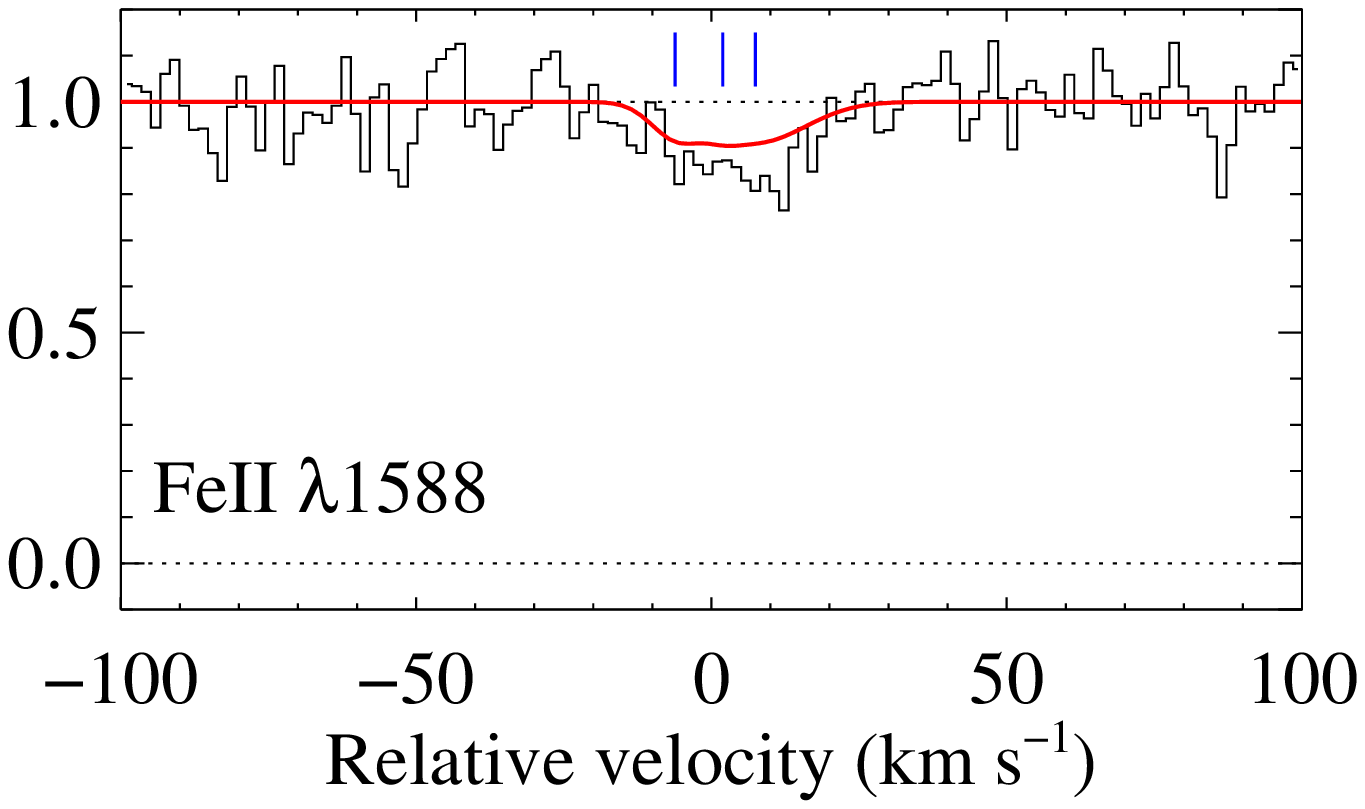}\\
 \includegraphics[bb=219 228 393 617,clip=,angle=90,width=0.15\hsize]{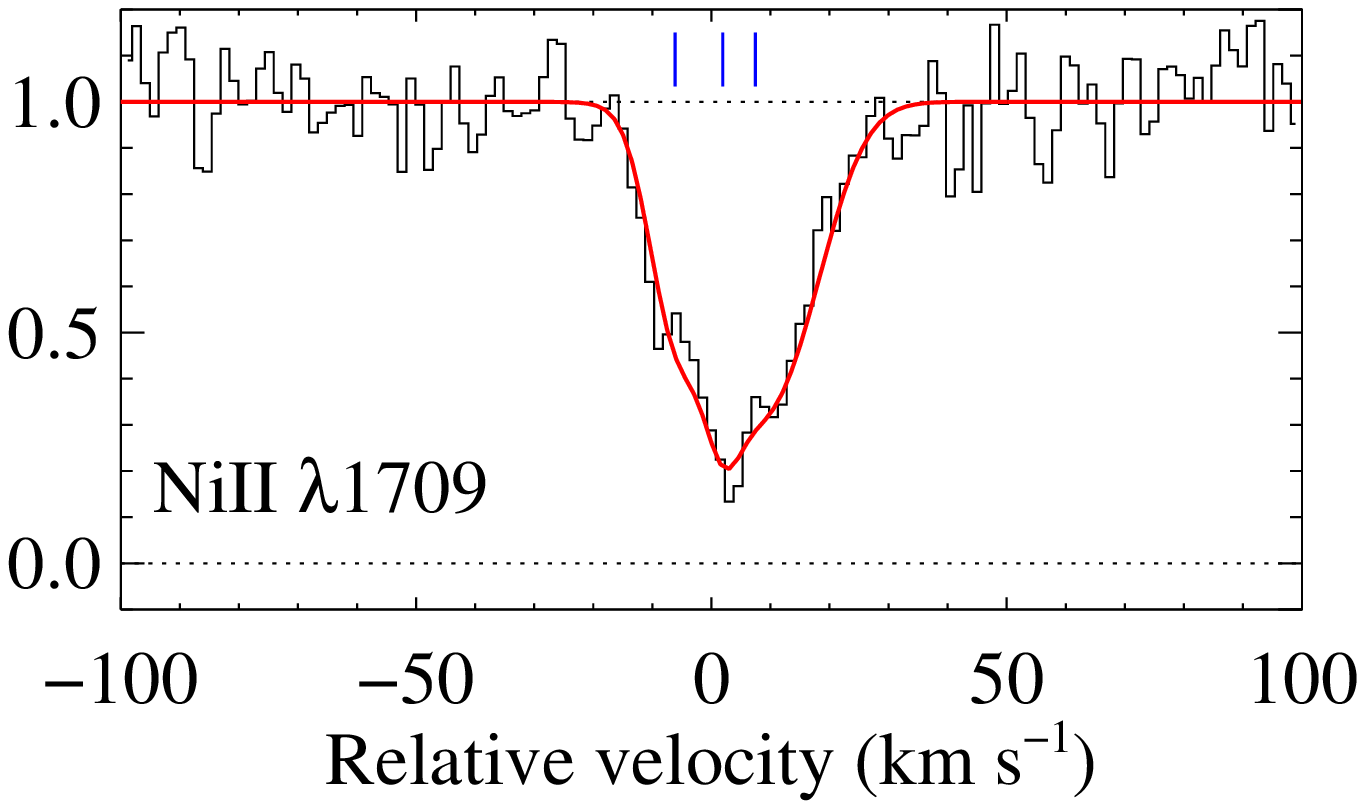}&
 \includegraphics[bb=219 228 393 617,clip=,angle=90,width=0.15\hsize]{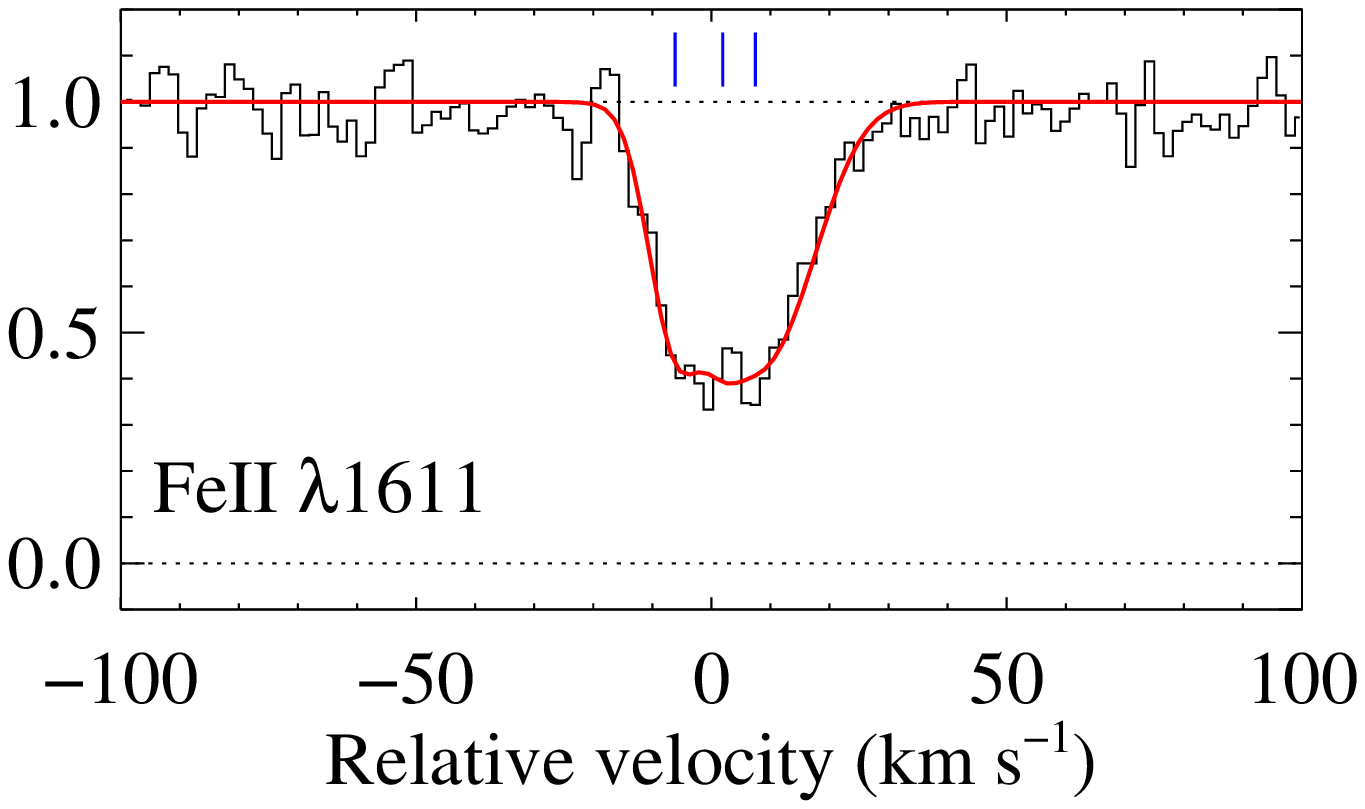}\\
 \includegraphics[bb=219 228 393 617,clip=,angle=90,width=0.15\hsize]{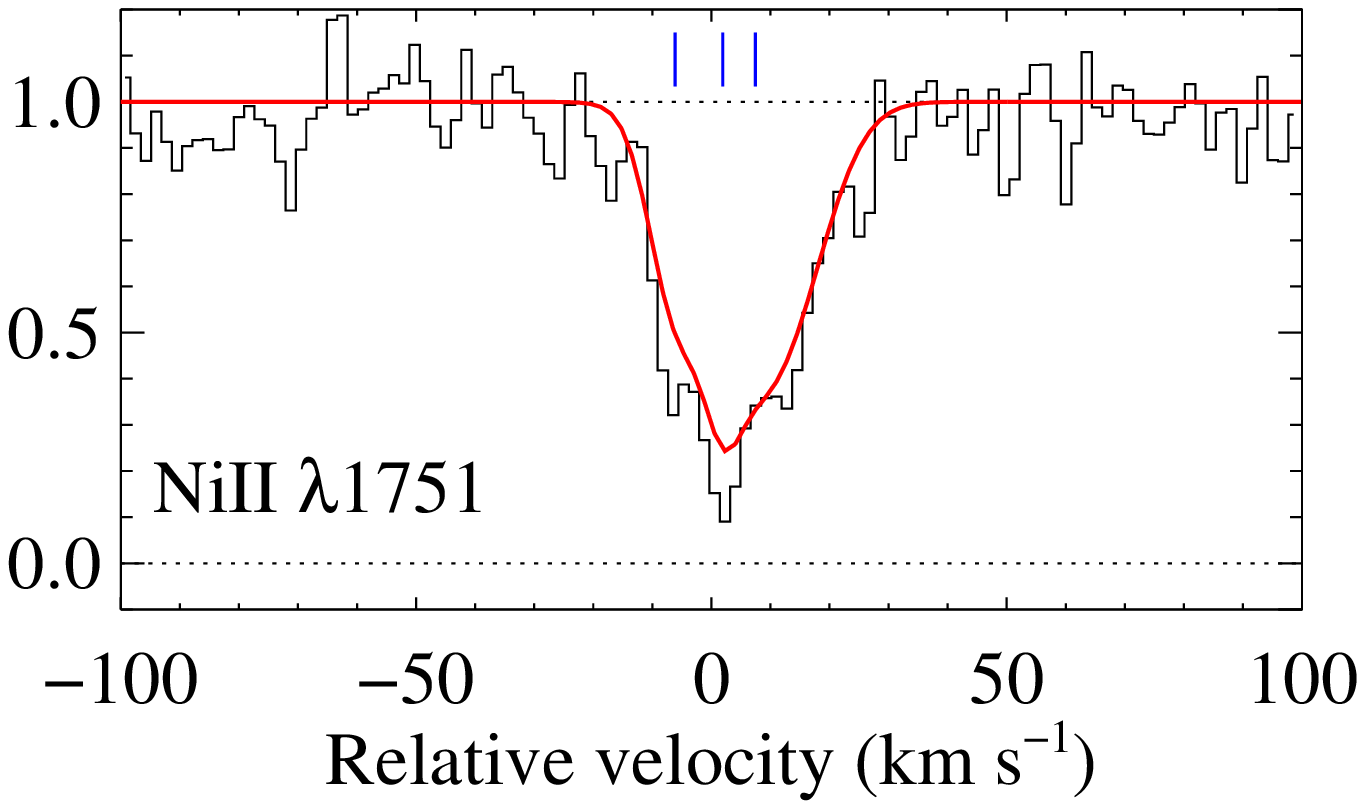}&
 \includegraphics[bb=219 228 393 617,clip=,angle=90,width=0.15\hsize]{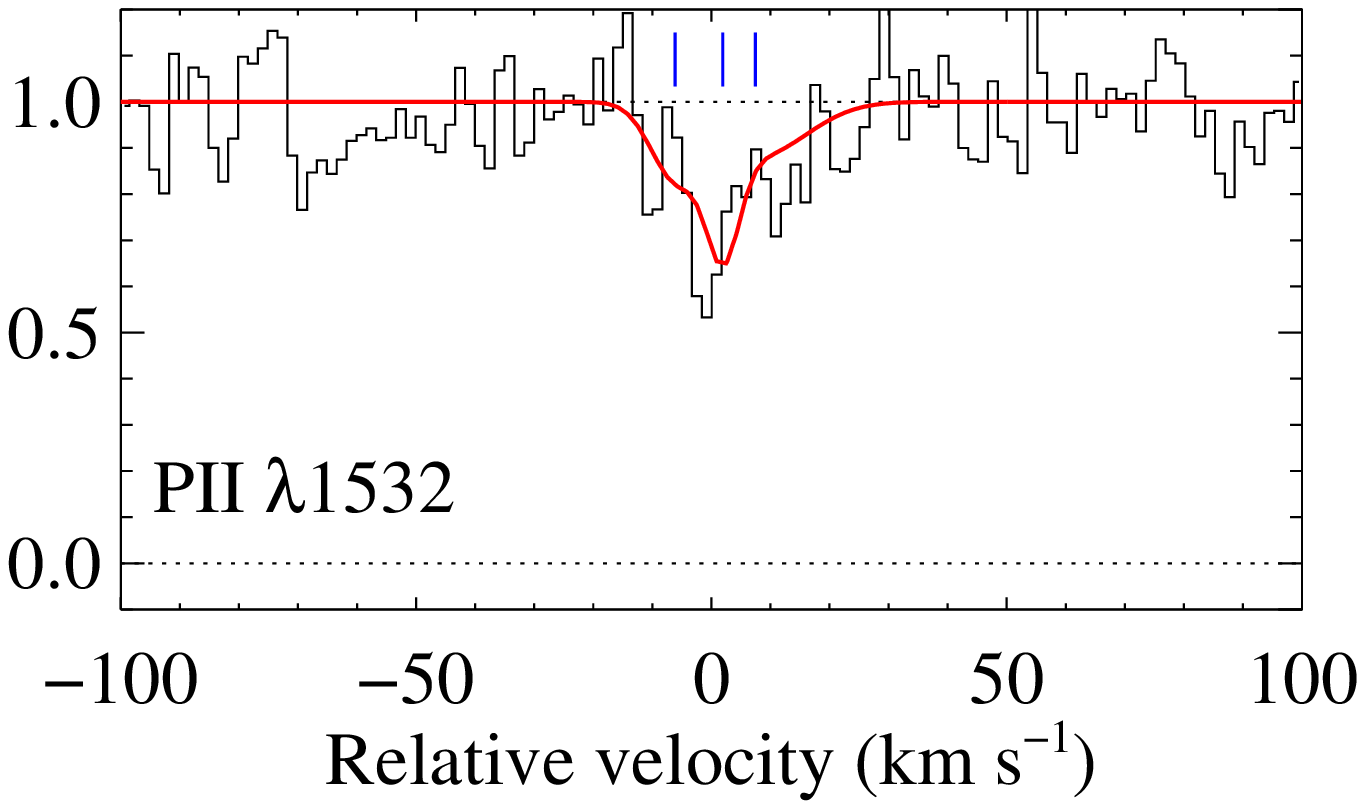}\\
 \includegraphics[bb=219 228 393 617,clip=,angle=90,width=0.15\hsize]{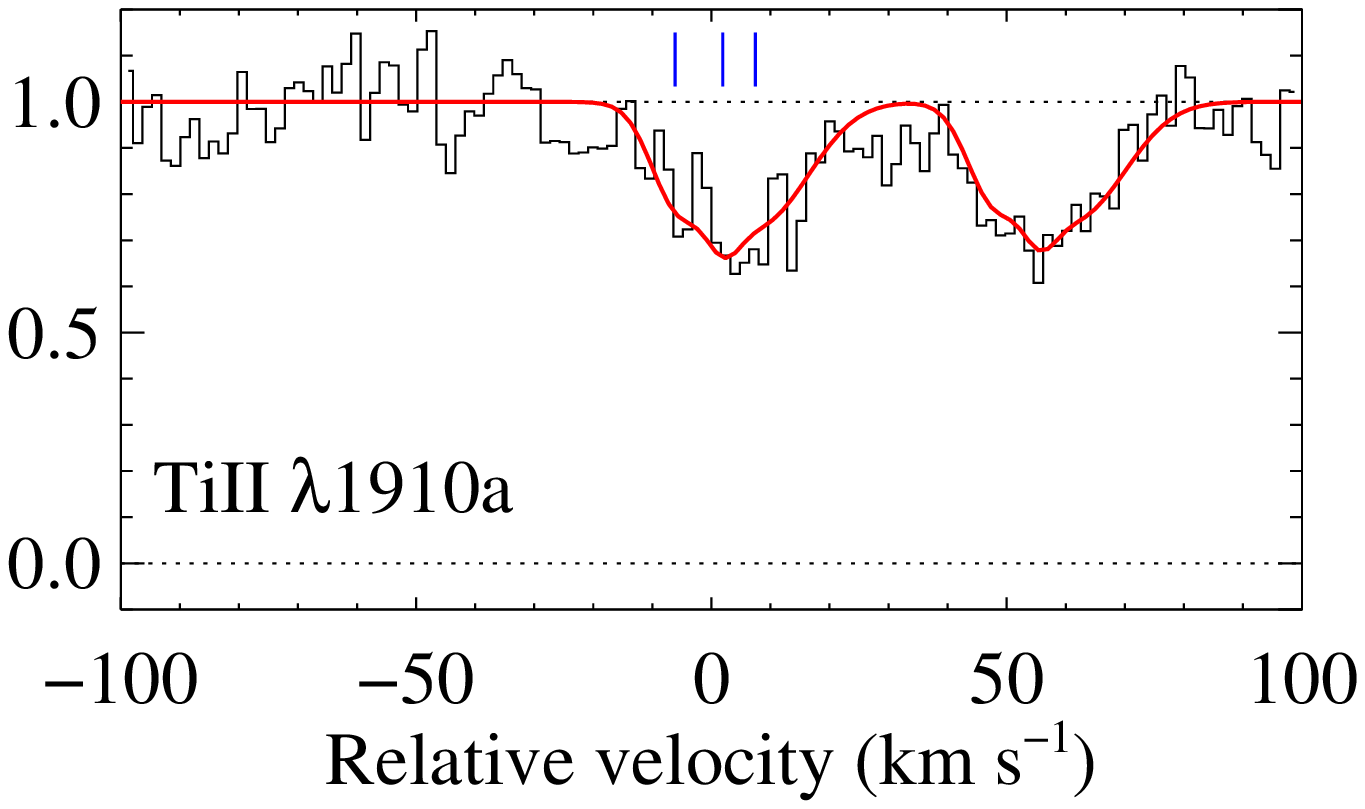}&
 \includegraphics[bb=219 228 393 617,clip=,angle=90,width=0.15\hsize]{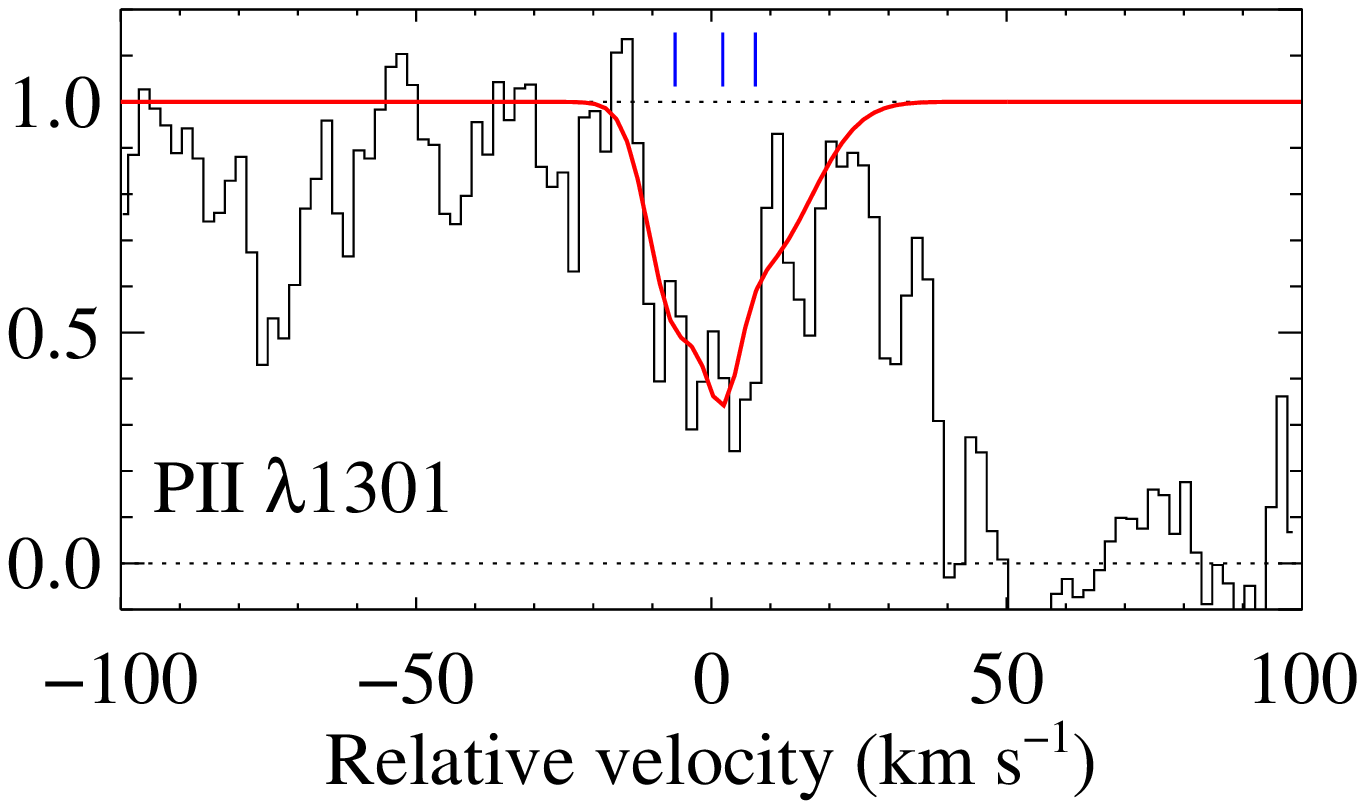}\\
 \includegraphics[bb=164 228 393 617,clip=,angle=90,width=0.15\hsize]{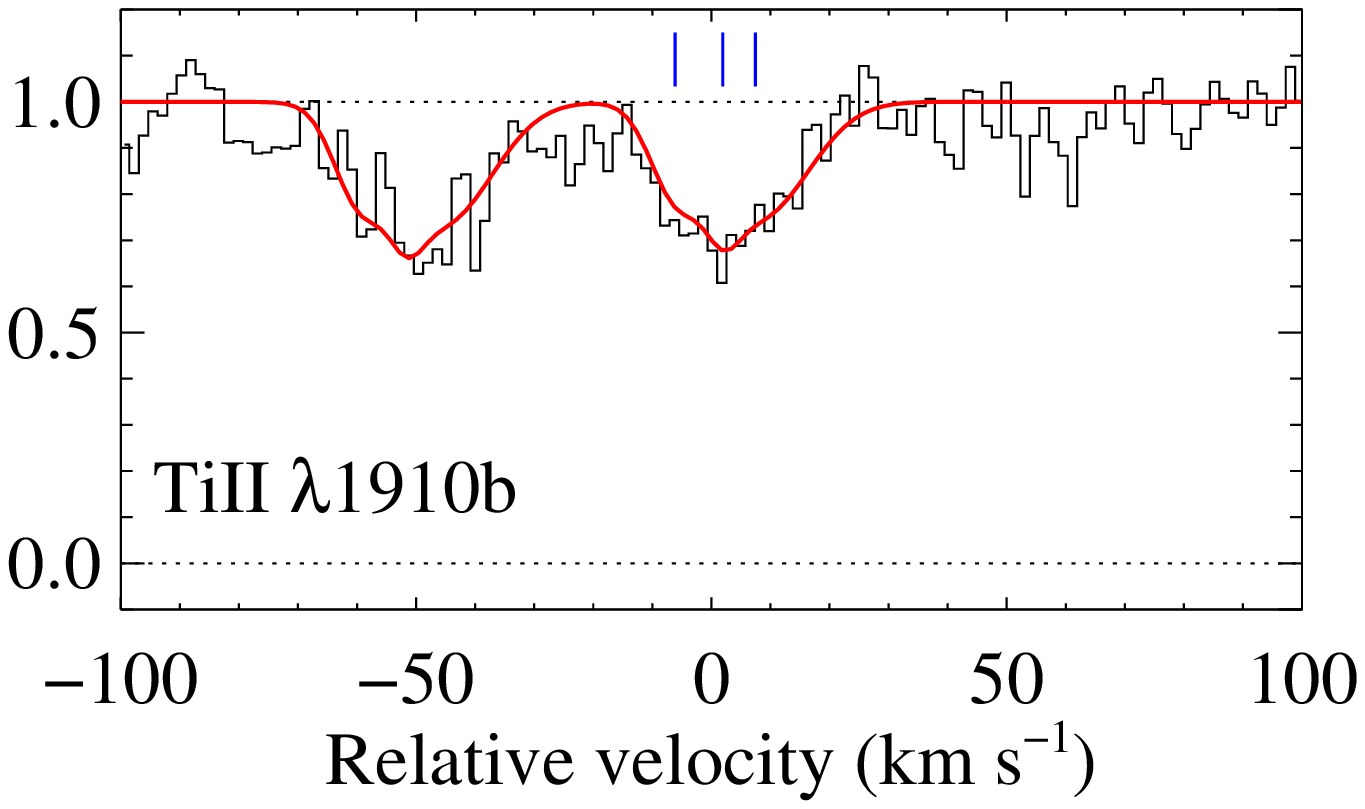}&
 \includegraphics[bb=164 228 393 617,clip=,angle=90,width=0.15\hsize]{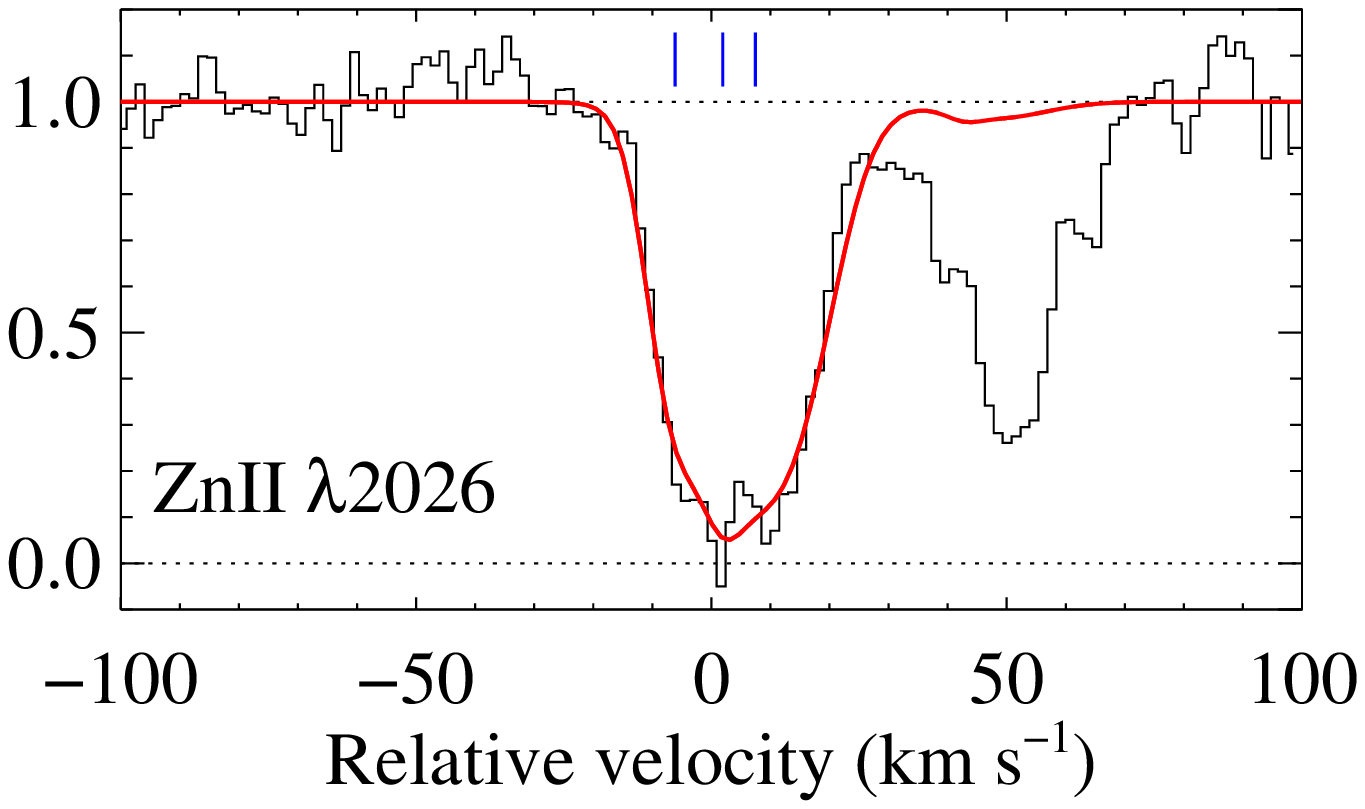}\\
 \end{tabular} & ~~ ~~&

 \begin{tabular}{cc}
 \includegraphics[bb=219 228 393 617,clip=,angle=90,width=0.15\hsize]{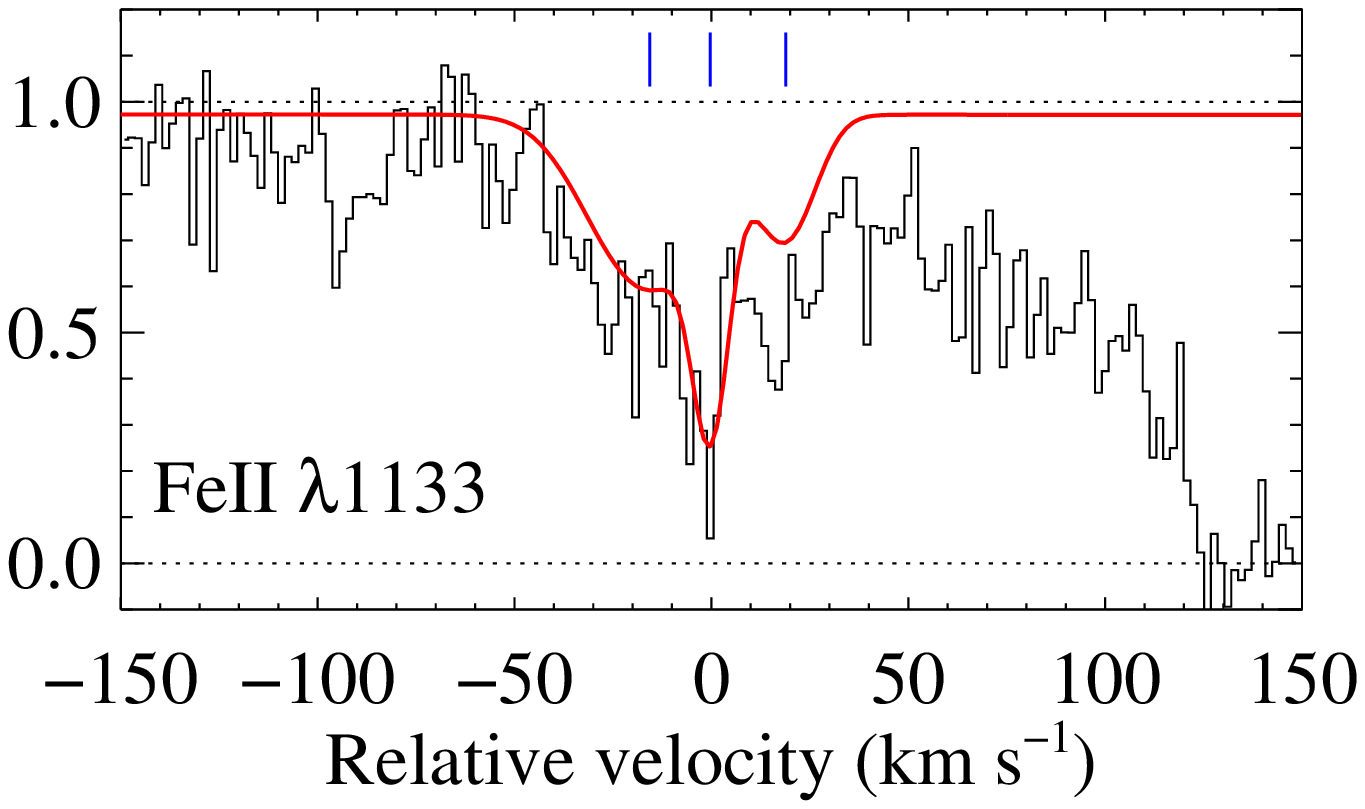} & 
 \includegraphics[bb=219 228 393 617,clip=,angle=90,width=0.15\hsize]{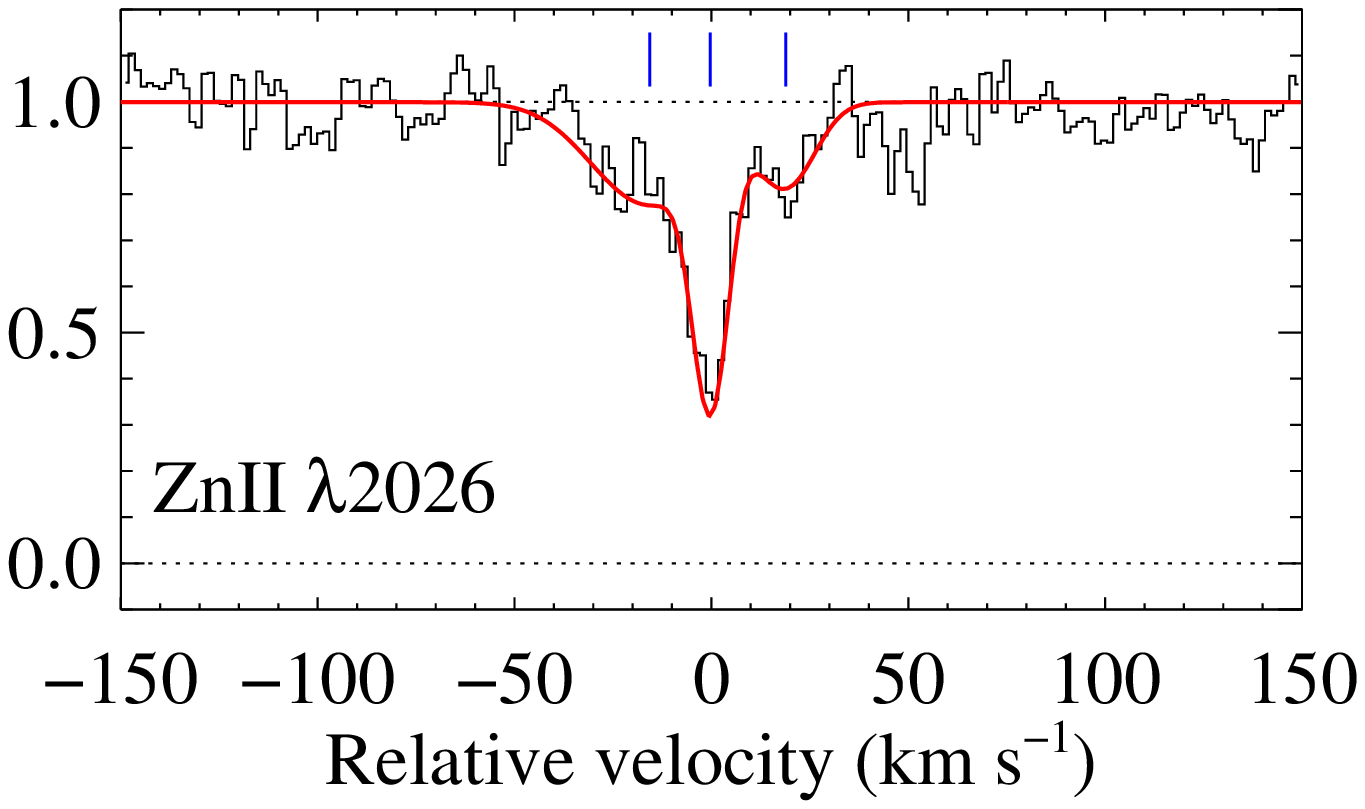} \\
 \includegraphics[bb=219 228 393 617,clip=,angle=90,width=0.15\hsize]{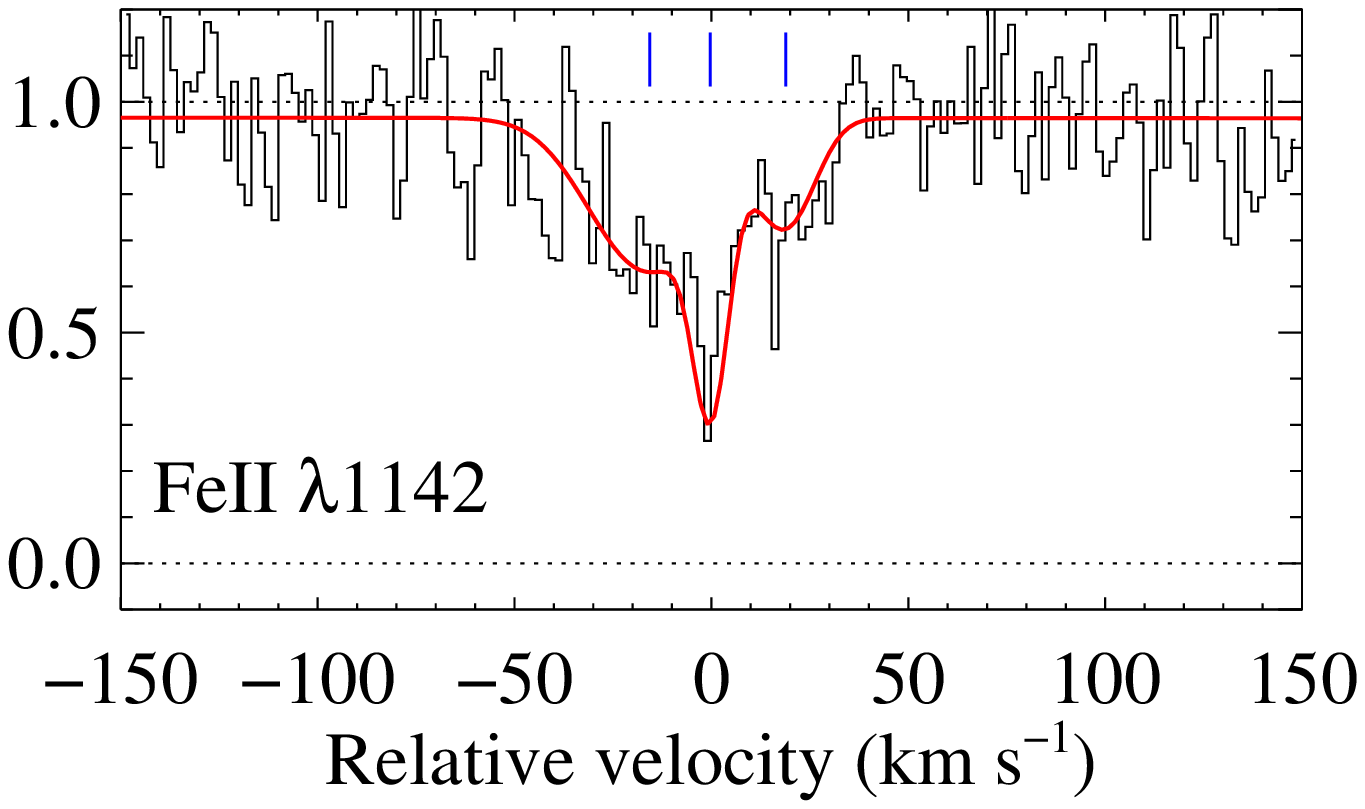} &
 \includegraphics[bb=219 228 393 617,clip=,angle=90,width=0.15\hsize]{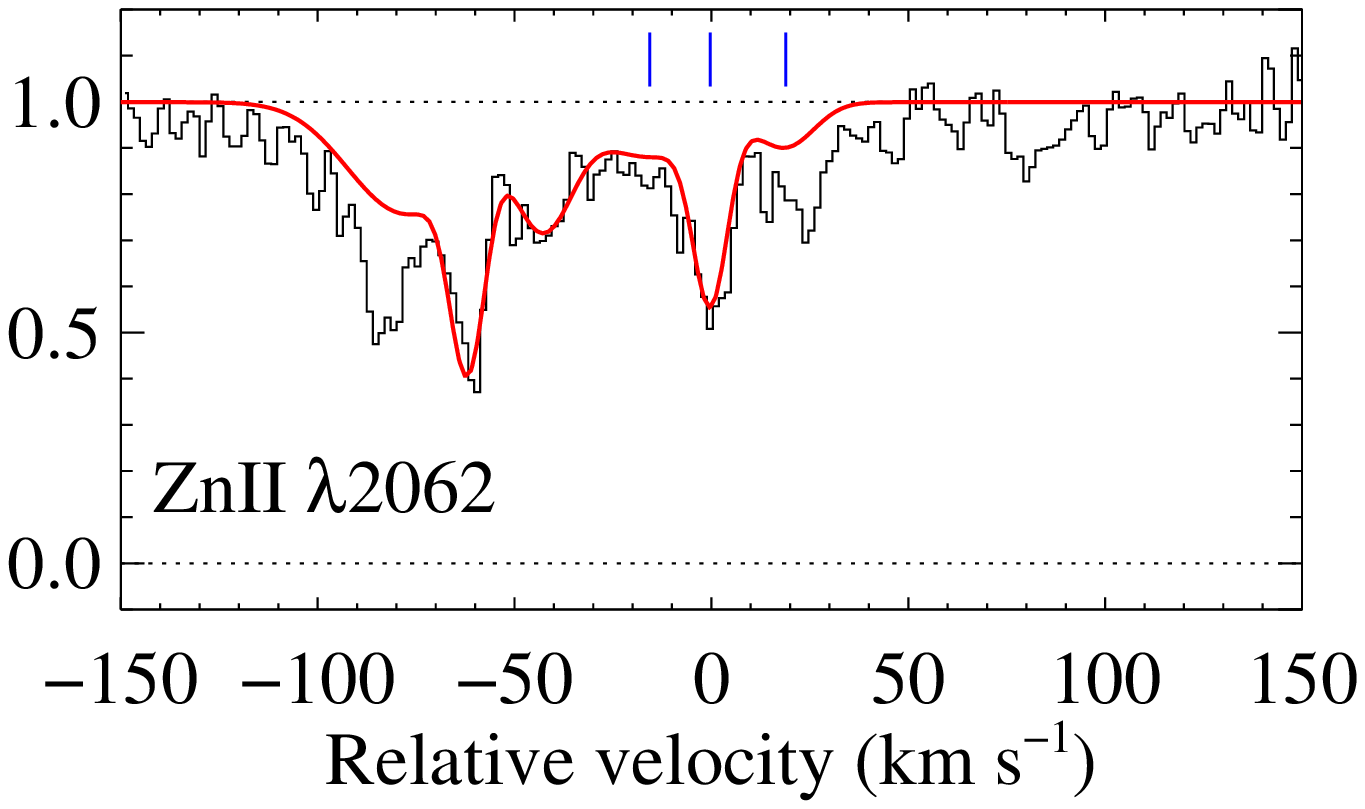} \\
 \includegraphics[bb=219 228 393 617,clip=,angle=90,width=0.15\hsize]{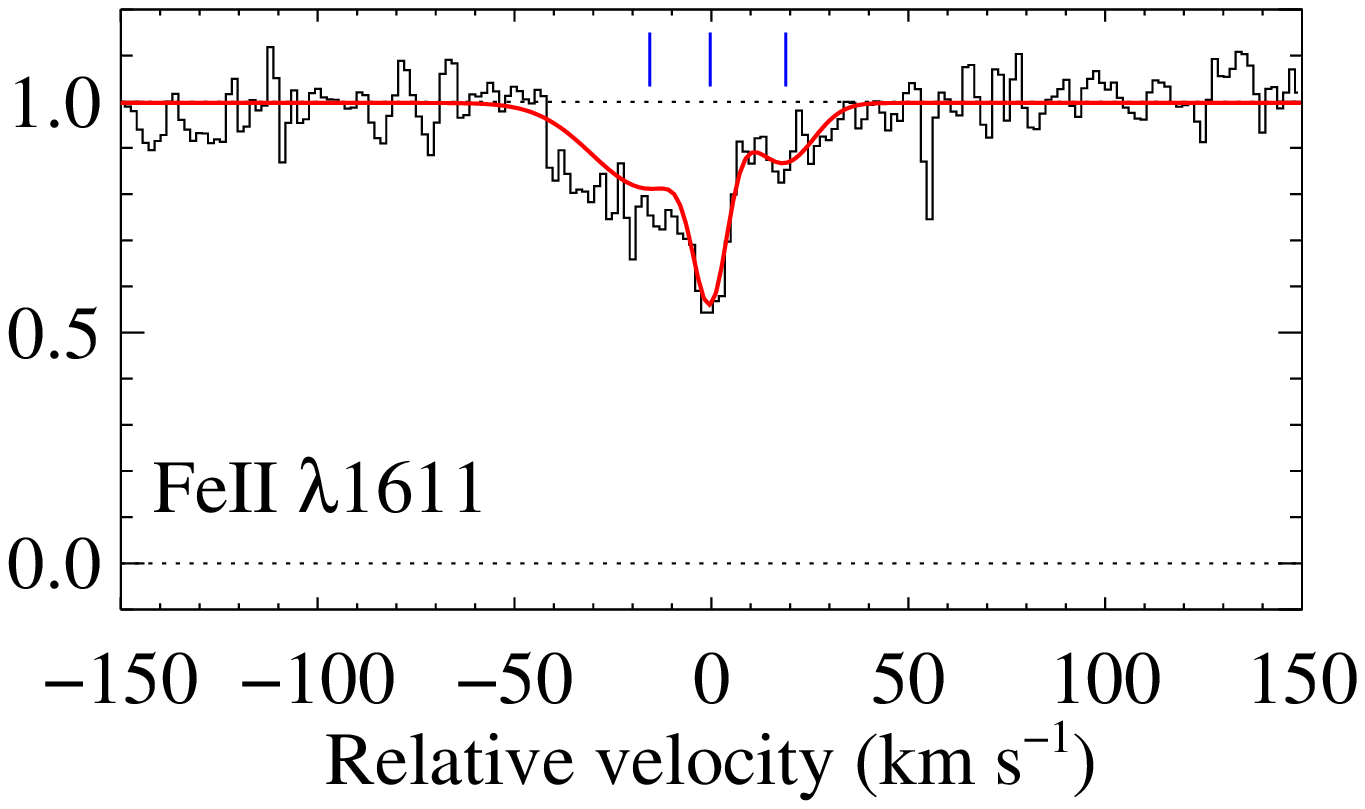} &
 \includegraphics[bb=219 228 393 617,clip=,angle=90,width=0.15\hsize]{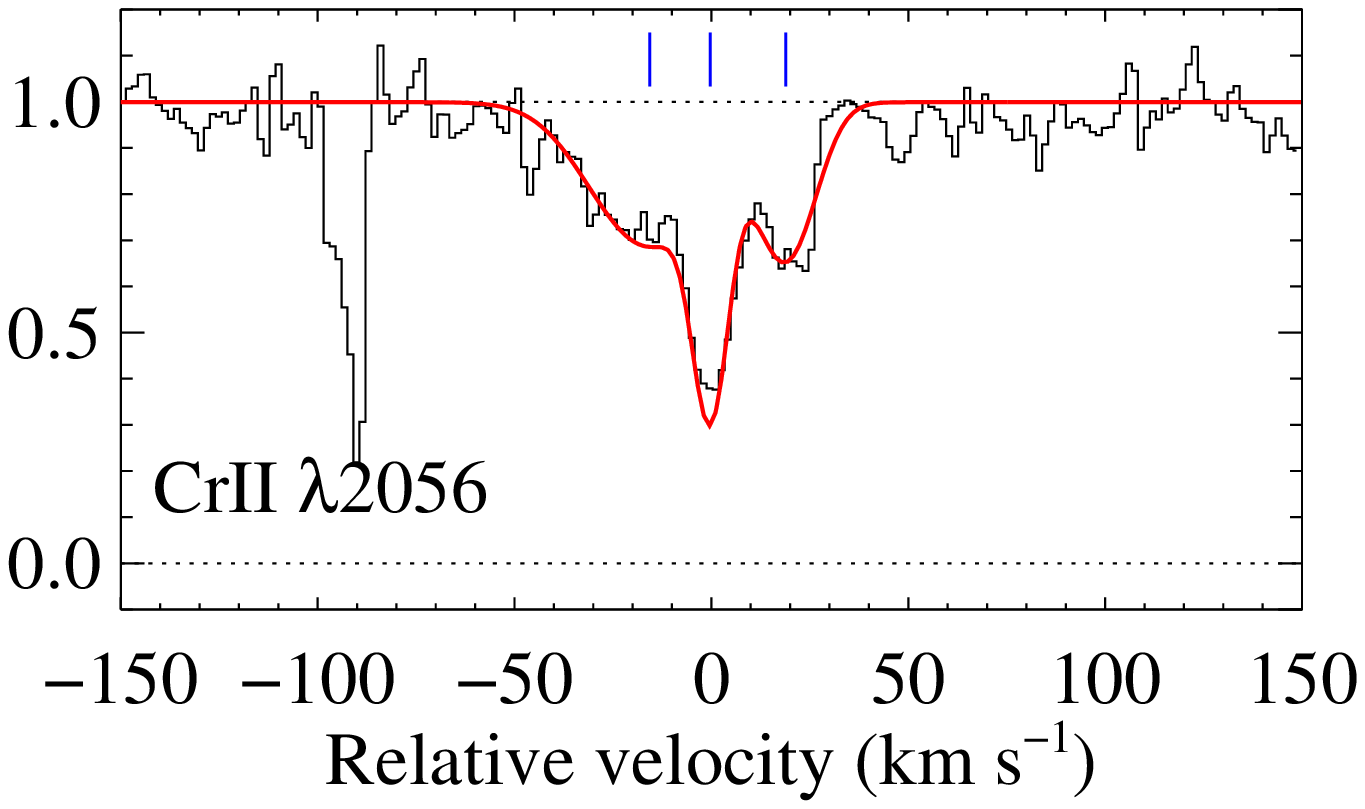} \\
 \includegraphics[bb=219 228 393 617,clip=,angle=90,width=0.15\hsize]{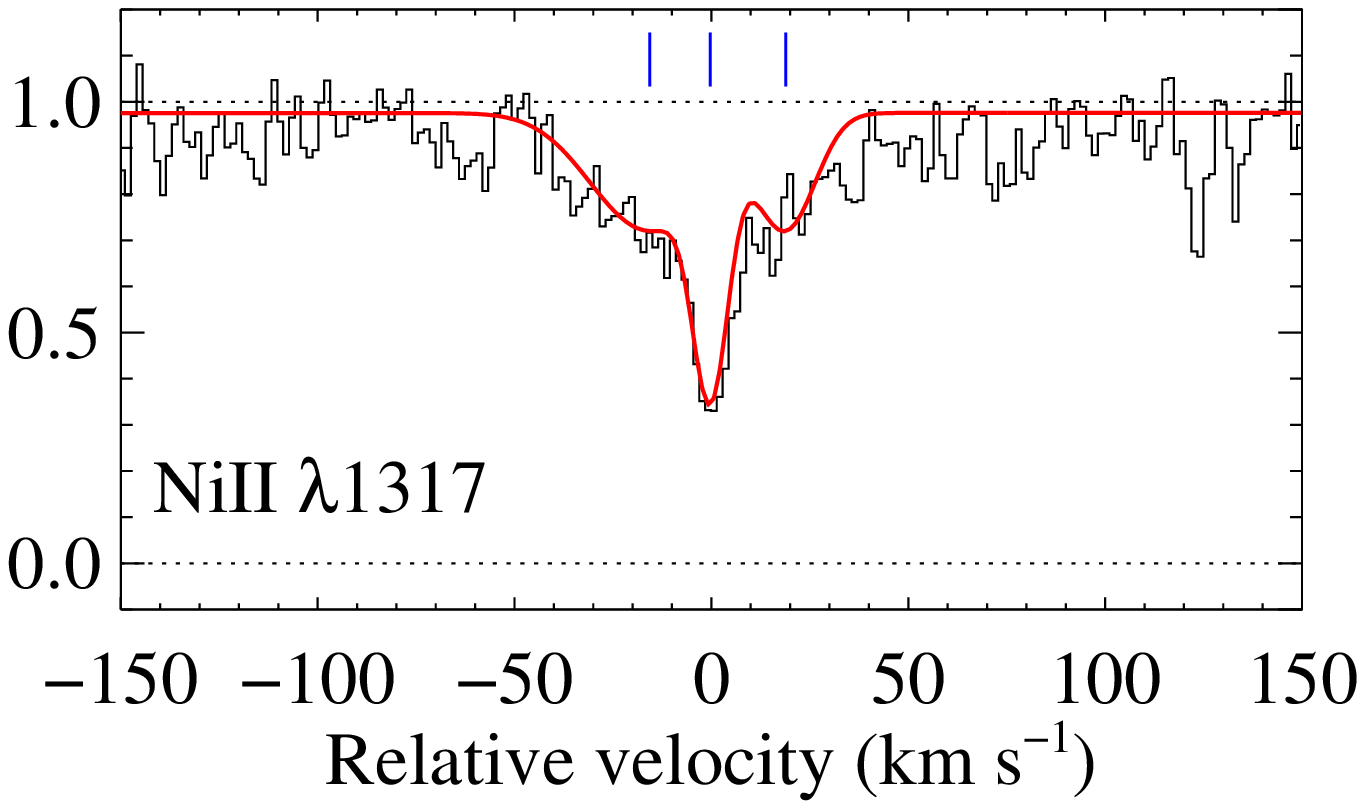} &
 \includegraphics[bb=219 228 393 617,clip=,angle=90,width=0.15\hsize]{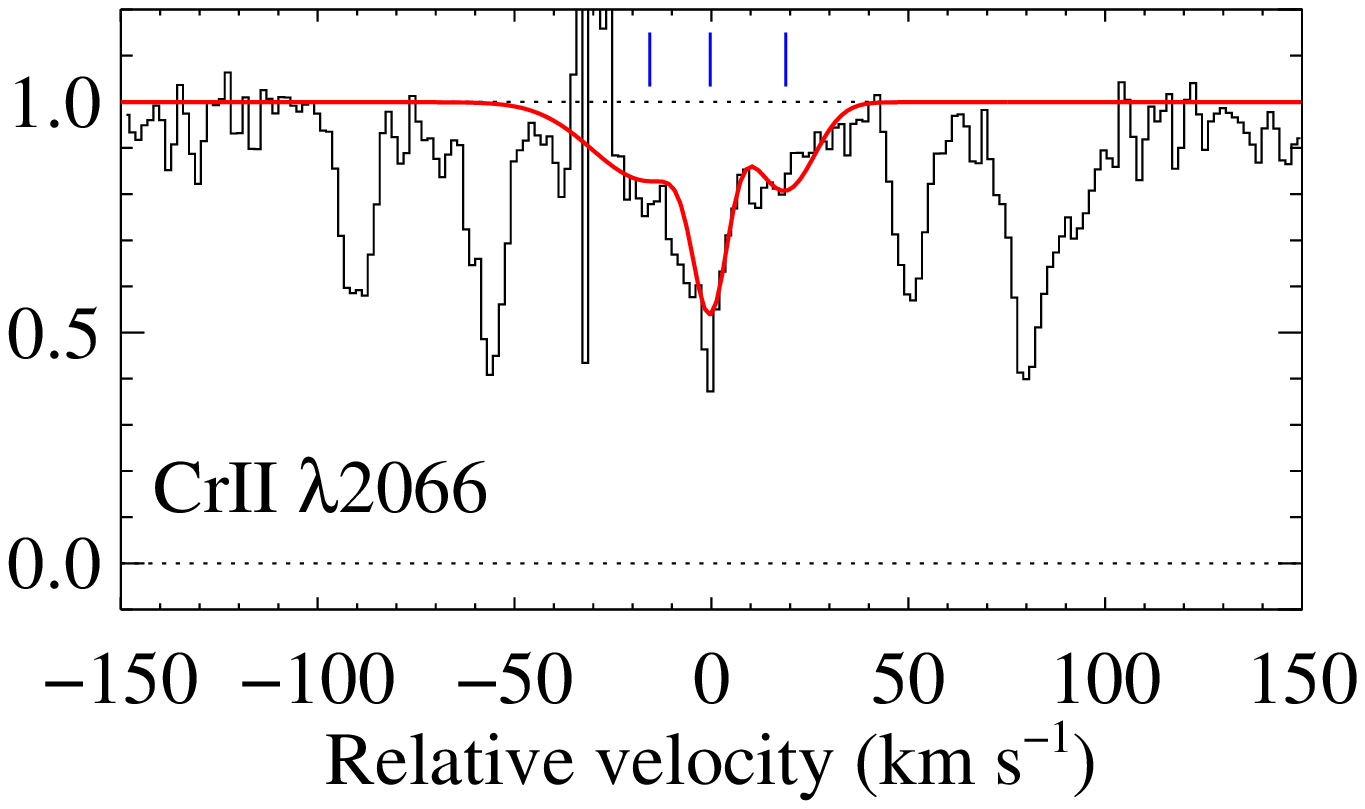} \\
 \includegraphics[bb=164 228 393 617,clip=,angle=90,width=0.15\hsize]{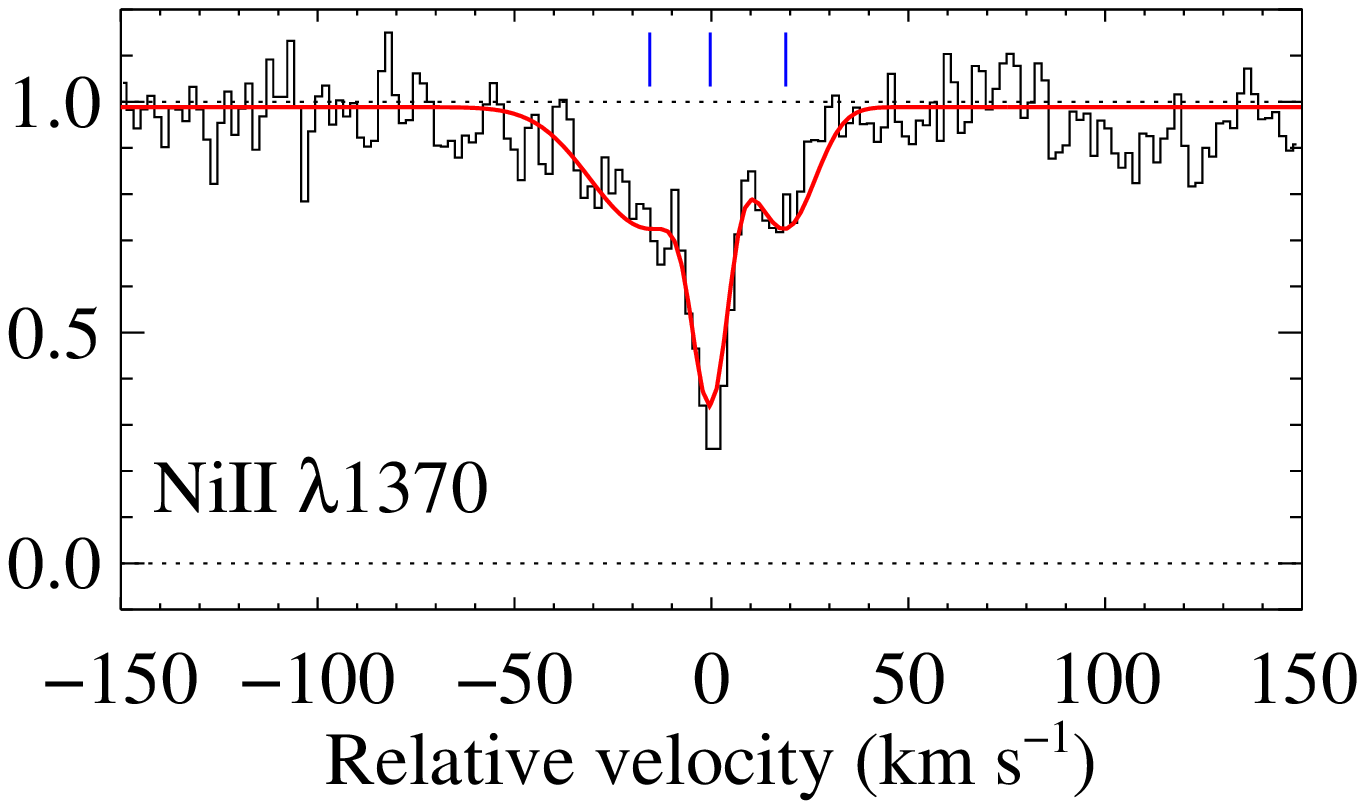} &
 \includegraphics[bb=164 228 393 617,clip=,angle=90,width=0.15\hsize]{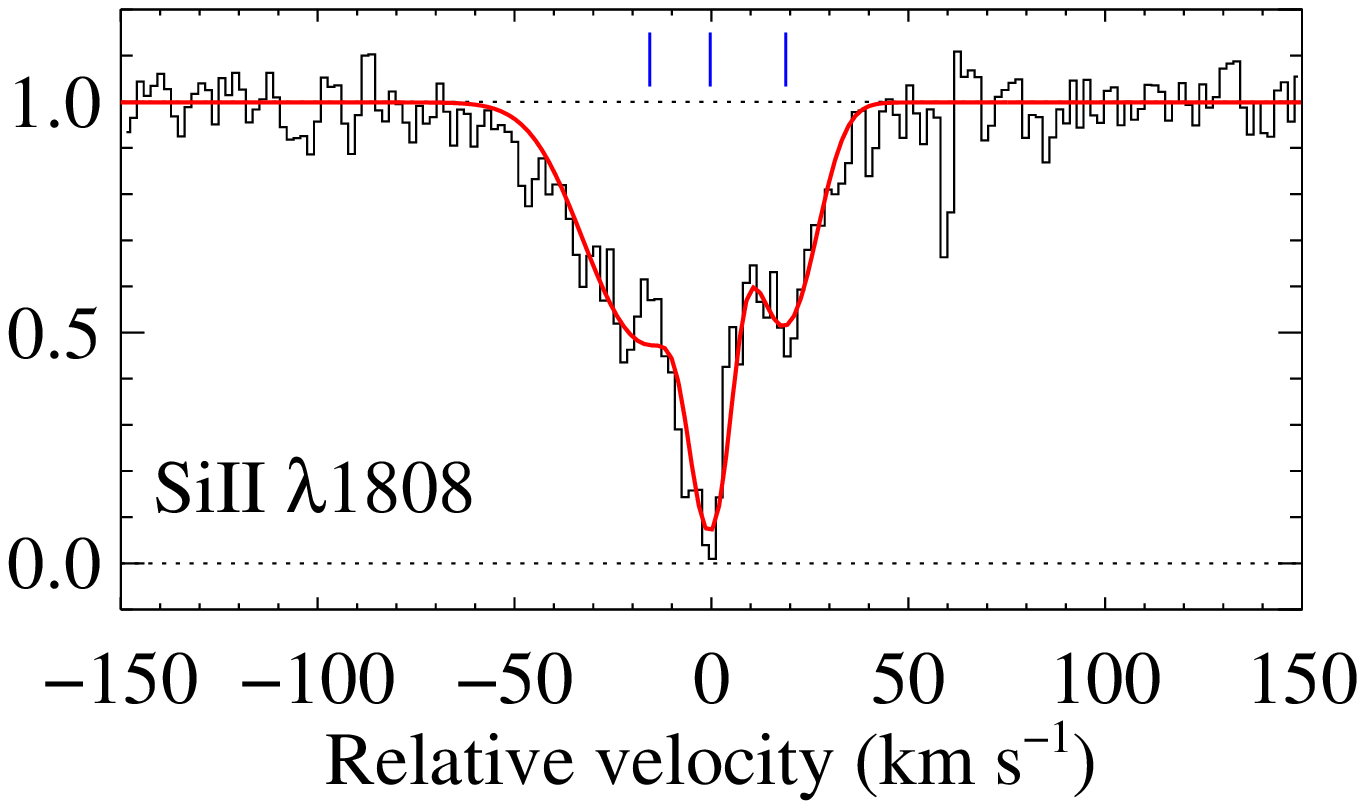} \\
 \end{tabular}   & ~~ ~~ &

 \begin{tabular}{cc}
 \includegraphics[bb=219 228 393 617,clip=,angle=90,width=0.15\hsize]{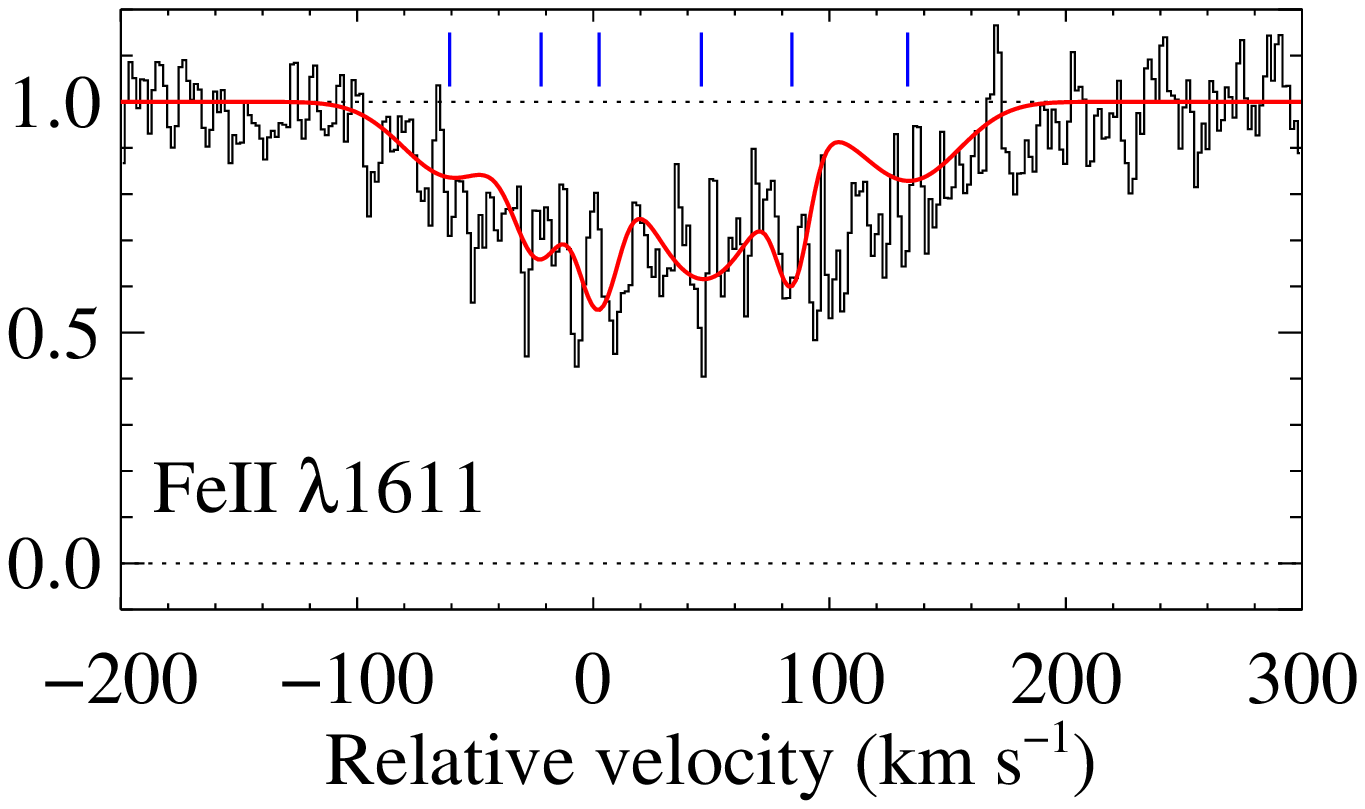} & 
 \includegraphics[bb=219 228 393 617,clip=,angle=90,width=0.15\hsize]{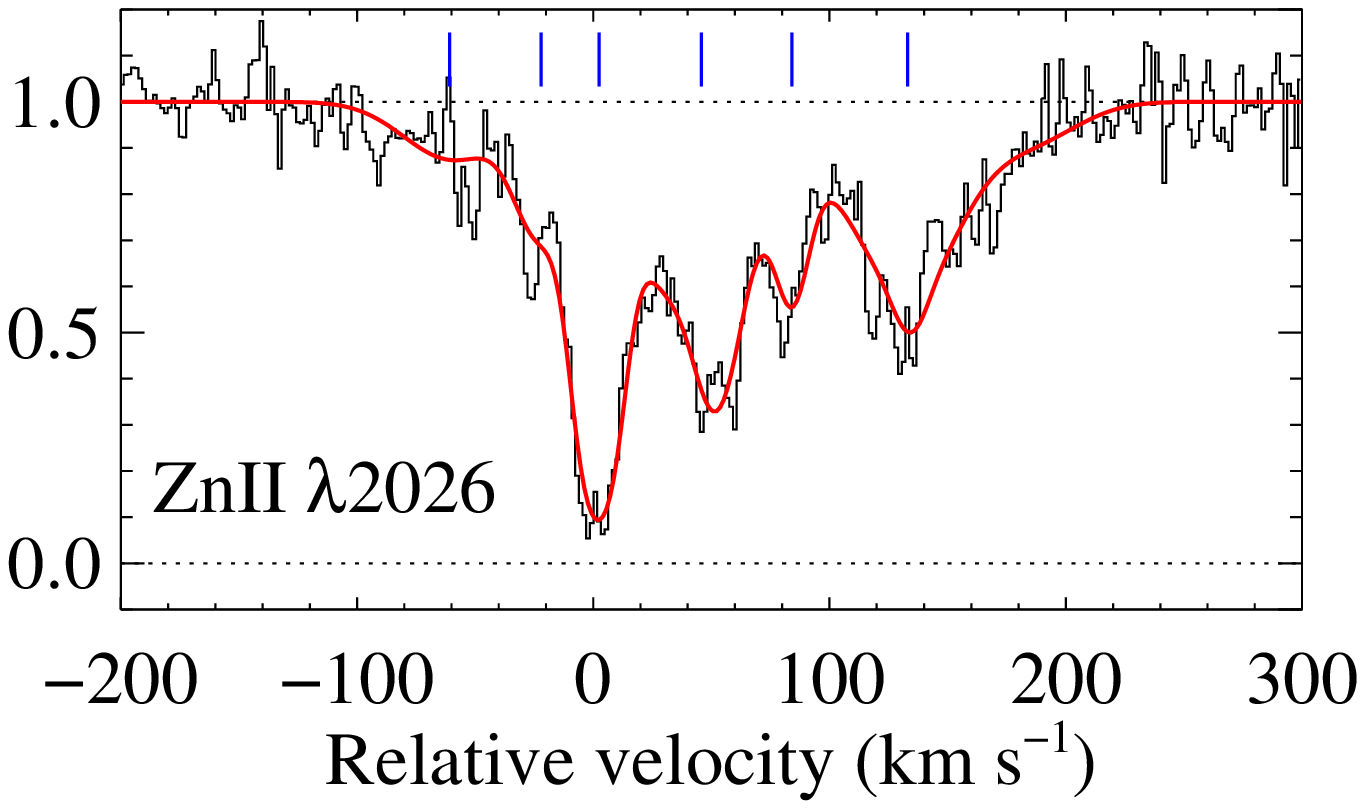} \\
 \includegraphics[bb=219 228 393 617,clip=,angle=90,width=0.15\hsize]{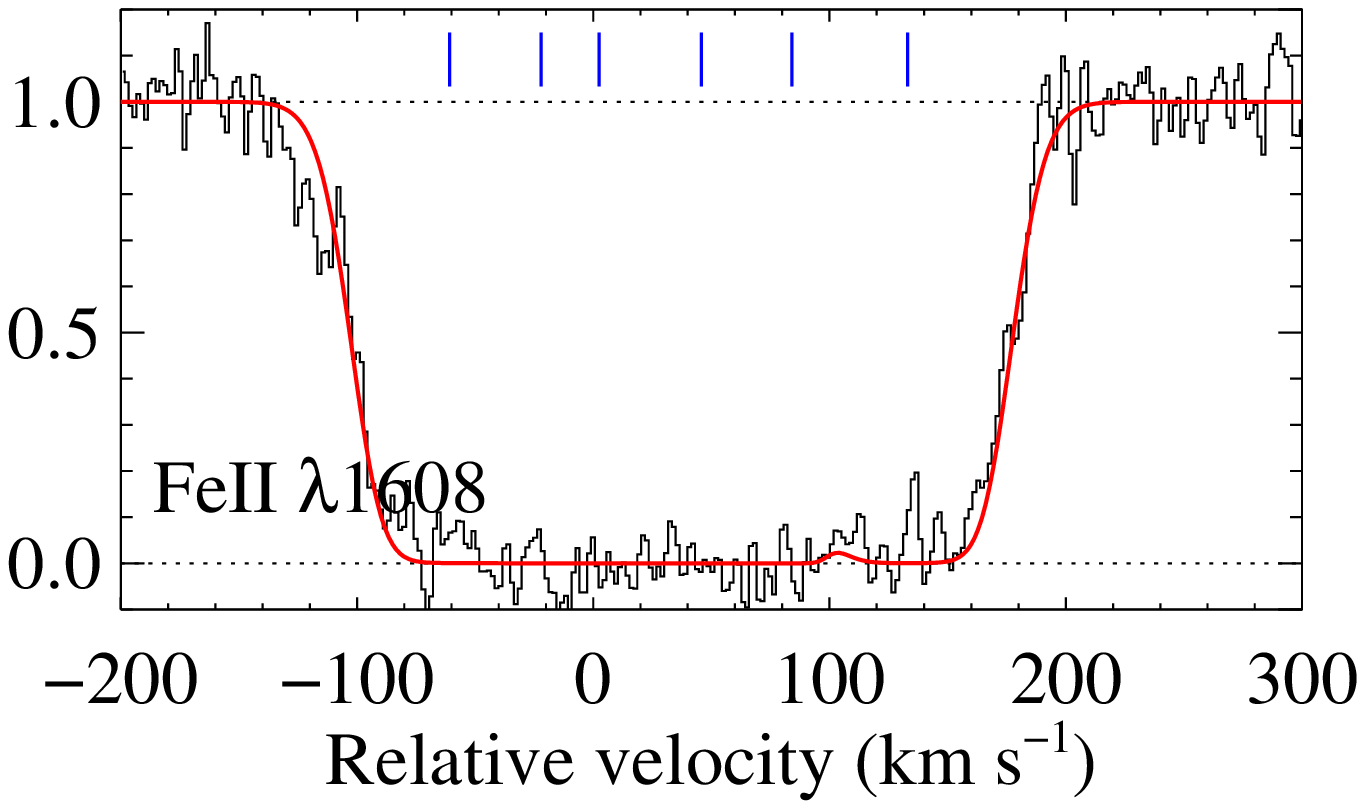} &
 \includegraphics[bb=219 228 393 617,clip=,angle=90,width=0.15\hsize]{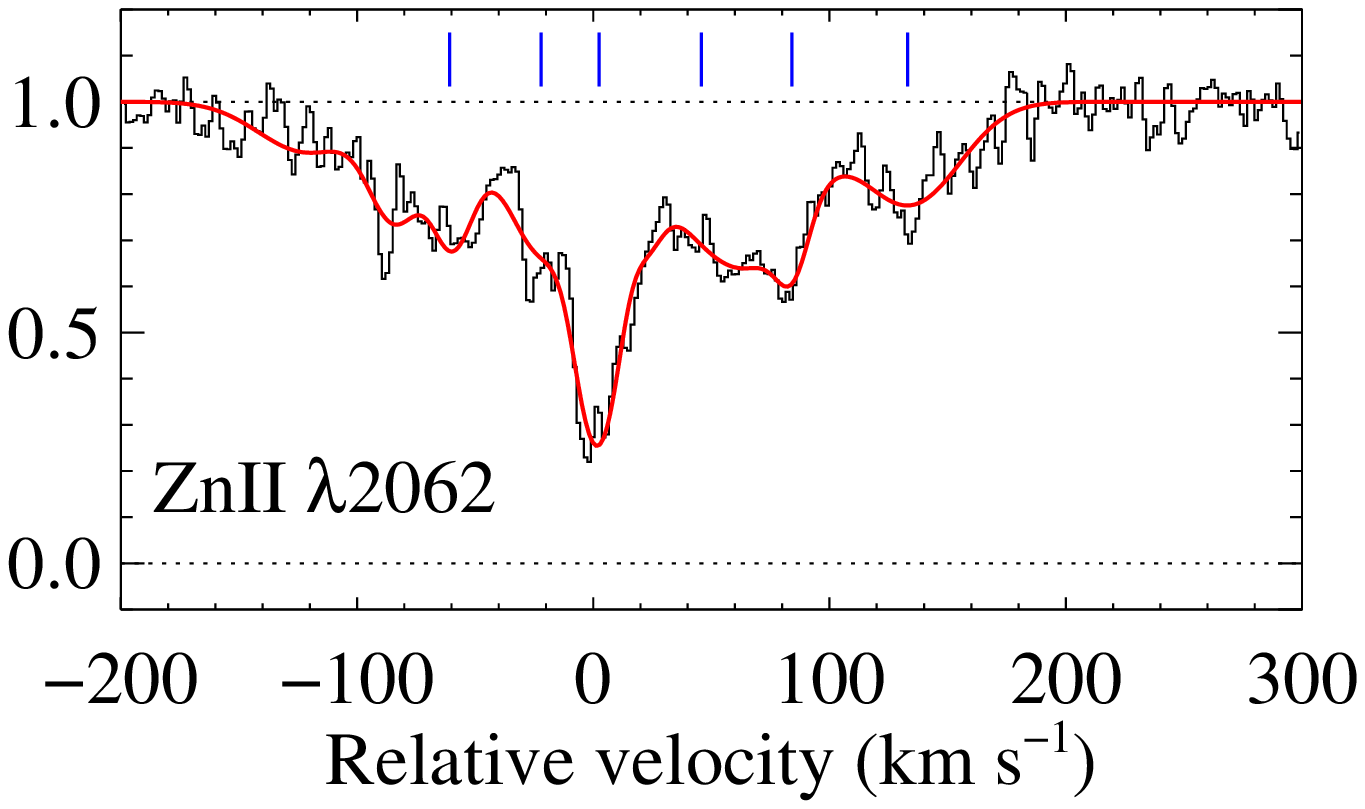} \\
 \includegraphics[bb=219 228 393 617,clip=,angle=90,width=0.15\hsize]{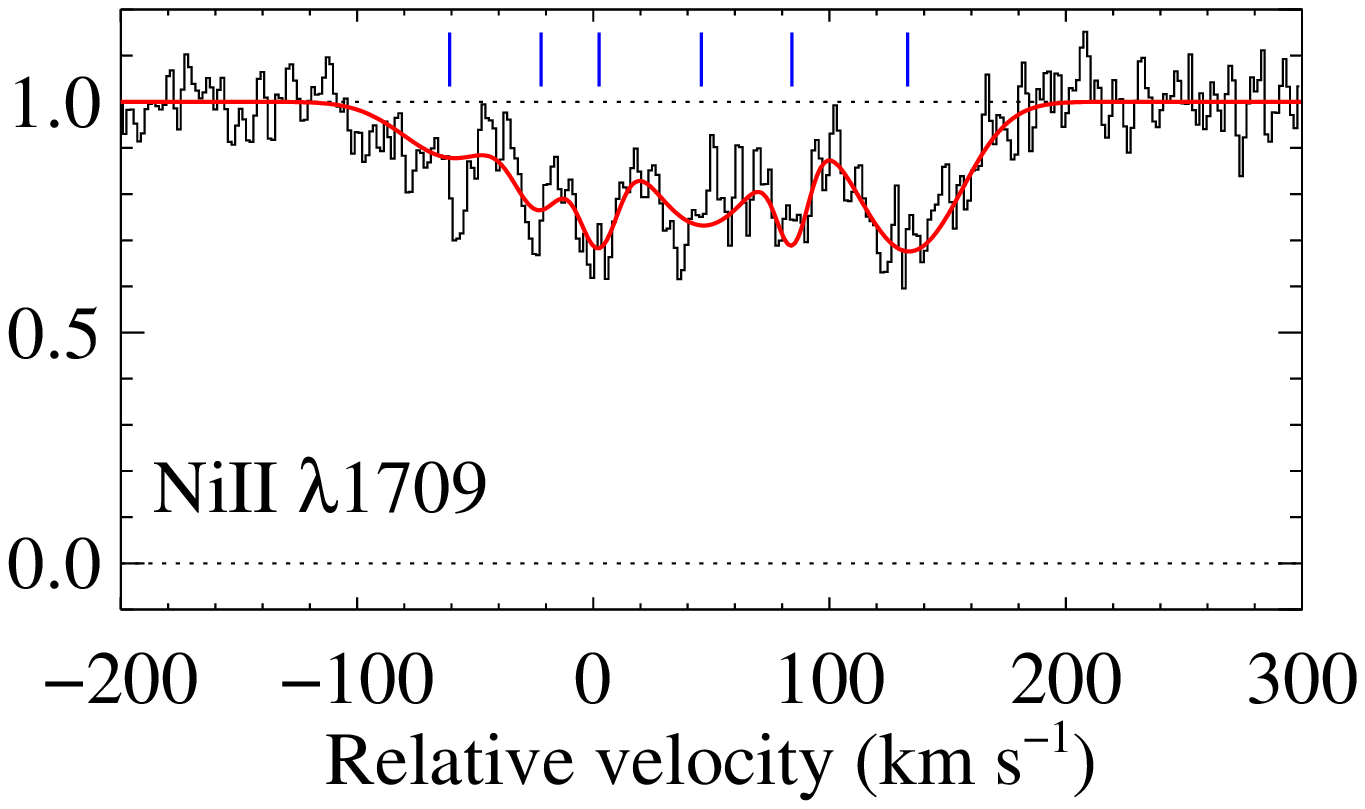} &
 \includegraphics[bb=219 228 393 617,clip=,angle=90,width=0.15\hsize]{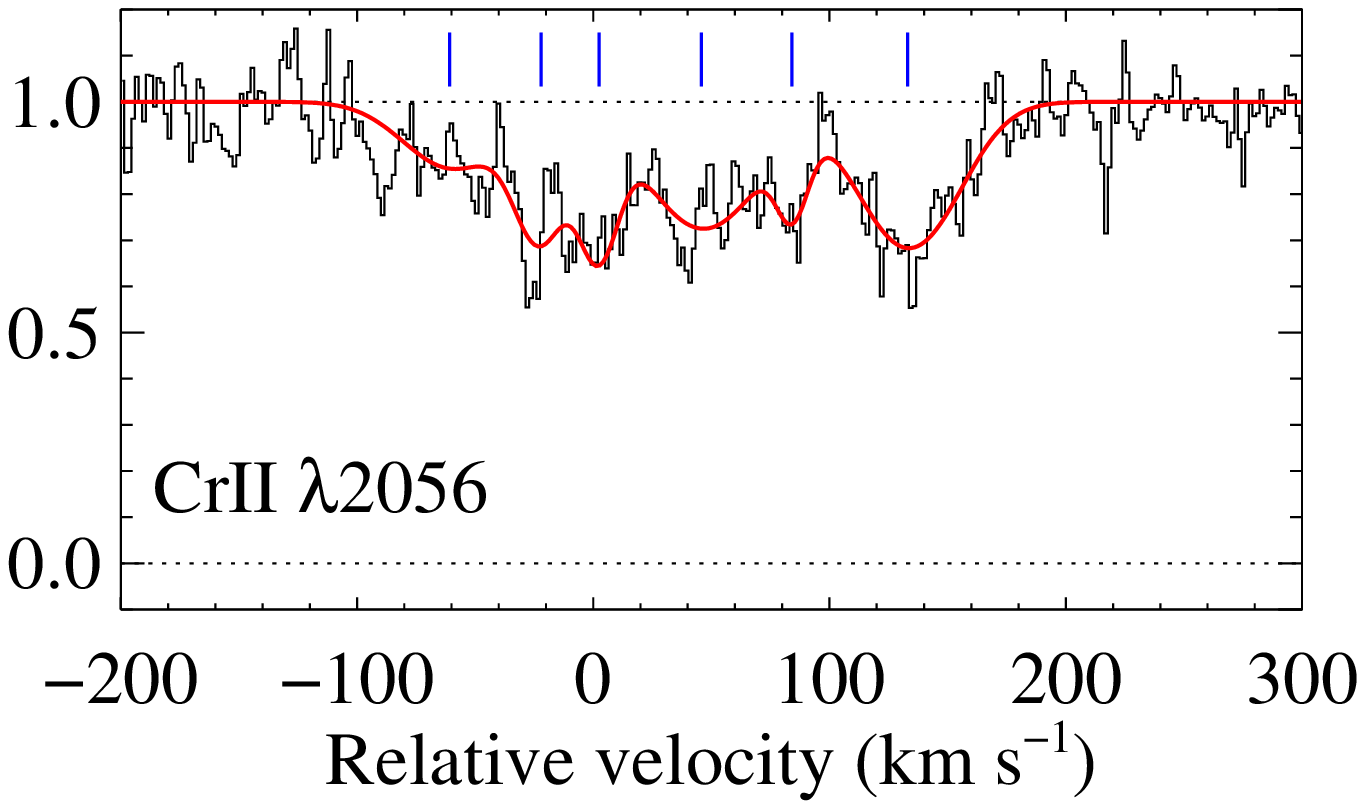} \\
 \includegraphics[bb=219 228 393 617,clip=,angle=90,width=0.15\hsize]{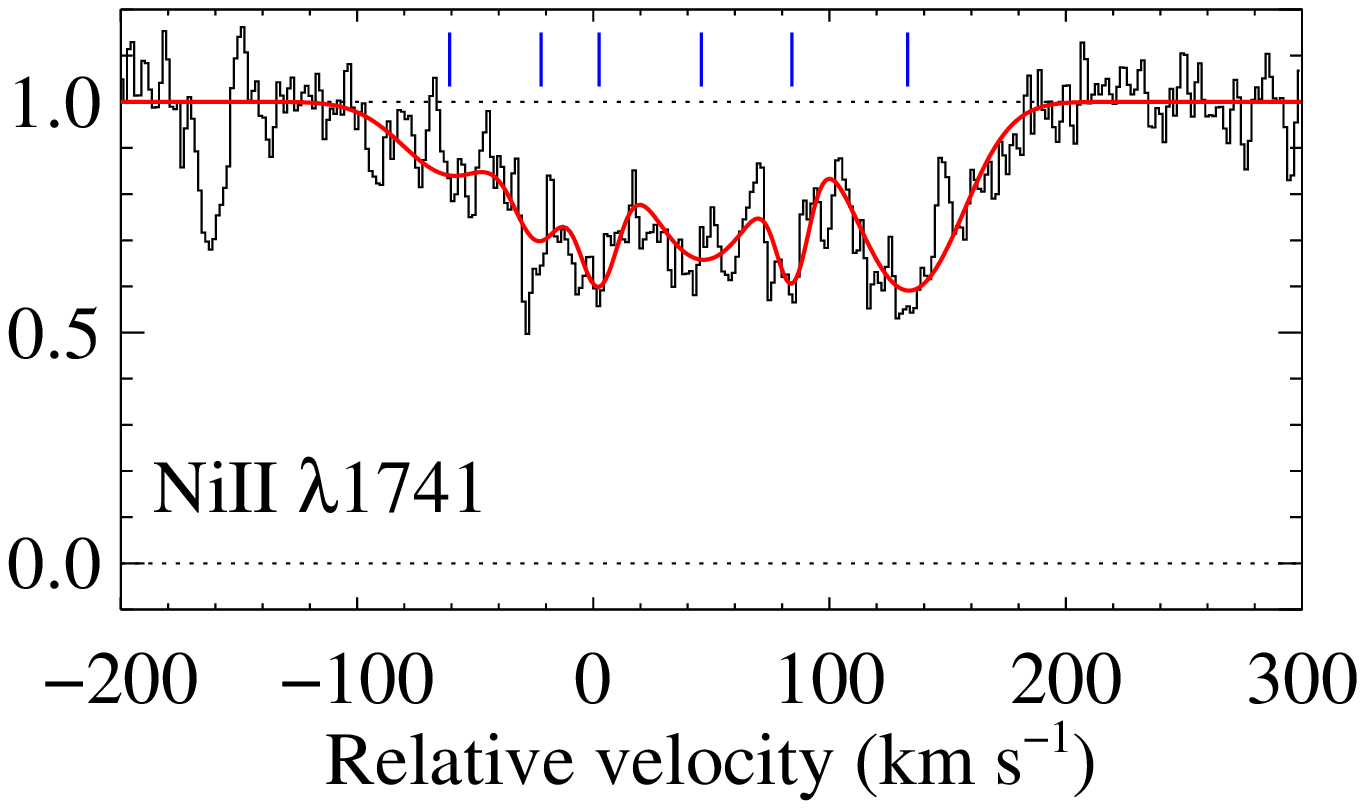} &
 \includegraphics[bb=219 228 393 617,clip=,angle=90,width=0.15\hsize]{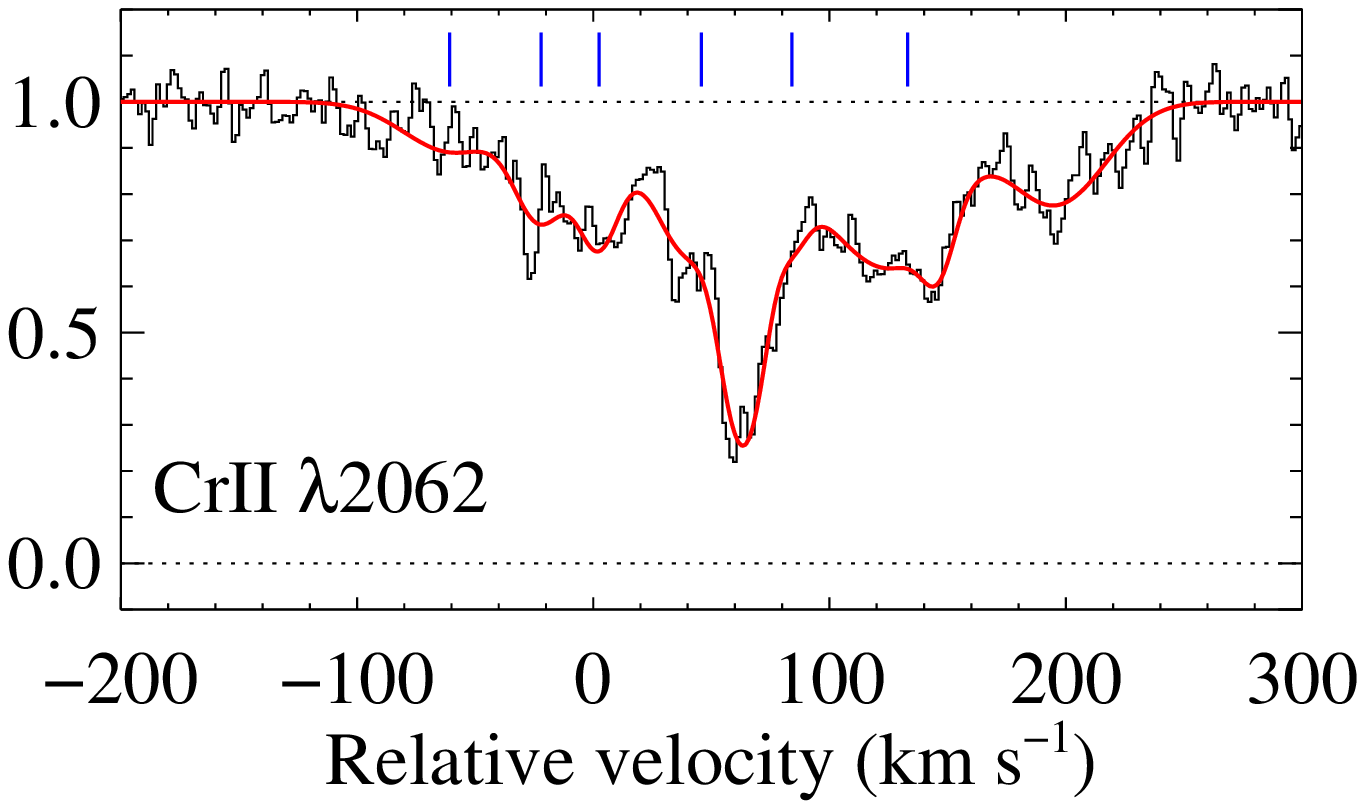} \\
 \includegraphics[bb=164 228 393 617,clip=,angle=90,width=0.15\hsize]{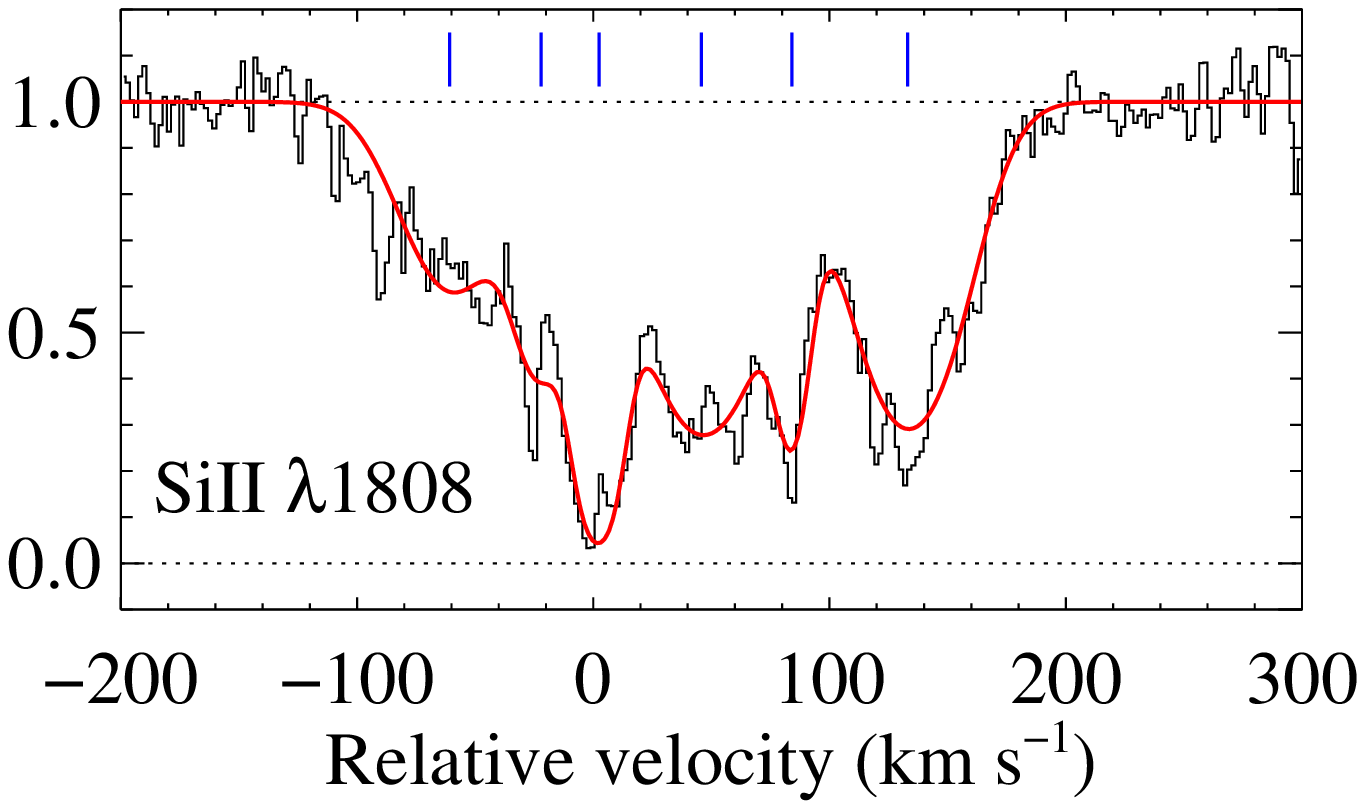} &
 \includegraphics[bb=164 228 393 617,clip=,angle=90,width=0.15\hsize]{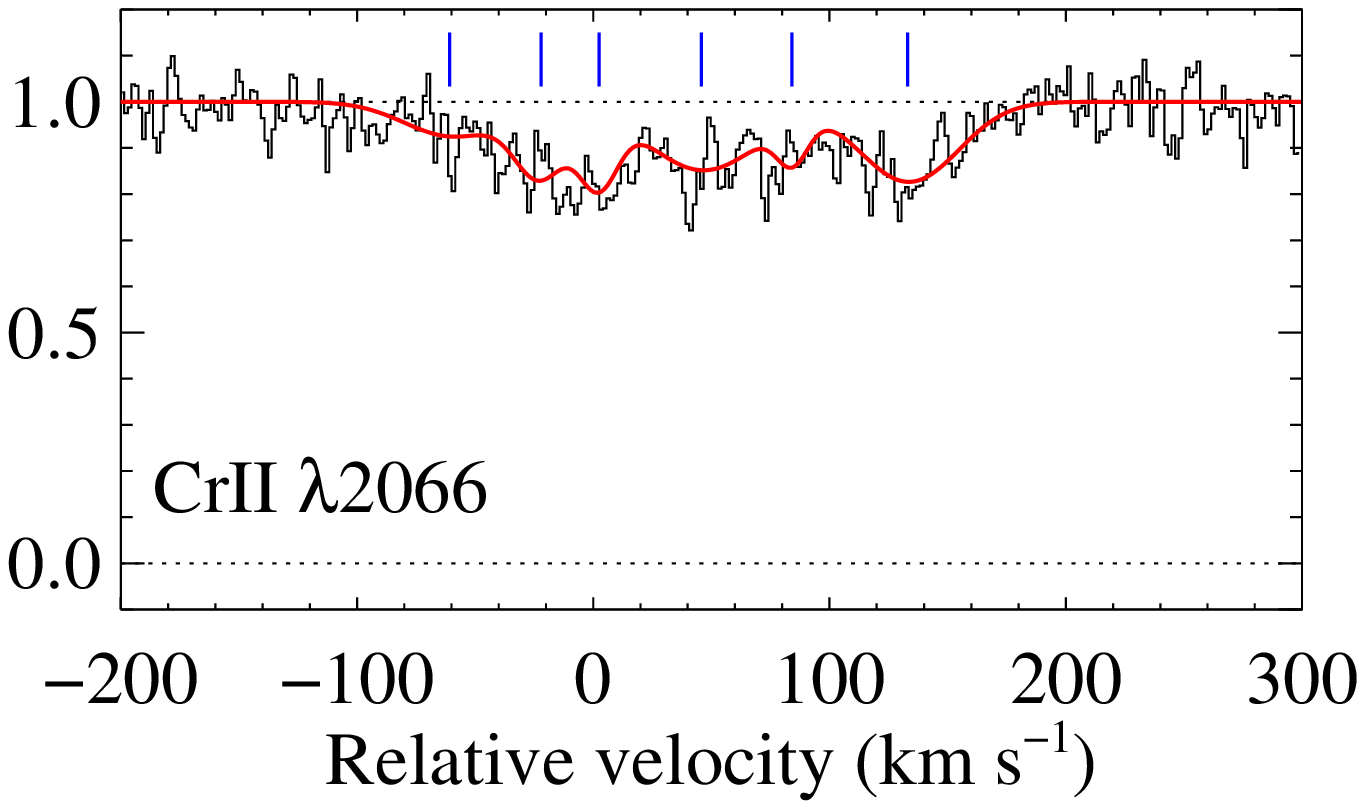} \\
 \end{tabular}   \\                                                      
\end{tabular}   
               \addtolength{\tabcolsep}{+5pt}
\caption{Voigt-profile fit to the metal absorption lines in the three ESDLAs. The redshift that defines 
the origin of the velocity scale, indicated on the top, correspond to that of the cold gas (from \CI\ for \jt, \jz\ and from H$_2$ for \jf). 
The blue tick marks indicate the position of the individual velocity components in the fit.
\label{fig:met}.}                      
\end{figure*}

We analyse the three spectra using standard $\chi^2$-minimisation of multiple Voigt-profiles performed 
with vpfit v9.5\footnote{Carswell \url{http://www.ast.cam.ac.uk/~rfc/vpfit.html}}. 
The velocity profiles of the metals and the corresponding best Voigt-profile fitting are shown in 
Fig.~\ref{fig:met}.
For the three systems, the metal 
profiles are rather compact and the decomposition into individual components is not unique. Therefore, we quote the 
total column densities in Table~\ref{metsum}, i.e. summed over all velocity components, which should in principle 
be more reliable and depend less on the exact velocity component decomposition.
We also independently checked our results using the apparent optical depth method 
\citep[AOD, see][]{Savage91,Fox05}. The weighted-mean values obtained by applying this 
technique to several unblended 
transitions are in very good agreement 
with those based on Voigt-profile fitting.

Since \ZnII\ is a volatile species, it is expected to be undepleted onto dust grains and to be 
one of the best indicator of 
the true metallicity. We do not apply any ionisation correction, which 
should actually be very small given the very large \HI\ column densities in the clouds \citep[see e.g.][]{Peroux07}. 

The metallicity measurement is particularly difficult for the system towards \jt: a two-component 
model with  $\log N(\ZnII) \sim 13.4$ provides a good fit to the bulk of the profile. 
However, \ZnII$\lambda$2026 is very strong and unfortunately the only zinc line covered by our spectrum. 
This makes the column density highly uncertain, despite a satisfactory fit.
A more careful look at the metal profile, in particular \NiII\ lines, reveals a central narrow component. 
We find a fitting solution that converges with a very narrow component ($b \sim 1.2$~\kms) that  
contains a significant 
amount of \ZnII: we then get a total $\log N(\ZnII) \sim 13.7$, i.e. a factor of two higher than above. 

The uncertainty on the \ZnII\ column density becomes also much larger and we get [Zn/H]~$\sim -1.24 \pm 0.93$. 
Unfortunately the AOD method cannot be used to derive $N(\ZnII)$ from \ZnII$\lambda$2026 because 
the flux in this line reaches zero values.
A higher resolution spectrum covering also \ZnII$\lambda$2062 is therefore required to get a more precise 
abundance of zinc. 
Our spectrum covers the position of \PII$\lambda\lambda$1301,1532 which are fortunately not saturated, and from 
which we derive [P/H]~$\sim -1.05$. Because phosphorus is likely a non-refractory element 
\citep[see e.g.][and references therein]{Molaro01}, we can use it as an indicator of the metallicity for this system.
For the DLAs towards \jf\ and \jz, we find overall metallicities of respectively 
$[{\rm Zn/H}]~=~-1.3$ and $-0.7$ based on unsaturated \ZnII$\lambda$2026 and \ZnII$\lambda$2062 absorption lines.

\begin{figure}[!ht]
\centering
\includegraphics[bb=72 197 482 396,width=\hsize]{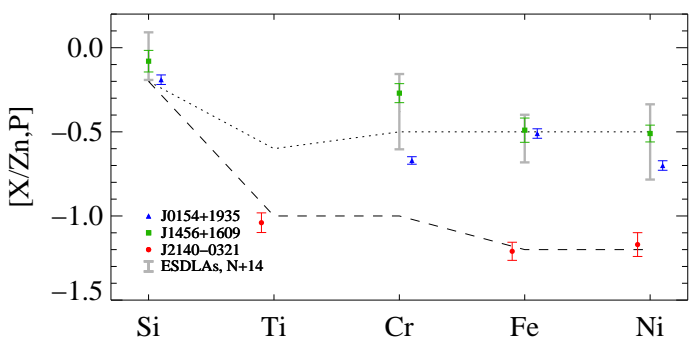}
\caption{Depletion patterns of different elements. For \jt\, we use the phosphorus abundance 
(relative to solar) as a proxy for that of zinc.  
The grey bars correspond to the value obtained from averaging low-resolution BOSS spectra of ESDLAs \citep{Noterdaeme14}. The dotted (resp. dashed) 
line corresponds to the Milky-Way Halo (resp. Warm disc) depletion pattern. \label{f:deple}
}
\end{figure}

\TiII, \CrII, \NiII\ and \FeII\ are refractory elements that deplete easily onto dust grains 
\citep[e.g.][]{Pettini97}. 
We observe iron 
depleted by a factor about 15, 3 and 3 compared to non-depleted 
elements (Zn, P) in the systems towards respectively, \jt, \jf\ and \jz.  The depletion patterns (Fig.~\ref{f:deple})
for the systems towards \jf\ and \jz\ are similar to that derived from stacking BOSS spectra of ESDLAs \citep{Noterdaeme14}.
This is typical of what is seen in the Small Magellanic Cloud \citep{Welty97}, which also presents high \HI\ column 
densities and relatively low metallicities. 
In turn, the depletion pattern in the DLA towards \jt\ is more similar to that seen in the warm ISM in the Milky-Way disc.

\begin{table*}
\centering
\caption{Overall column densities and abundances \label{metsum}}
\begin{tabular}{c c c c}
\hline \hline
{\Large \strut}  Species   &                           \multicolumn{3}{c}{$\log N$ [\cmsq]}                       \\
                           &     \jt, $\zabs=2.34$                        &      \jf, $\zabs=3.35$               &      \jz, $\zabs=2.25$               \\
\hline                                                   
{\large \strut} \HI\       &  $22.40 \pm 0.10$              & $21.70\pm 0.10$          & $21.75\pm 0.15$          \smallskip \\
               {\SiII}     &  $\ldots$                      & $15.81 \pm 0.05$ {(15.83)} & $16.42 \pm 0.02$         \\                           
                           &  $\ldots$                      & [Si/H] = $-1.40\pm 0.11$ & [Si/H] = $-0.84\pm 0.15$ \smallskip \\
               {\PII}      &  $14.76 \pm 0.08$ {(14.74)}      & $\ldots$                 &   $\ldots$               \\
                           &  [P/H] = $-1.05\pm 0.13$       & $\ldots$                 &   $\ldots$             \smallskip \\ 
               {\TiII}     &  $13.26 \pm 0.05$ {(13.27)}      &      $\ldots$            &     $\ldots$                     \\             
                           &  [Ti/H] = $-2.09\pm 0.11$      &      $\ldots$            &     $\ldots$                     \\
               {\CrII}     &   $\ldots$                     & $13.75 \pm 0.04$ {(13.68)} & $14.07 \pm 0.01$ {(14.09)}        \\
                           &   $\ldots$                     & [Cr/H] = $-2.22\pm 0.11$ & [Cr/H] = $-1.32\pm 0.15$ \smallskip \\
               {\FeII}     &  $15.64 \pm 0.03$ {(15.60)}      & $15.39 \pm 0.06$ {(15.44)} & $16.09 \pm 0.02$ {(16.13)}        \\
                           &  [Fe/H] = $-2.26\pm 0.10$      & [Fe/H] = $-1.81\pm 0.12$ & [Fe/H] = $-1.16\pm 0.15$ \smallskip \\
               {\NiII}     &  $14.40 \pm 0.05$ {(14.36)}      & $14.09 \pm 0.03$ {(14.11)} & $14.62 \pm 0.02$ {(14.67)}        \\
                           &  [Ni/H] = $-2.22\pm 0.11$      & [Ni/H] = $-1.83\pm 0.10$ & [Ni/H] = $-1.35\pm 0.15$ \smallskip \\
               {\ZnII}     &  $13.72 \pm 0.92$              & $12.94 \pm 0.04$ {(12.87)} & $13.66 \pm 0.02$ {(13.75)}        \\
                           &  [Zn/H] = $-1.24\pm 0.93$      & [Zn/H] = $-1.32\pm 0.11$ & [Zn/H] = $-0.65\pm 0.15$ \\ 
\hline
\end{tabular}
\tablefoot{Column densities derived from the Apparent Optical Depth method are indicated in parenthesis (see text). 
Abundances are expressed with respect to photospheric solar values from \citet{Asplund09}.}
\end{table*}

\subsection{Molecular hydrogen}

\begin{figure*}
\centering
\includegraphics[bb=76 54 580 755,clip=,angle=90,width=\hsize]{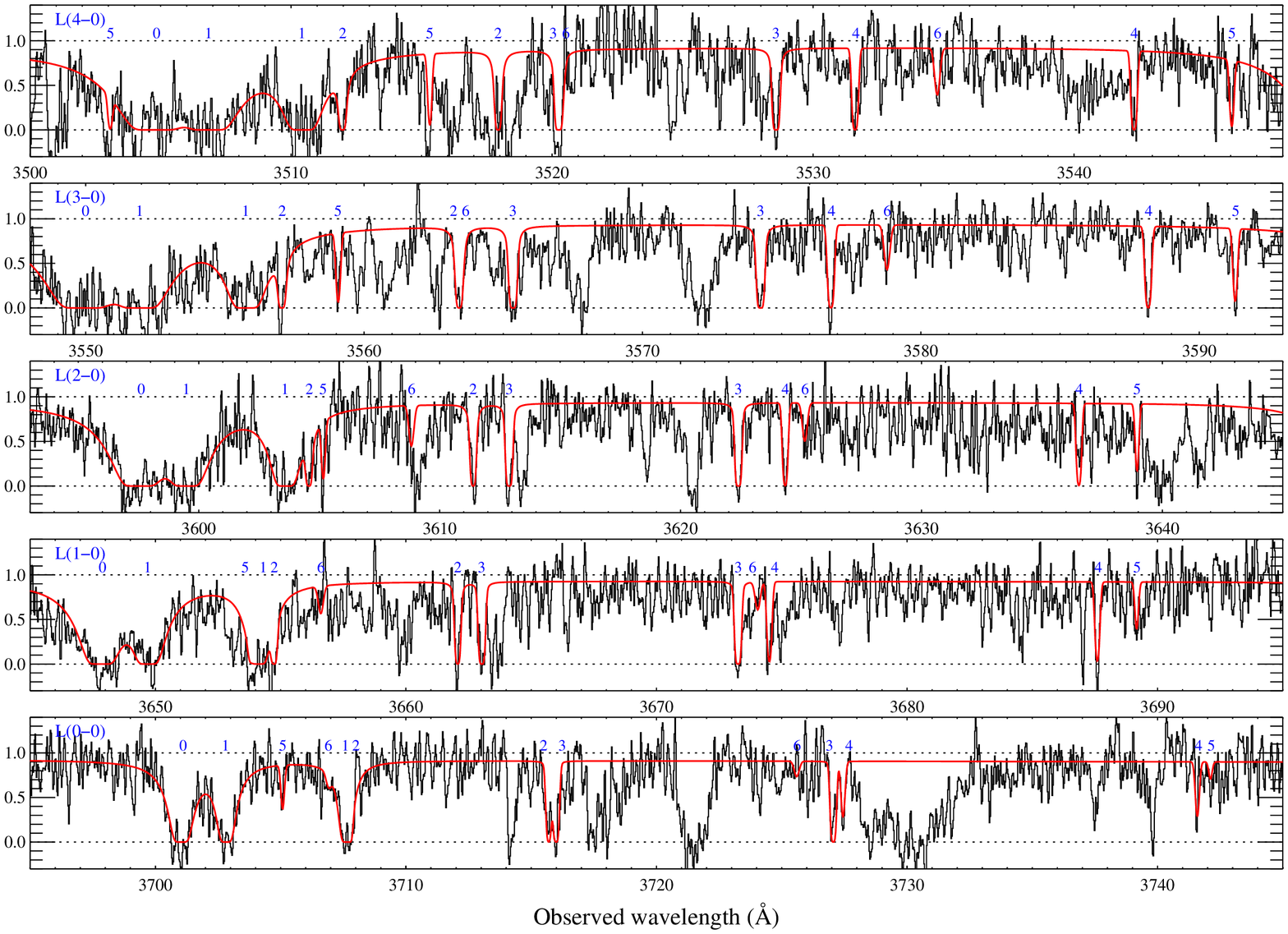}
\caption{Portions of the UVES spectrum of \jt\ covering H$_2$ absorption lines at $\zabs = 2.34$. 
The data has been boxcar-smoothed by 3-pixels for the sake of visual clarity only.
Each wavelength region covers a Lyman band, indicated in the top left corner of each panel. 
The numbers in blue above each absorption line indicates the rotational level to which the line belongs. \label{2140:f:h2}}
\end{figure*}

Molecular hydrogen absorption lines are unambiguously detected  
in the systems towards \jt\ and \jf\ from 
the Lyman ($B^1\Sigma_u^+ \leftarrow X^1\Sigma_g^+$) and/or Werner ($C^1\Pi_u \leftarrow X^1\Sigma_g^+$) 
electronic bands.
In the former case, very strong Lyman-band absorptions 
are seen up to rotational level $J=5$ 
(Fig.~\ref{2140:f:h2}), with damping wings for the first two rotational levels in all transitions 
and for several $J=2$ lines. This allows for accurate column density measurements in these low 
rotational levels. 
In turn, all $J=3$ lines are on the flat 
part of the curve of growth, implying a higher measurement uncertainty. Some $J\ge 4$ lines are 
unsaturated, making again the column density measurement in principle more reliable.
We note that the H$_2$ absorption is well modelled by a single component but 
with slightly different Doppler parameters and redshifts for the different rotational levels. 
This indicates that the different rotational populations of H$_2$ 
may not be fully co-spatial with a possible velocity structure in the 
H$_2$ gas that is not resolved into distinct components with the current data.
The column densities 
in the individual rotational levels of H$_2$ are provided in Table~\ref{2140:t:h2}. We measure a total H$_2$ column density 
of $\log N($H$_2) = 20.13 \pm 0.07$, from which we derive the overall molecular fraction $\log f = -2.14 \pm 0.17$, 
where $f=2N($H$_2)/(2N($H$_2)+N(\HI))$. 

\begin{table}
\centering
\caption{Molecular hydrogen in the $\zabs=2.34$ DLA towards \jt. \label{2140:t:h2}}
\begin{tabular}{c c c c}
\hline
\hline
{\Large \strut} Rot. level      & $\zabs$  & $b$            & $\log N$ \\
                                &          & $[\kms]$       &  $[\cmsq]$          \\  
 
\hline
J=0   &  2.33990  & --            & 19.84 $\pm$ 0.09           \\
J=1   &  2.33999  & --            & 19.81 $\pm$ 0.04           \\
J=2   &  2.33997  & 4.0 $\pm$ 0.7 & 17.96 $\pm$ 0.14           \\
J=3   &  2.33997  & 4.9 $\pm$ 0.7 & 17.76 $\pm$ 0.40           \\
J=4   &  2.33996  & 6.5 $\pm$ 0.5 & 15.88 $\pm$ 0.26           \\
J=5   &  2.33998  & 5.7 $\pm$ 1.8 & 15.17 $\pm$ 0.16           \\
J=6   &  2.33998  &               & $\le$ 14.72                \\
total &           &               & 20.13 $\pm$ 0.07           \\
      &           &               & $\log f$ = -1.97 $\pm$ 0.17           \\
\hline                            
\end{tabular}
\end{table}

\begin{figure*}
\centering
\includegraphics[bb=46 54 580 755,clip=,angle=90,width=\hsize]{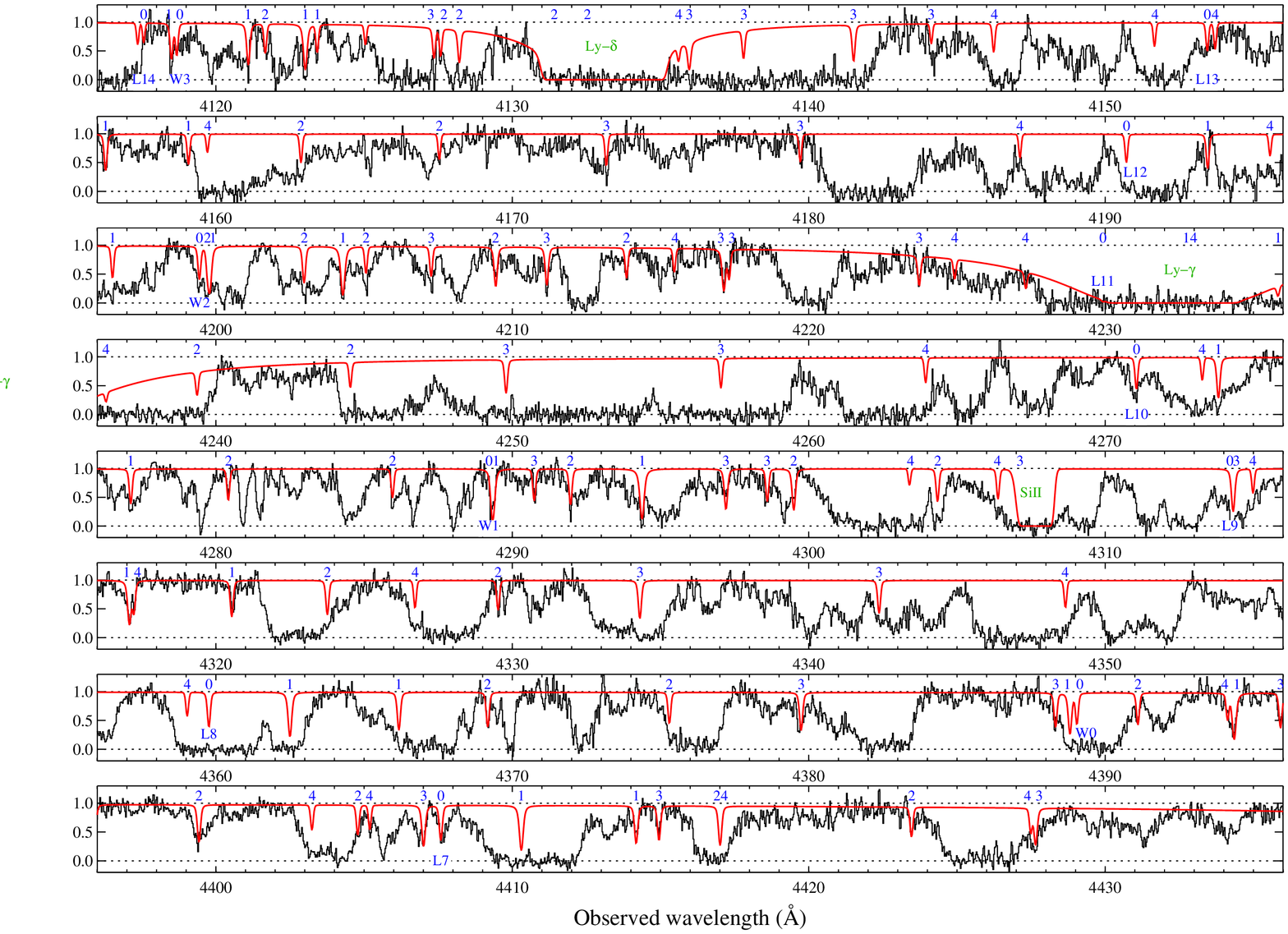}
\caption{Portion of the J1456$+$1609 spectrum covering H$_2$ absorption lines at $\zabs = 3.35$. The number in blue 
above each line correspond to its rotational level. The Lyman and Werner bands are indicated as blue label below each $J=0$ H$_2$ line. 
Other absorption lines from the same system are labelled in green (\lyg\, \lyd\ and \SiII). \label{1456:f:h2}}
\end{figure*}

H$_2$ lines are much weaker towards \jf, but still detected in a single component up to $J=4$ thanks to the large oscillator 
strengths of the lines in the covered Lyman and Werner bands 
(Fig.~\ref{1456:f:h2} and Table~\ref{1456:t:h2}). We measure 
$\log N($H$_2)=17.10 \pm 0.09$, which corresponds to 
$\log f = -4.30 \pm 0.20$.
Since the errors in each rotational level are not independent, the total error 
is taken from the extremum values in each level. 

\begin{table}
\centering
\caption{Molecular hydrogen in the $\zabs=3.35$ DLA towards \jf. \label{1456:t:h2}}
\begin{tabular}{c c c c}
\hline \hline
{\Large \strut} H$_2$  & $\zabs$    &      $b$ & $\log N$        \\
      Rot. level                 &            & $[\kms]$ & $[\cmsq]$      \\
\hline
J=0   &  3.351829 & 0.6$\pm$0.1 & 16.09$\pm$0.26 \\
J=1   &   ''      &             & 16.75$\pm$0.05 \\
J=2   &   ''      &             & 16.29$\pm$0.10 \\
J=3   &   ''      &             & 16.49$\pm$0.06 \\
J=4   &   ''      &             & 15.78$\pm$0.16 \\
\hline
total &   ''      &             & 17.10$\pm$0.09 \\
      &           &             & $\log f$=-4.30$\pm$0.20 \\
\hline
\end{tabular}
\end{table}

Unfortunately, the signal-to-noise ratio of our UVES spectrum of \jz\ is too low in the blue 
to firmly detect H$_2$. This makes any $\chi^2$-minimisation unreliable as the total 
$\chi^2$ changes little irrespective of the assumed column density. 
Rebinning the UVES spectrum by 4 pixels, we can see however that the data is 
consistent with the presence of H$_2$ with column density in the range $\log N($H$_2$)=17-19, assuming 
a thermal excitation of $T=100$~K and a Doppler parameter $b=3$~\kms, typical of what is seen in other 
H$_2$ absorbers.  Neutral carbon absorption 
lines, in turn, are unambiguously detected in two components in this system, with $N(\CI) \simeq 10^{14}$~\cmsq, see 
Fig.~\ref{0154:f:ci} and Table~\ref{0154:t:ci}. 
Given the first ionisation potential of carbon (11.2\,eV), the presence of strong \CI\ absorption 
is a reliable indicator of the presence of H$_2$ in the cloud \citep[][]{Srianand05}.

\begin{figure}
\centering
\includegraphics[bb=27 90 576 756,clip=,width=\hsize]{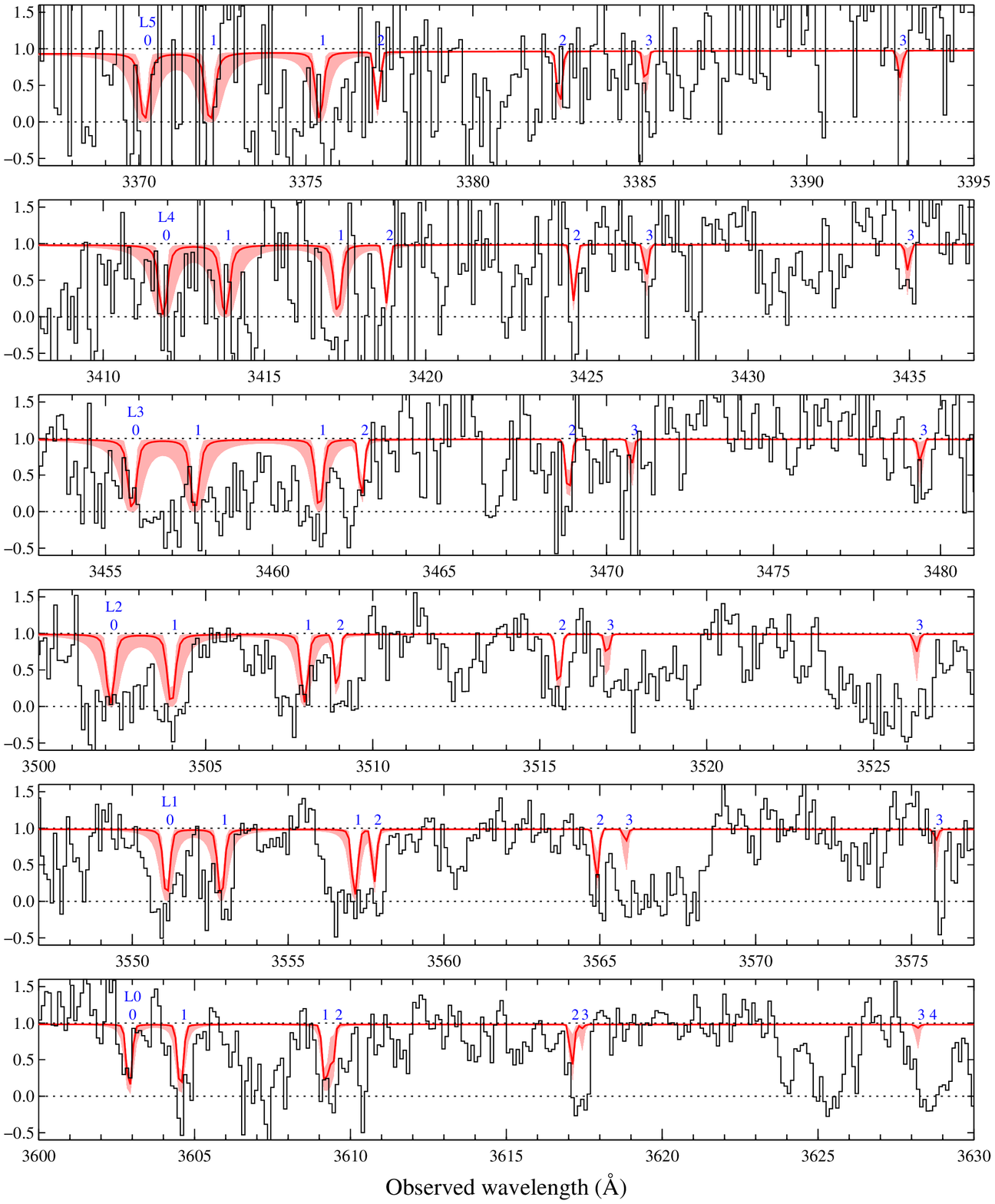}
\caption{Portion of the UVES spectrum of \jz\ covering the location of Lyman-bands of H$_2$ (from L(0-0) to L(5-0)). 
The spectrum has been rebinned by 4 pixel. The red 
synthetic profile corresponds to $N($H$_2) = 10^{18}$\cmsq\ with a single excitation temperature of 100~K. The shaded region 
corresponds to $\pm$1~dex range in $N($H$_2$) around this column density.}
\end{figure}

\begin{figure}
\centering
\begin{tabular}{c}
\includegraphics[bb=188 13 394 755,clip=,angle=90,width=\hsize]{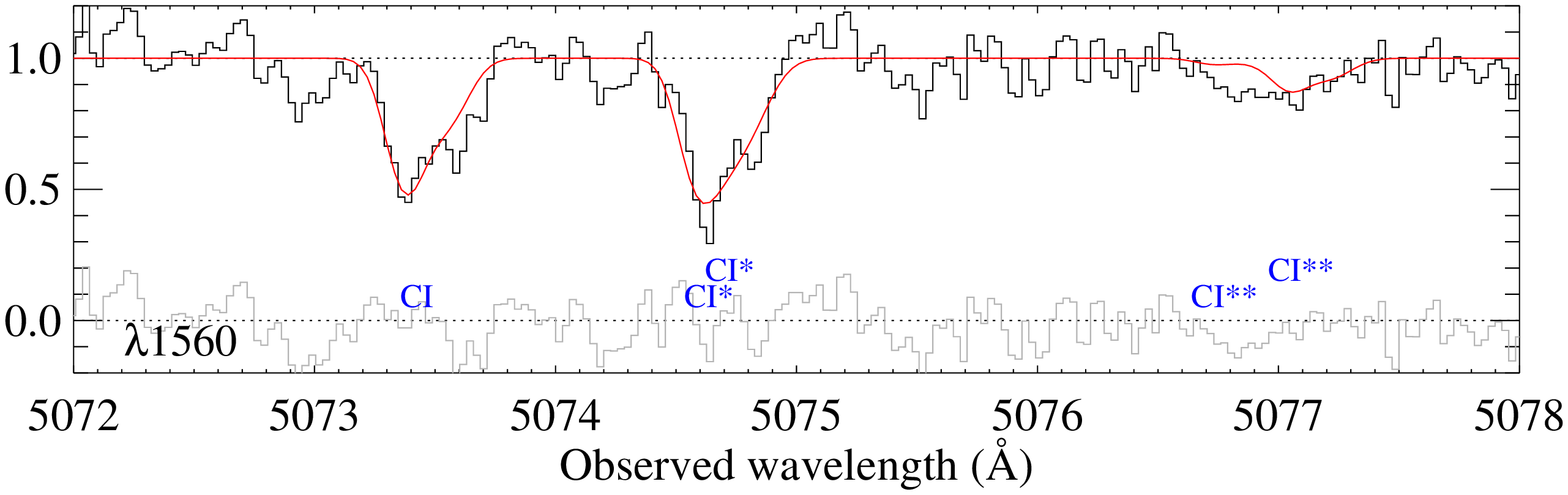}\\
\includegraphics[bb=164 13 394 755,clip=,angle=90,width=\hsize]{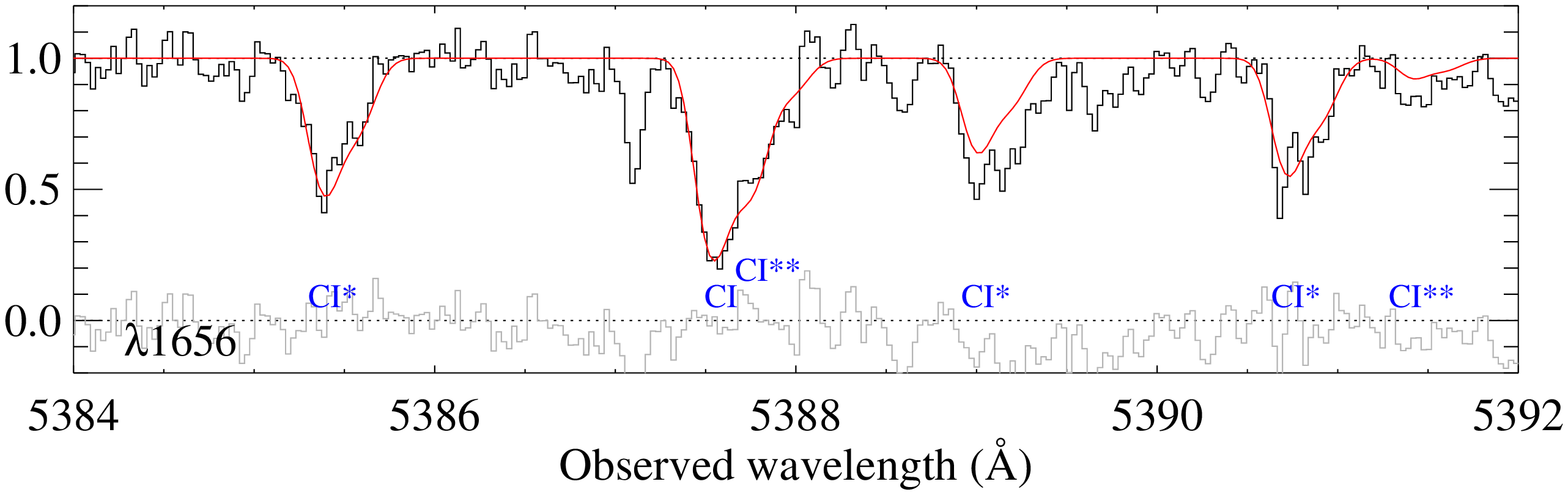}\\
\end{tabular}
\caption{C\,{\sc i} absorption lines in the $\zabs=2.25$ DLA towards \jz. \label{0154:f:ci}}
\end{figure}

\begin{table}
\centering
\caption{Neutral carbon in the $\zabs=2.25$ DLA towards \jz. \label{0154:t:ci}}
\begin{tabular}{c c c c}
\hline
\hline
{\Large \strut} \CI\     & $\zabs$ & $\log N$          & $b$      \\ %Asplund 09, photo
     fine-structure level                        &         & $[\cmsq]$        & $[\kms]$   \\   
\hline                                            
{\large \strut} J=0  (g.s.) &   2.25152  & 13.39 $\pm$ 0.06  & 5.2 $\pm$ 0.7  \\
          &   2.25163  & 13.06 $\pm$ 0.21  & 6.3 $\pm$ 2.0  \\
J=1  ($^{*}$)  &   2.25152  & 13.47 $\pm$ 0.06  & 5.2 $\pm$ 0.7  \\
          &   2.25163  & 13.19 $\pm$ 0.14  & 6.3 $\pm$ 2.0  \\
J=2   ($^{**}$) &   2.25152  & 12.71 $\pm$ 0.10  & 5.2 $\pm$ 0.7  \\
          &   2.25163  & 12.56 $\pm$ 0.21  & 6.3 $\pm$ 2.0  \\
total     &            & 13.95 $\pm$ 0.05  &                \\
\hline
\end{tabular}
\end{table}

\section{The DLA towards \jt: chemical and physical conditions \label{s:phys}}

In this section, we study the $\zabs = 2.34$ system towards \jt\ in more details, 
as it is particularly interesting for having the largest \HI\ and H$_2$ column 
densities ever measured in a high-$z$ QSO-DLA. In addition, the presence of chlorine, as 
well as fine-structure lines of \CI, \OI\ and \SiII\ allows us to probe the chemical and physical conditions
prevailing in the absorbing gas as well as the multicomponent nature of the absorber. 
In particular, we apply standard
procedures used in Galactic ISM studies to extract physical parameters like
density, temperature and radiation field and compare them with those derived for
the \h2\ clouds in different environments. We also consider photo-ionisation
models using the code {\sc cloudy} to confirm our findings.  

\subsection{Heavy elements}

Different elements are released by stars through several channels, depending on the stellar masses 
and life-times. 
The very high column densities towards \jt\ provide a unique opportunity to detect rare heavy 
elements, bringing additional clues to the chemical enrichment history of the cloud. 

\begin{figure}[!t]
\centering
\begin{tabular}{c}
\includegraphics[bb=219 228 393 617,clip=,angle=90,width=0.48\hsize]{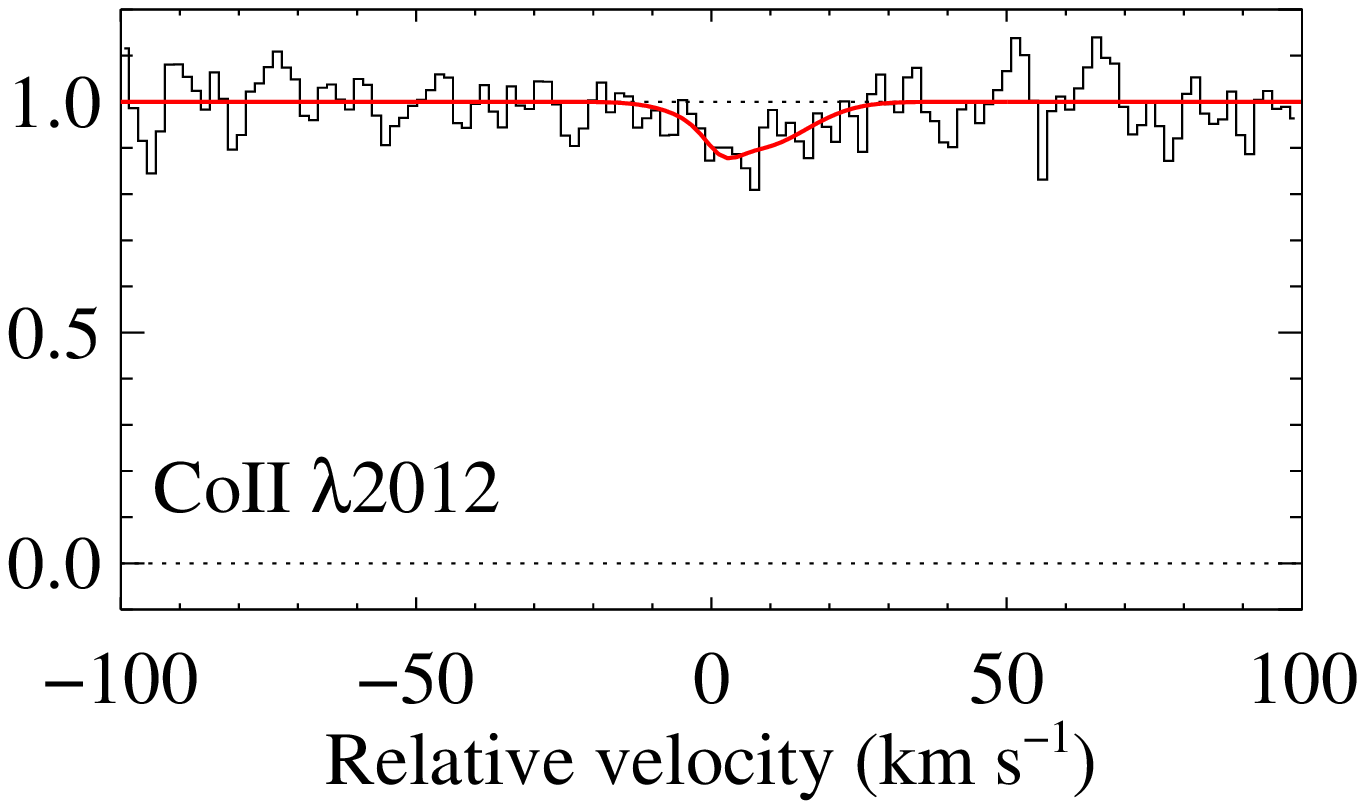}\\
\includegraphics[bb=164 228 393 617,clip=,angle=90,width=0.48\hsize]{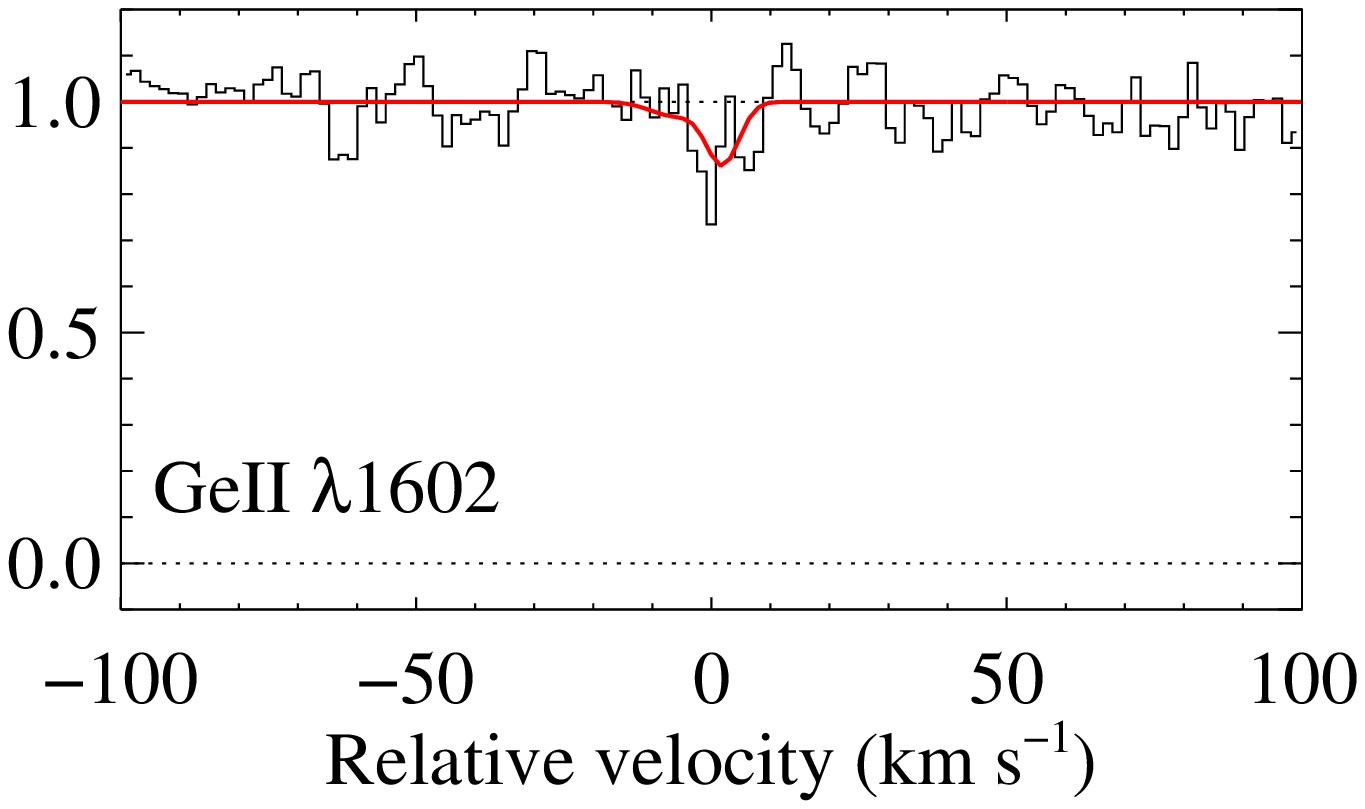}\\
\end{tabular}
\caption{Possible detection of \CoII\ and \GeII\ in the DLA towards \jt. \label{2140:f:GeCo}}
\end{figure}

Absorption lines from \GeII\ and \CoII\ are probably detected in our spectrum, albeit at less than 3$\sigma$ significance level 
and from single lines (Fig~\ref{2140:f:GeCo}). We estimate the corresponding column densities to be $\log N(\GeII) \sim 12.25$ and $\log N(\CoII) \sim 13.05$ from fitting these lines with the redshifts and 
Doppler parameters fixed to those derived from the other metals. We then derive  
[Ge/H]$ \sim -1.8$ and [Co/H]$ \sim -2.3$. 
Germanium is the heaviest element detected in our spectrum, and has very rarely been detected in DLAs to date 
\citep{Prochaska03,Dessauges-Zavadsky06}. Measuring the abundance of germanium is of particular interest because its nucleosynthesis
resides at the transition between explosive synthesis of iron-peak elements and neutron-capture synthesis of heavier 
elements. In metal-poor stars, the abundance of germanium seems to follow well that of iron \citep{Cowan05}, 
favouring an explosive origin of this element at low metallicity, while the increase of [Ge/Fe] with metallicity 
\citep{Roederer12} indicates the onset of $s$-process production of germanium at high metallicity. Here, the observed 
ratio is super-solar ([Ge/Fe]~$\sim$~0.46), similar to what is seen in other DLAs \citep{Dessauges-Zavadsky06}. However, iron 
is a refractory element that very easily depletes onto dust grains, making it difficult to interpret the observed ratio in terms of nucleosynthesis.
Unfortunately, the abundance 
of undepleted elements zinc and sulphur are not well constrained in the DLA towards \jt\ while Ni is also depleted.
A possible reference element could be phosphorus, which seems to be little depleted. However, its 
exact nucleosynthesis is still not fully understood \citep[][]{Molaro01}. The observed sub-solar [Ge/P] ratio ([Ge/P]~$\sim$-0.75) could partly be 
due to depletion of germanium but could also reflect an enrichment dominated by massive stars.  
Cobalt is also rarely seen in DLAs, with the first detection reported by \citet{Ellison01b}. As argued by these authors, 
the similar depletion of cobalt and iron should make the [Co/Fe] ratio little dependent on the actual amount of dust. Here, 
we measure a solar cobalt-to-iron ratio, similar to what is seen in the DLA towards Q\,0948$+$43 \citep{Rao05}, and 
in the Galactic disc \citep[e.g.][]{Prochaska00}. 
The DLA towards \jt\ therefore appears to be an excellent place to study the nucleosynthesis processes at the origin of 
the chemical enrichment, although further observations, with higher S/N ratio, better wavelength 
coverage (in particular for \CuII\ that falls in a gap of our spectrum and for \ZnII, for which only 
one transition is covered) and higher spectral resolution are 
required to put strong constraint on the abundances.

\subsection{Neutral chlorine \label{sect:cl1}}

Chlorine is a peculiar species whose ionisation potential (12.97 eV) should make \ClII\ 
its dominant form in atomic gas. However, in the presence of H$_2$, chemical reactions 
make \ClI\ the dominant form of chlorine 
\citep{Jura74a}:  
Cl$^0$ results from the exothermic reaction Cl$^+ + $H$_2 \rightarrow $HCl$^+$ that 
has a very high rate \citep{Fehsenfeld74}. Several channels then allow the quick 
release of \ClI, in particular from collisions with H$_2$ or electrons. 

\begin{figure}
\centering
\begin{tabular}{c}
\includegraphics[bb=164 228 393 617,clip=,angle=90,width=0.5\hsize]{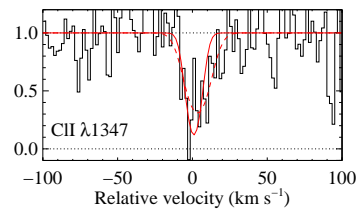} \\
\end{tabular}
\caption{Neutral chlorine in the DLA at $\zabs = 2.34$ towards \jt. The dashed profile corresponds 
to a fit will all parameters free, while $b$ is fixed to that of \CI\ for the solid profile. \label{2140:f:cl}}
\end{figure}

In addition, 
\ClI\ is little depleted onto dust grains and should be an excellent tracer of clouds 
with high molecular fractions \citep[e.g.][]{Jura78, Sonnentrucker02, Balashev15}. \ClI\ has been 
detected in the translucent cloud at $\zabs=2.7$ towards the quasar SDSS\,J1237$+$0647 
\citep{Noterdaeme10co}, that has high molecular fraction $f_{\rm H2}>0.3$ and also harbours 
CO molecules. 
Here, we detect a strong \ClI$\lambda$1347 line (see Fig.~\ref{2140:f:cl}), 
from which we derive $\log N(\ClI) = 13.61 \pm 0.10$~\cmsq\ when all parameters 
are left free. However, the data is noisy and that $b$ may be overestimated. Since 
\ClI\ is expected to be related to the cold gas phase, we rather 
expect a velocity broadening more similar to that of H$_2$ (or \CI) and a good velocity match between 
these species. We therefore perform a second fit, fixing $b$ to 5\,\kms. This 
gives $\log N(\ClI) = 13.69 \pm 0.13$~\cmsq.
We therefore adopt the value $\log N(\ClI) = 13.67 \pm 0.15$, where the error 
conservatively represent the full range allowed by the two fits and their associated 
uncertainties. 
Because \ClI\ is little depleted into dust and arises in the H$_2$-bearing gas, its 
column density can provide a rough idea of the amount of \HI\ present in that cold gas, 
provided its metallicity is known. Assuming a value equal to the overall metallicity, we 
estimate that the H$_2$-bearing gas contains a few times 10$^{21}$~\cmsq\ of 
the total hydrogen column density (10$^{22.4}$~\cmsq). The molecular fraction in the 
remaining gas is likely too low to make chemical reactions with H$_2$ counterbalance 
the \ClI\ photoionisation rate. 

\subsection{Carbon monoxide \label{s:cl}}

\begin{figure}
\centering
\includegraphics[bb=70 175 490 400,clip=,width=0.8\hsize]{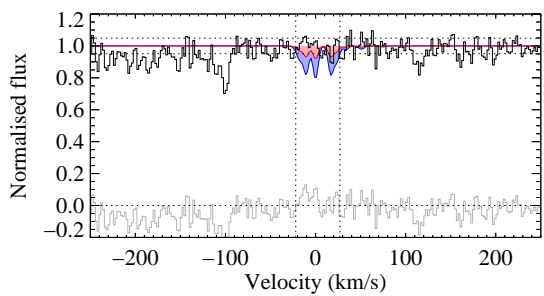}
\caption{Stack of covered CO\,AX absorption bands (from (0-0) to (3-0)) towards J2140. The red 
(resp. blue) profiles 
correspond to the expected profile having 1 time (resp. 3 times) the area sustained by noise (horizontal 
dotted lines) over the velocity range covered by the band (vertical lines) and assuming $T_{\rm ex}=T_{\rm CMB}$. 
\label{2140:f:co}}
\end{figure}

We do not detect CO absorption lines. In order to get a stringent limit, we stacked the covered 
CO\,AX absorption bands (from (0-0) to (3-0)), weighting each band by $f/\sigma_c^2$, where $f$ is the 
oscillator strength and $\sigma_c$ the noise in the continuum. This way, the signal-to-noise ratio in 
the stack is maximised\footnote{The signal here refers to the amplitude of the absorption, 
not the continuum.}. We assume an excitation temperature corresponding to radiative equilibrium with the 
cosmic microwave background, as seen in high-$z$ CO absorption systems \citep[][]{Noterdaeme11} and 
derive $N($CO$)< 5\times 10^{13}$\cmsq, see Fig.~\ref{2140:f:co}. This corresponds to 
CO/H$_2 \sim 0.4\times10^{-6}$, more than an order of magnitude less than what has been measured in 
CO-bearing DLAs \citep{Srianand08,Noterdaeme10co}, but consistent with the ratio observed towards 
Galactic stars with the same H$_2$ column density \citep[e.g. Fig.~4 of][]{Burgh07}. This shows that high H$_2$ column 
density is not a sufficient condition for the presence of CO \citep[see also][]{Bolatto13}.

\subsection{Excitation of H$_2$: Gas kinetic temperature and UV pumping \label{s:h2diag}}

\begin{figure}
\centering
\includegraphics[bb=70 175 490 575,clip=,width=\hsize]{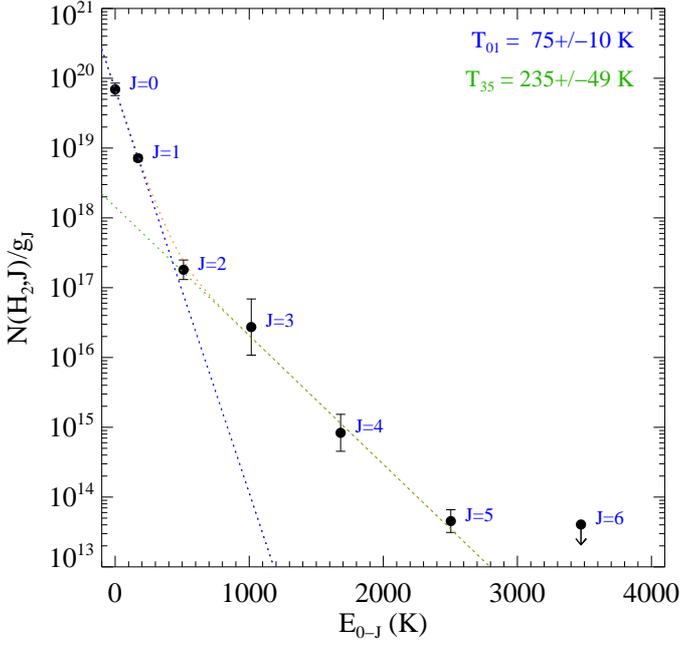}
\caption{Excitation diagram of H$_2$ towards \jt. The straight blue (resp. green) 
dotted line corresponds to the excitation of the low (resp. high) rotational levels. 
The orange curve correspond to the sum of the two populations in Boltzmann equilibrium at 
the two different temperatures.  \label{2140:f:h2ext}}
\end{figure}

At $\log N$(\h2) $\ge$~16, 
the low rotational level populations are 
most likely to be in thermal equilibrium. Therefore, the 
corresponding excitation temperature provides a good estimate of the gas kinetic 
temperature \citep[e.g.][]{Roy06,LePetit06}. 
In Fig.~\ref{2140:f:h2ext}, we show the population of the different rotational levels of H$_2$ 
as a function of the energy in these levels. The J=0 to J=2 levels can be explained by a single excitation 
temperature of $T_{\rm ex} \sim$~75K, which is lower than what is generally observed in both high-$z$ and low-$z$ 
H$_2$-bearing DLAs \citep[$\avg{T_{01}} \sim 150$~K,][]{Srianand05, Muzahid14} or at high Galactic latitude \citep[$\avg{T_{01}} \sim 125$~K,][]{Gillmon06} but 
very similar to what is observed in the Galactic disc or in the Magellanic clouds 
\citep[${T_{01}} \sim 80$~K,][]{Savage77, Tumlinson02, Welty12}. 
The higher excitation temperature (T$_{\rm ex} \sim$~240\,K) inferred for J$\ge$3 
rotational levels shows that they are not thermalised if arising from the same region and that an additional excitation process is at play. 
Generally, such a situation is considered to be due to UV pumping, although H$_2$-formation on 
dust grains could contribute to populate the high energy levels \citep{Wagenblast92}. 
Another possibility could be that the high-J levels are also populated by collisional excitation, 
but in  a warmer, presumably external layer of the cloud. The observed excitation 
diagram would then be the sum of a low temperature component (having a temperature of 
$\sim$75~K), contributing to most of the low-J column density and a warmer component (having 
a temperature of $\sim$240~K), 
responsible for the high-J column densities. If we assume such an excitation temperature
of the warm gas then the measured column densities at $J=4$ and $J=5$ are consistent with
a total $\log N$(\h2)$\sim$ 19 in this component.

\subsection{Neutral carbon: volumic density of hydrogen and electrons in the cold phase}

\begin{figure}
\centering
\begin{tabular}{c}
\includegraphics[bb=188 13 394 755,clip=,angle=90,width=\hsize]{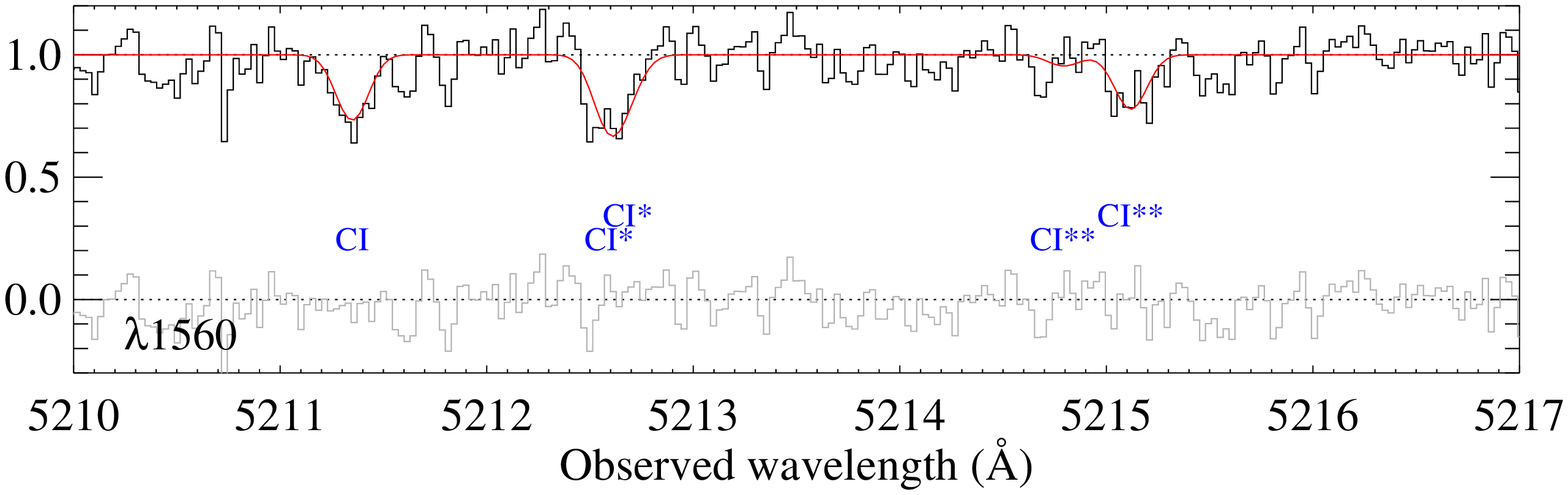}\\
\includegraphics[bb=164 13 394 755,clip=,angle=90,width=\hsize]{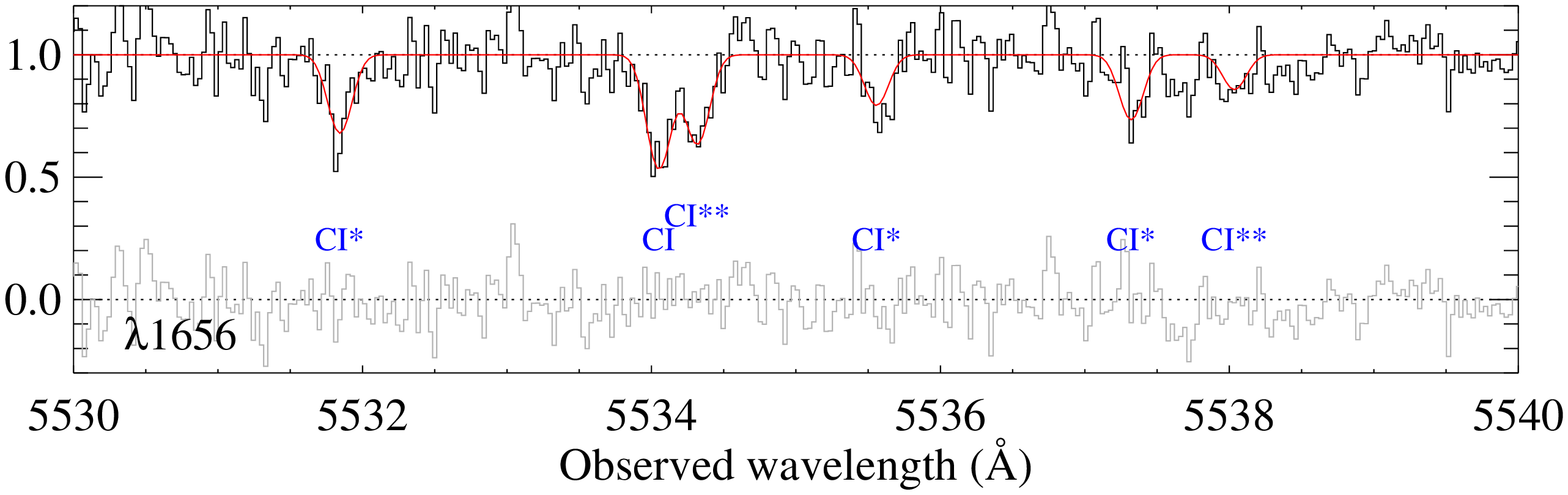}\\
\end{tabular}
\caption{Neutral carbon in the DLA at $\zabs = 2.34$ towards \jt. \label{2140:f:ci}}
\end{figure}

Absorption lines from neutral carbon in the ground state ($J=0$, \CI), first 
($J=1$, \CI$^{*}$) and second ($J=2$, \CI$^{**}$) excited levels are detected 
in our spectrum (Fig.~\ref{2140:f:ci}). 
From fitting the absorption lines around 1560 and 1656~{\AA} (in the DLA's rest-frame), 
that are located outside the \lya-forest, we derive the total (ground and excited 
states) \CI\ column density to be $\log N(\CI)=13.57 \pm 0.03$. Both the $\CI^*/\CI$ 
and $\CI^{**}/\CI$ ratios indicate high particle density in the cloud. 
From Fig.~2 of \citet{Silva02} and Fig.~12 of \citet{Srianand05}, we can already 
see that the density in the \CI-bearing gas should be of the order of $n_{\rm H} \sim 200$\,cm$^{-3}$. 
We note that at such density, the contribution from radiative excitation (UV pumping and 
cosmic microwave background) becomes negligible compared to collisions. While 
\CI\ and H$_2$ are not necessarily fully co-spatial, to the first order, we can 
consider this density to be that of the molecular gas. We note that this implies that 
the cloud is of pc-scale, when considering about 10\% of \HI\ in this phase (Sect. ~\ref{s:cl} 
and \ref{s:mod}).

\begin{table}
\centering
\caption{Additional species at $\zabs=2.34$ towards \jt. \label{J2140:t:ab}}
\begin{tabular}{l c c}
\hline
\hline
{\Large \strut} Species      & $\log N$          & $b$    \\    
             & $[\cmsq]$        & $[\kms]$      \\ %&           \\   

\hline
$\CI$, J=0 (g.s)               & 13.03 $\pm$ 0.04  & 5.0 $\pm$ 0.8 \\%& \\
$\CI$, J=1 ($^{*}$)         & 13.20 $\pm$ 0.04  & ''            \\%& \\
$\CI$, J=2 ($^{**}$)    & 13.02 $\pm$ 0.05  & ''            \\%& \\
total                         & 13.57 $\pm$ 0.03  &               \\%& \\
\hline
$\ClI$         & 13.67 $\pm$ 0.15  & 5-10\tablefootmark{a} \\
CO           & $<13.73$          &              \\     
\hline
$\OI$, J=2 (g.s.)                     & 17.9 \tablefootmark{b}      & 11 \\%& \\
$\OI$, J=1 ($^{*}$)         & 13.89 $\pm$ 0.11  & 9.5 $\pm$ 1.0\tablefootmark{c}             \\%& \\
$\OI$, J=0 ($^{**}$)    & 13.85 $\pm$ 0.08  & 9.5 $\pm$ 1.0          \\% & \\
$\SiII$, J=1/2 (g.s.)           & $\ge 16.16$ & \\
$\SiII$, J=3/2 ($^{\star}$) & 12.81 $\pm$ 0.03  & 9.5 $\pm$ 0.8 \\ %&                                 \\             
\hline
\end{tabular}
\tablefoot{
\tablefoottext{a}{The two values correspond to $b$ fixed to that of \CI\ and that obtained when left as free parameter.}
\tablefoottext{b}{Value from fitting the line, consistent with the [Zn/H] and [P/H] ratios. See text.}
\tablefoottext{c}{Because \OI$^{*}\lambda$1304 is noisy, we fixed $b$ to the value obtained 
from fitting \OI$^{**}\lambda$1306 alone.}
}
\end{table}

For the inferred metallicity, if we ignore the depletion of C into dust grains
and a possible under-abundance of carbon compared to other elements,
we expect $N(\CII) \simeq 6\times 10^{17}$~\cmsq. This implies $N$(\CI)/$N$(\CII)=10$^{-4.2}$.
Using Eq.~5 of \citet{Srianand05} we derive $n_e/\Gamma$ = 1.4$\times10^6$ when
we assume T$\sim$100 K. Here, $\Gamma$ is the photo-ionisation rate of C$^0$.
If we take $\Gamma \sim 2 \times 10^{-10}$ found in our galaxy \citep{Welty03} then
we get $n_e \simeq 3 \times 10^{-4}$. This value is much lower than the inferred
range (0.7$-$4.7) $\times 10^{-2}$ in other high-z H$_2$ absorbers using similar
approach \citep[see Table~4 of][]{Srianand05}. 
We note however that the lower inferred electron density could also result from 
a much higher actual photo-ionisation rate than the Galactic mean value or the inferred
$\log N(\CII)$ is overestimated. Indeed, carbon could actually be under-abundant, similarly to 
what has been suggested for the lack of C\,{\sc i} in a low-$z$ \h2\ DLA \citep{Srianand14}.
We will come back to this while discussing photo-ionisation models.

\subsection{Fine-structure levels of \OI}

\begin{figure}
\centering
\includegraphics[bb=164 13 394 755,clip=,angle=90,width=\hsize]{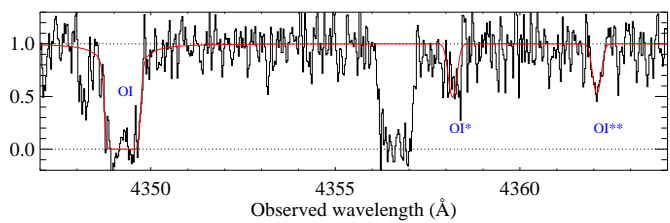}
\caption{Neutral oxygen fine-structure lines in the DLA at $\zabs = 2.34$ towards \jt. The 
absorption lines seen at $\lambda \approx$ 4348 and 4356.5~{\AA} correspond to \PII$\lambda$1301 
and \SiII$\lambda$1304, respectively.  \label{2140:f:oi}}
\end{figure}

We detect very strong \OI$\lambda$1302 absorption together with 
the absorption from the two fine-structure levels of 
the ground state (\OI$^{*}\lambda$1304 and \OI$^{**}\lambda$1306, see Fig.~\ref{2140:f:oi})
\footnote{Note that the ground state of neutral oxygen is comprised of the 2s$^2$2p$^4$\,$^3$P$^e_{\rm J=2,1,0}$ 
fine-structure levels, the ground-state of which correspond to $J=2$}. This is, to our knowledge, 
the first detection of excited neutral oxygen in an intervening absorption system. However, 
\OI$^{*}$ and \OI$^{**}$ are more commonly detected in DLAs associated to GRB afterglows 
\citep{Vreeswijk04,Chen05,Fynbo06}, where it is generally interpreted as an indicator of high volumic density.
Unfortunately, the weaker \OI$\lambda$1355 line is not covered by our spectrum and other \OI\ lines 
are far in the blue and strongly blended in the \lya\ forest. Fitting the \OI$\lambda$1302 
absorption line with a single component gives $\log N(\OI)= 17.9 \pm 0.2$. The Doppler parameter 
is left free and we find $b \simeq 11$~\kms, surprisingly close to the $b\sim 12$~\kms\ expected from 
the velocity width of the profile ($\Delta_v \simeq 28~\kms$, Sect.~\ref{s:kin}), using the relation between these 
quantities from \citet{Noterdaeme14}. The derived oxygen abundance would then be [O/H]~$\sim -1.2$, 
i.e. consistent with that derived from zinc and phosphorus.

We derive $\log N(\OI^{*}) = 13.89 \pm 0.11$ and $\log N(\OI^{**}) = 13.85 \pm 0.07$ 
from fitting \OI$^{*}\lambda$1304 and \OI$^{**}\lambda$1306. We note however that the former line 
is noisy and possibly affected by blends. We therefore fixed the Doppler parameter to the value from 
\OI$^{**}\lambda$1306 alone but note that this has little influence on the derived column density.

Using the \HI\ collisional rate constant from \citet{Abrahamsson07} and radiative decay rates from 
\citet{Galavis97},  
\citet{Jenkins11} have found that $N(\OI^{*})/N(\OI^{**})$ can be a good indicator of local
kinetic temperature (see their Fig~15). For T$\sim$80 K as suggested by $T_{01}$ we expect the 
ratio to be higher than 2. However, the observed ratio ($\sim$ 1) is rather consistent with a gas temperature of 1000 K.
It is clear from Fig~6 of \citet{Silva02} that the observed ratio of  $N(\OI^*)/N(\OI)$ 
and $N(\OI^{**})/N(\OI)$ cannot be reproduced by a single density for $T$~=~150\,K.
However, this ratio can be reproduced for $T=1000$\,K and $n_H \sim 300$\,cm$^{-3}$. It is also clear from Fig~6 of \citet{Silva02} 
that the UV pumping may not be very efficient for the observed range in ratios of column densities.
Therefore, the observed fine-structure level populations of \OI\
favours the presence of warm neutral gas in addition to the dense cold gas probed by
the low rotational level population of \h2. Direct excitation by infra-red photons is also a possibility.

\subsection{Excited \CII\ and \SiII}

Strong and saturated \CII$^{*}\lambda$1335 absorption is detected in this 
system, with an equivalent width of roughly 0.2~{\AA}. Because the \CII$^*$ 
column density is directly related to the gas cooling through \CII\,$158\,\mu m$ 
emission (${^2P}_{3/2}~(\CII^{*}) \rightarrow {^2P}_{1/2}~(\CII)$), it allows in 
principle to derive the surrounding 
UV flux by equating this cooling rate with the heating from photo-electric effect onto dust 
grains (\citealt{Wolfe03}), although cosmic-ray heating may contribute significantly 
as well \citep{Dutta14}. Unfortunately, the strong saturation of \CII$^{*}\lambda$1335 
prevents meaningful measurement or useful limit on $N(\CII^{*})$ and the weaker  
\CII$^{*}\lambda$1037 is blended with \lya\ forest absorption. Nonetheless, the very 
strong absorption should still indicate significant cooling and hence star-formation activity.

\begin{figure}
\centering
\begin{tabular}{c}
\includegraphics[bb=219 228 394 617,clip=,angle=90,width=0.48\hsize]{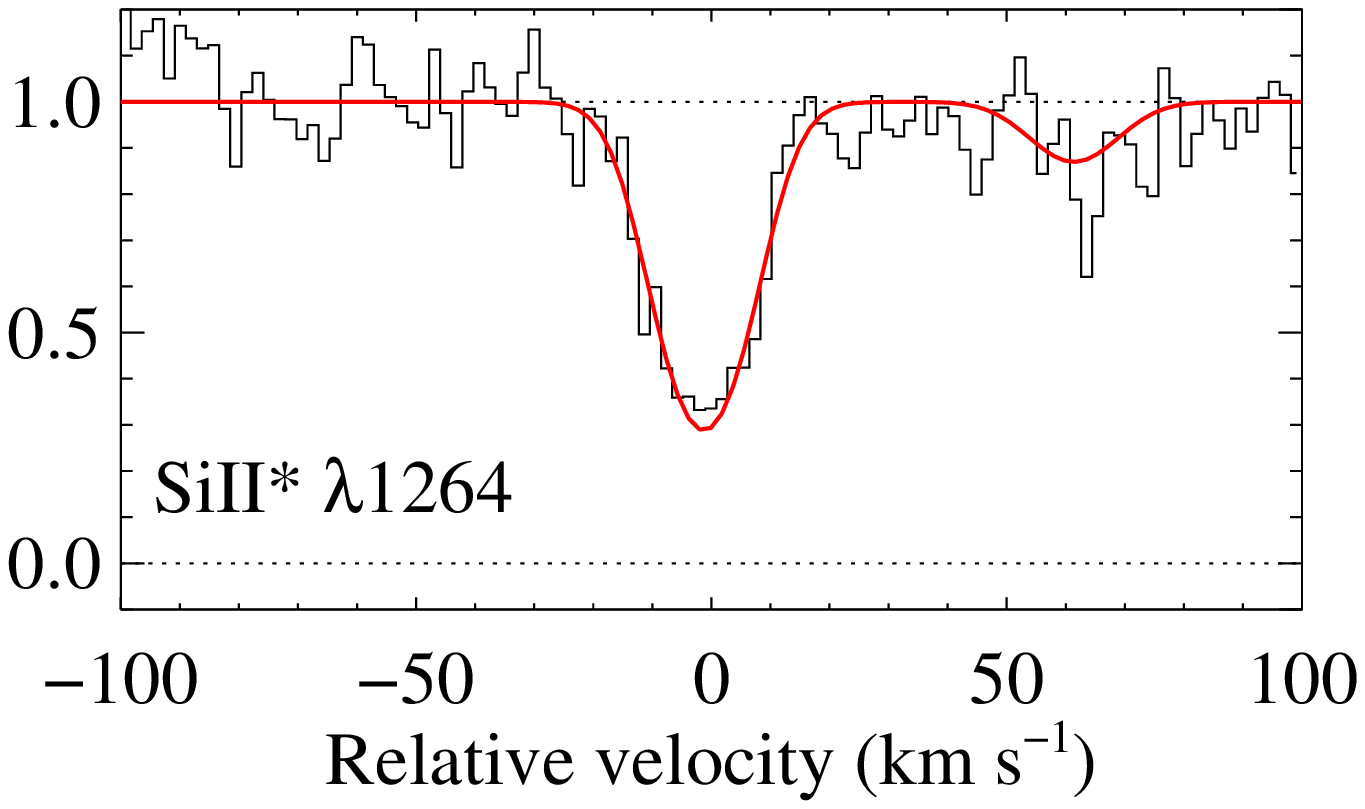}\\
\includegraphics[bb=164 228 394 617,clip=,angle=90,width=0.48\hsize]{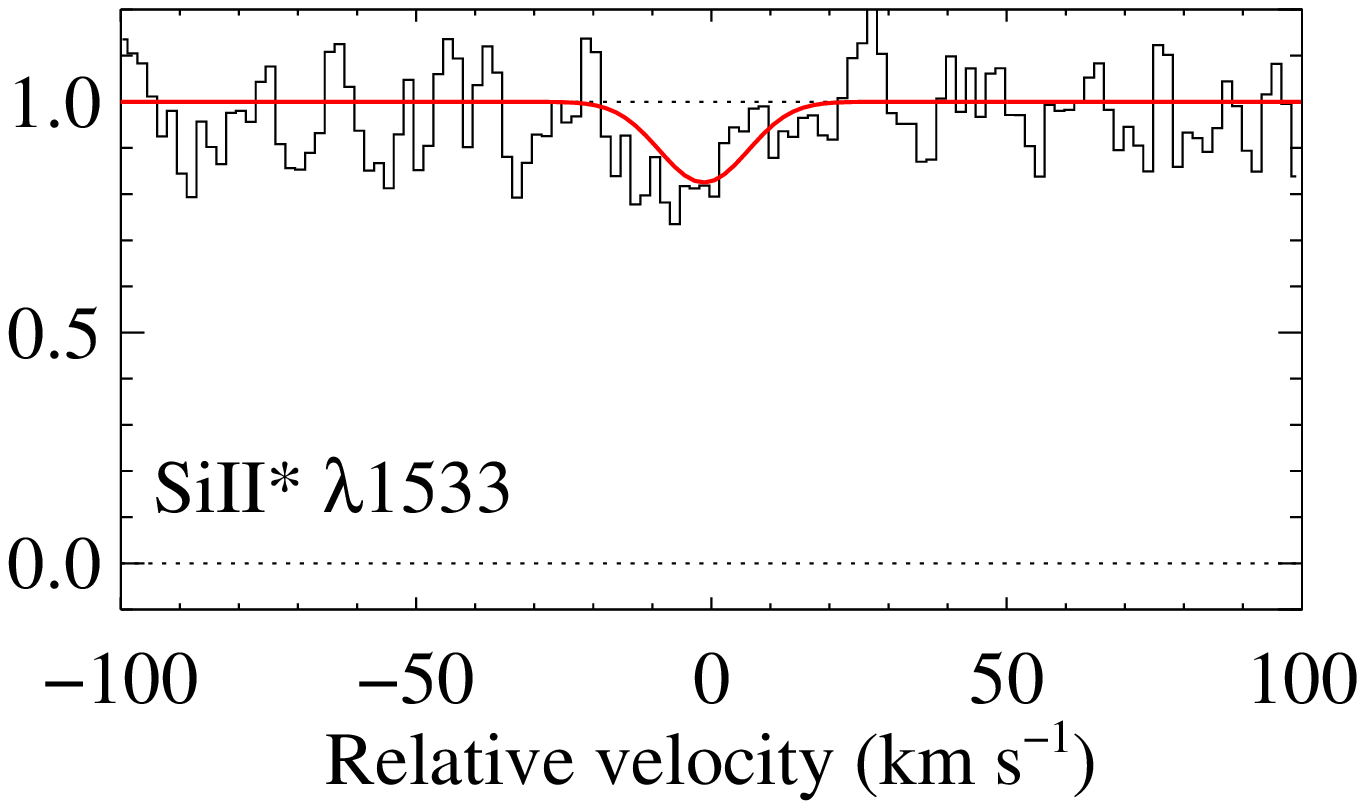} \\
\end{tabular}
\caption{\SiII$^{*}$ at $\zabs=2.34$ towards \jt. The absorption at $+50~\kms$ in the 
\SiII$^{*}\lambda1264$ panel is due to \SiII$^{*}\lambda1265$. \label{2140:f:siIIs}}
\end{figure}

The excited fine-structure level of ionised silicon ($\SiII({\rm J=3/2})$ or $\SiII^{*}$) 
is also detected in this system. The 
first detection of this kind in an intervening system was reported 
by \citet{Kulkarni12} in the extremely strong DLA towards \jonze, 
which arises from a young low-impact parameter galaxy with SFR~$\sim 25\,\msyr$ 
\citep[][]{Noterdaeme12a}. More recently, \citet{Neeleman15} suggested that this line 
arises from highly turbulent ISM in young star-forming galaxies.
Here, we unambiguously detect \SiII$^{*}\lambda$1264, 
\SiII$^{*}\lambda$1265 and \SiII$^{*}\lambda$1533 (Fig.~\ref{2140:f:siIIs}) 
from which we derive $\log N(\SiII^{*}) = 12.81 \pm 0.03$. Assuming an intrinsic 
solar abundance ratio of silicon with respect to zinc and 0.2~dex depletion 
onto dust as seen in different galactic and DLA environments (Fig.~\ref{f:deple}), 
we estimate, in the gas phase, $N(\SiII)$ to be about a few times $10^{16}$~\cmsq, implying 
$N(\SiII^{*})/N(\SiII) \sim 10^{-4}$. It is well known that the electron collisional
cross-section is three orders of magnitude higher than that of hydrogen for the 
fine-structure excitation of \SiII. If we consider only the collisions
by electrons followed by radiative de-excitation we get $n_e\sim 0.8$~cm$^{-3}$
if $\SiII^*$ absorption is associated to the \h2\ gas with T$\sim$100 K \citep[see][]{Srianand00b}. 
We get $n_e \sim 0.06$ cm$^{-3}$ if we assume the gas temperature to be 1000~K.

\subsection{Kinematics \label{s:kin}}

\begin{figure}
\centering
\includegraphics[bb=219 228 450 647,clip=,angle=90,width=0.65\hsize]{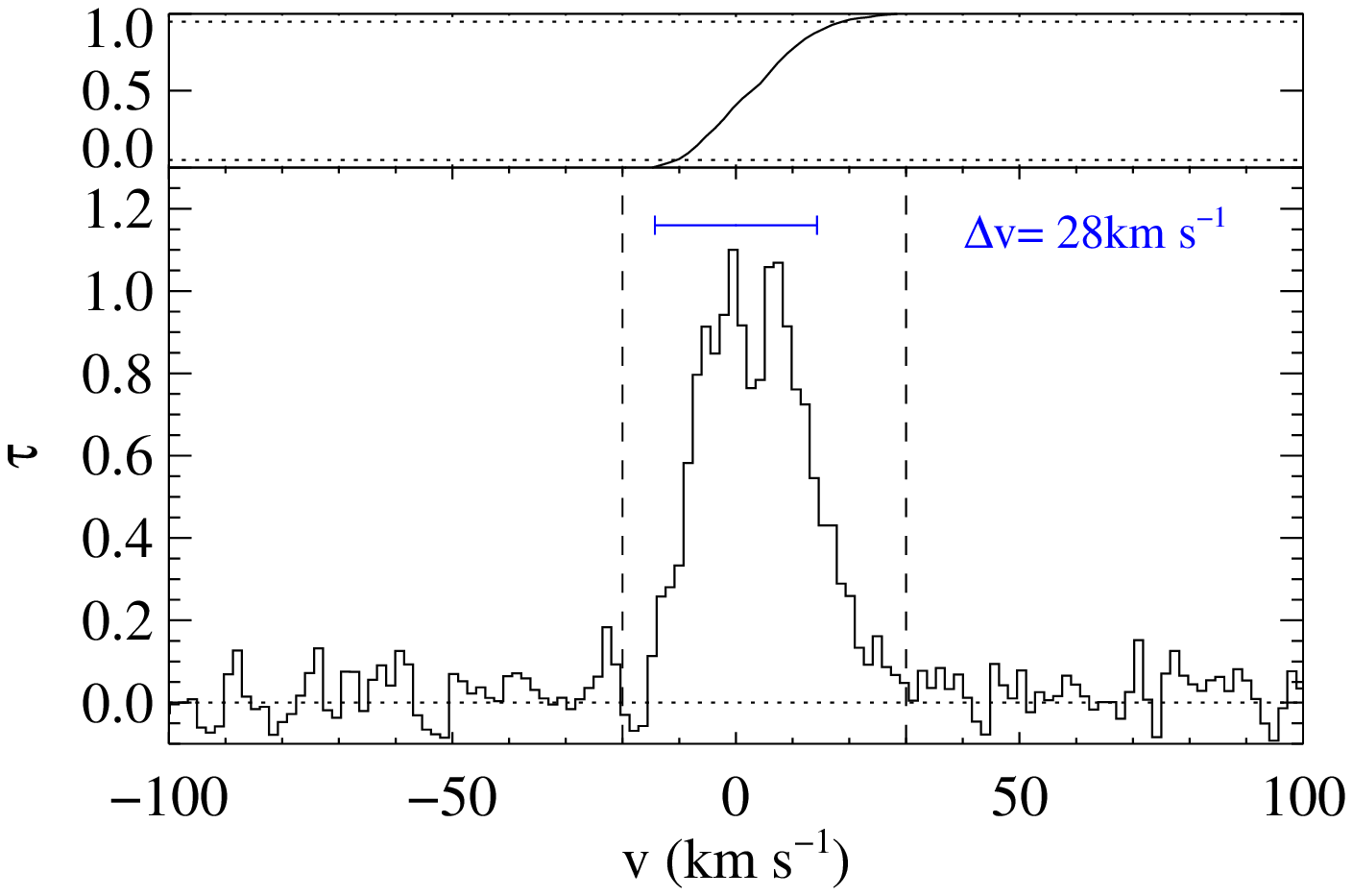}
\includegraphics[bb=219 228 394 647,clip=,angle=90,width=0.65\hsize]{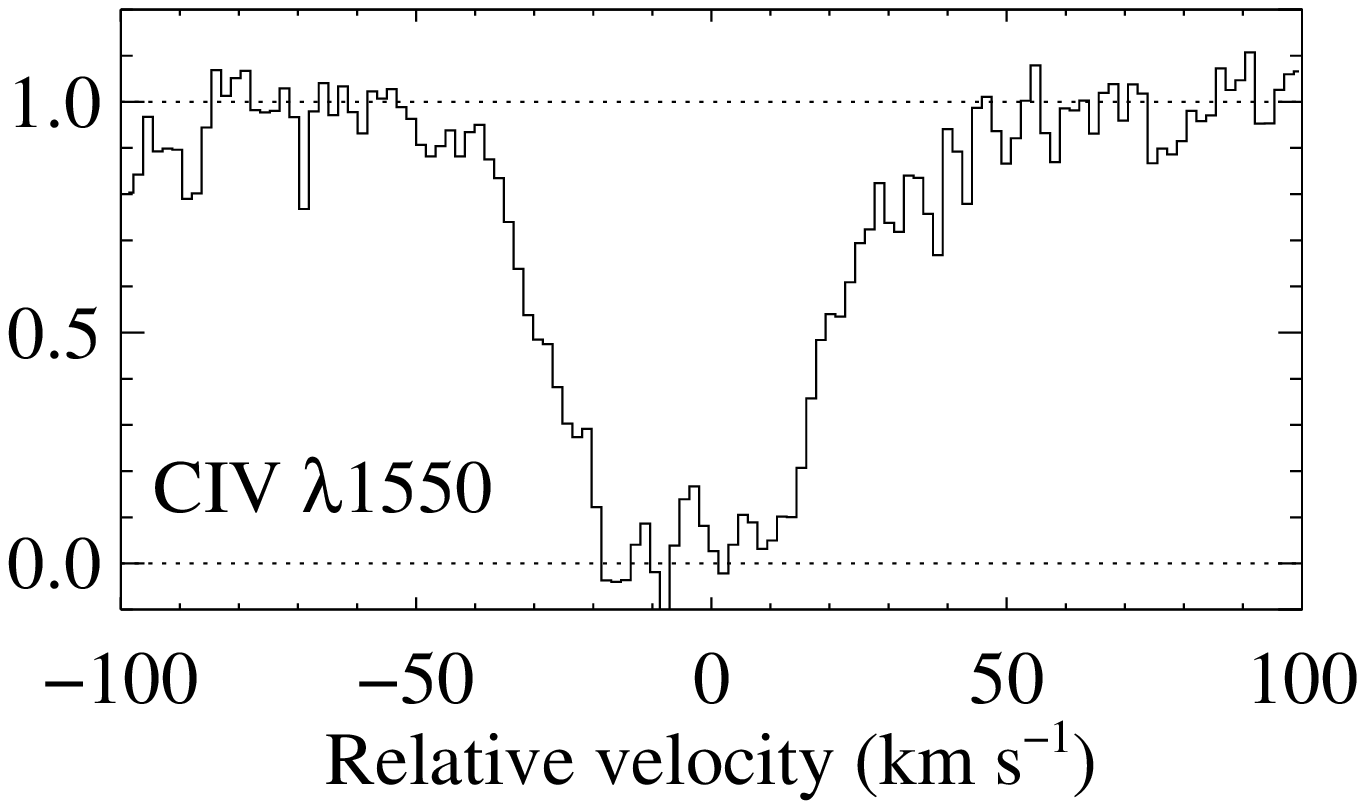}
\includegraphics[bb=219 228 394 647,clip=,angle=90,width=0.65\hsize]{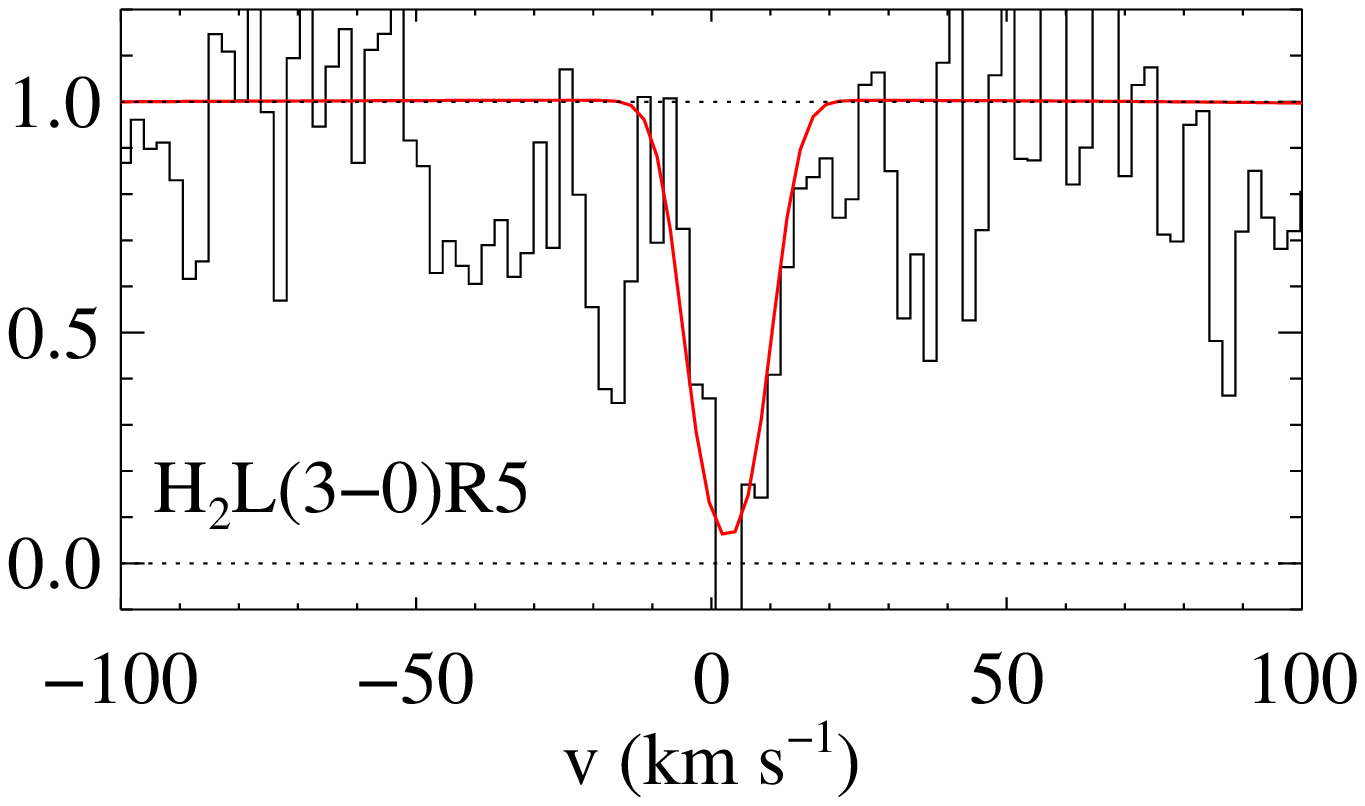}
\includegraphics[bb=219 228 394 647,clip=,angle=90,width=0.65\hsize]{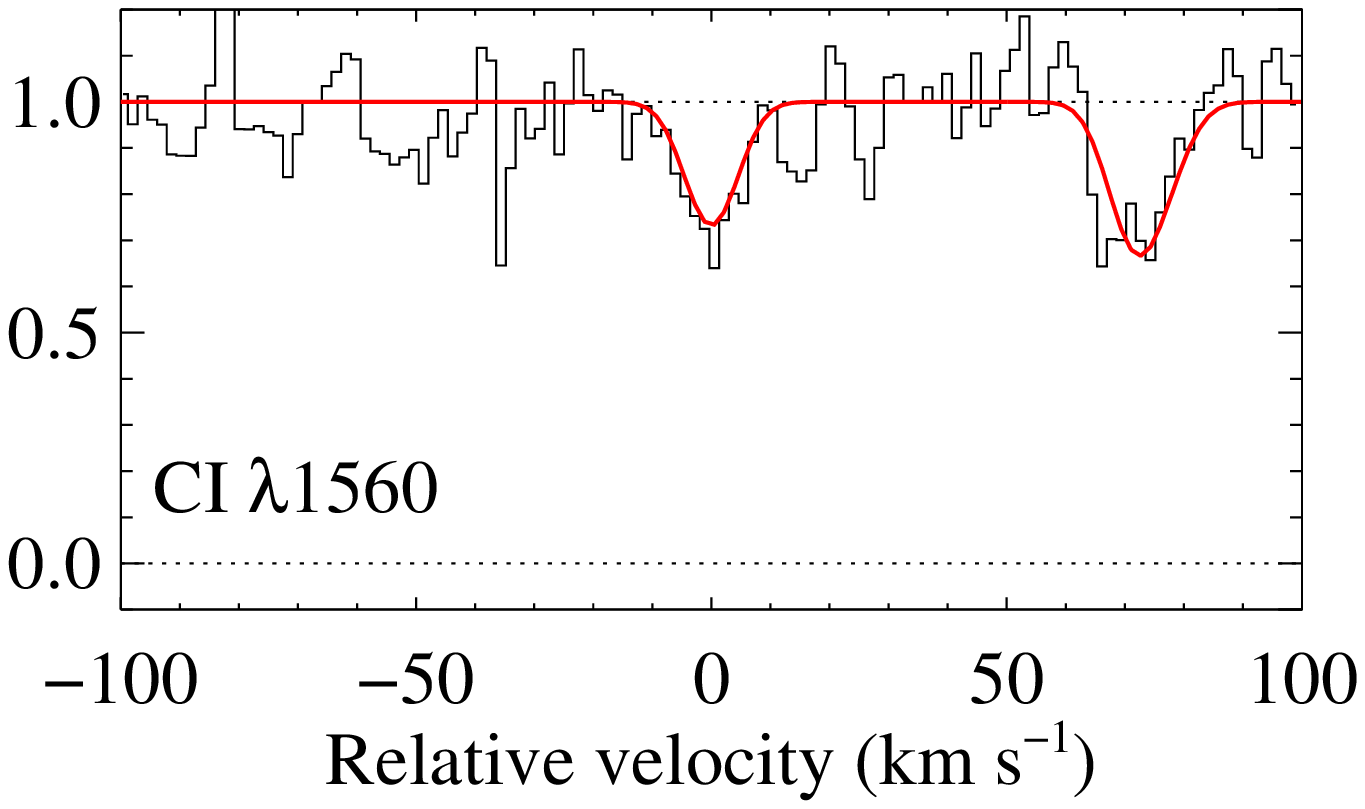}
\includegraphics[bb=219 228 394 647,clip=,angle=90,width=0.65\hsize]{J2140-ClI-1347-2.339946_alt.ps}
\includegraphics[bb=164 228 394 647,clip=,angle=90,width=0.65\hsize]{J2140-SiII1s-1264-2.339946.ps}
\caption{Measurement of metal-absorption kinematics (using $\FeII\lambda1611$, top panel) 
at $\zabs=2.34$ towards \jt. The other panels represent the normalised spectrum around 
several species on the same velocity scale.\label{2140:f:kin}}
\end{figure}

The study of the velocity extent of different species can provide additional view 
on the overall puzzle (Fig.~\ref{2140:f:kin}). 
The top panel presents the measurement of the velocity width defined as the interval 
comprised between 5\% and 95\% of the total optical depth of a low-ionisation species. 
We find $\delta v = 28$~\kms\ which is in the low range of the distribution for the overall 
population of DLAs. This may be indicative of a low-mass system \citep{Ledoux06a}. 
The width\footnote{Because 
\CIV\ is saturated, the kinematics cannot be quantified using the 5-95\% range of the 
optical depth.} of the \CIV\ profile (FWHM$\sim$40~\kms) is also among the lowest measured in DLAs \citep{Fox07a} which 
suggests a quiet environment: its origin could be due to photo-ionisation by nearby massive 
stars rather than collisional ionisation in a hot halo or in galactic winds \citep[e.g.][]{Oppenheimer06}. 
 We may 
anticipate that in-situ star formation could contribute to the high excitation of 
\SiII\ and \OI. These excited species also have $b \sim 10$~\kms, twice 
larger than what is seen in the cold phase ($b(\CI) \sim 5$~\kms), hence possibly 
arising from a warmer and more turbulent gas.
In conclusion, we are likely observing cold gas in a star-forming region of a 
possibly low-mass galaxy. The presence of nearby massive stars would result in warming and exciting the 
outer layers of the cloud, giving rise to the observed \SiII$^*$ and \OI$^*$.

\subsection{Comparisons with models \label{s:mod}}

In this section, we intent to understand the abundances and excitation 
of the different species in a single picture by comparing the observations 
with models using the spectral simulation code {\sc Cloudy} \citep{Ferland98}, 
including the micro-physics of H$_2$ \citep{Shaw05}.

\begin{figure*}
\centering
\begin{tabular}{c c c}
\includegraphics[bb=125 280 432 513,clip=,width=0.30\hsize]{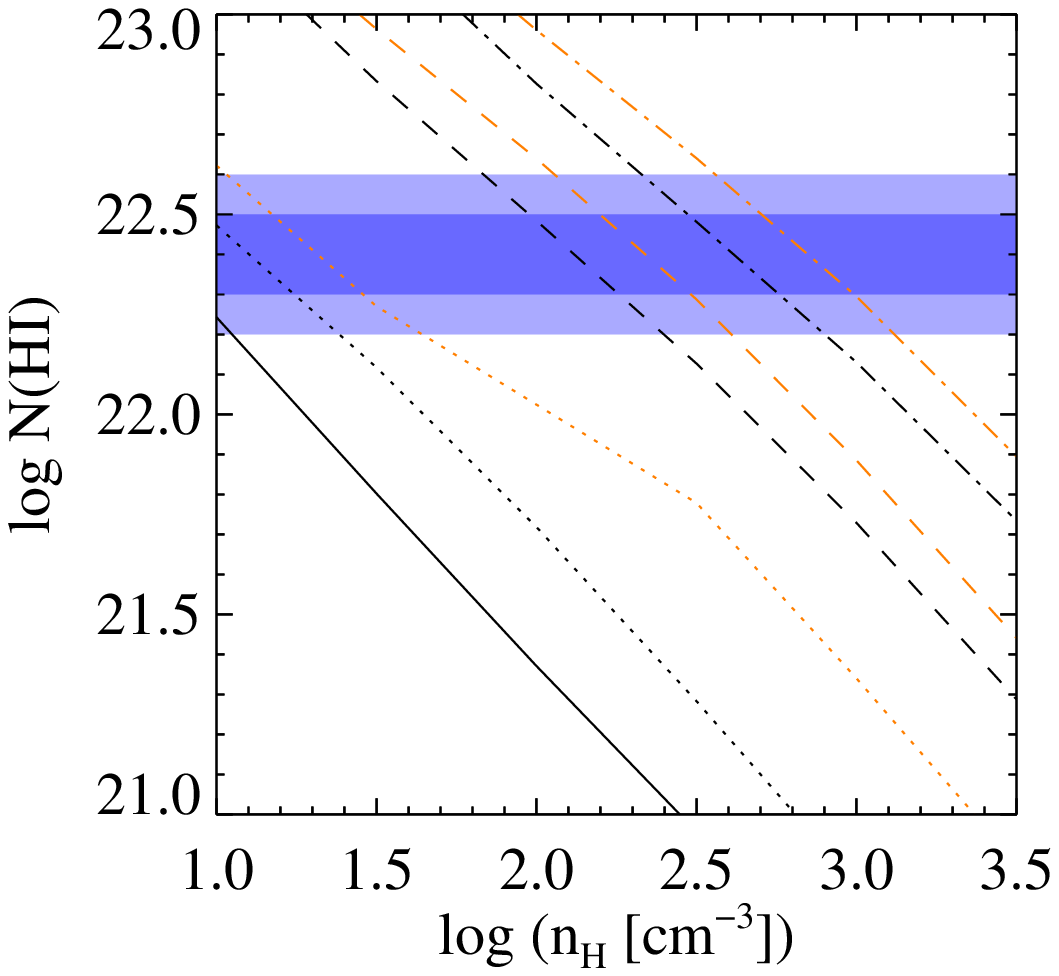} &
\includegraphics[bb=125 280 432 513,clip=,width=0.30\hsize]{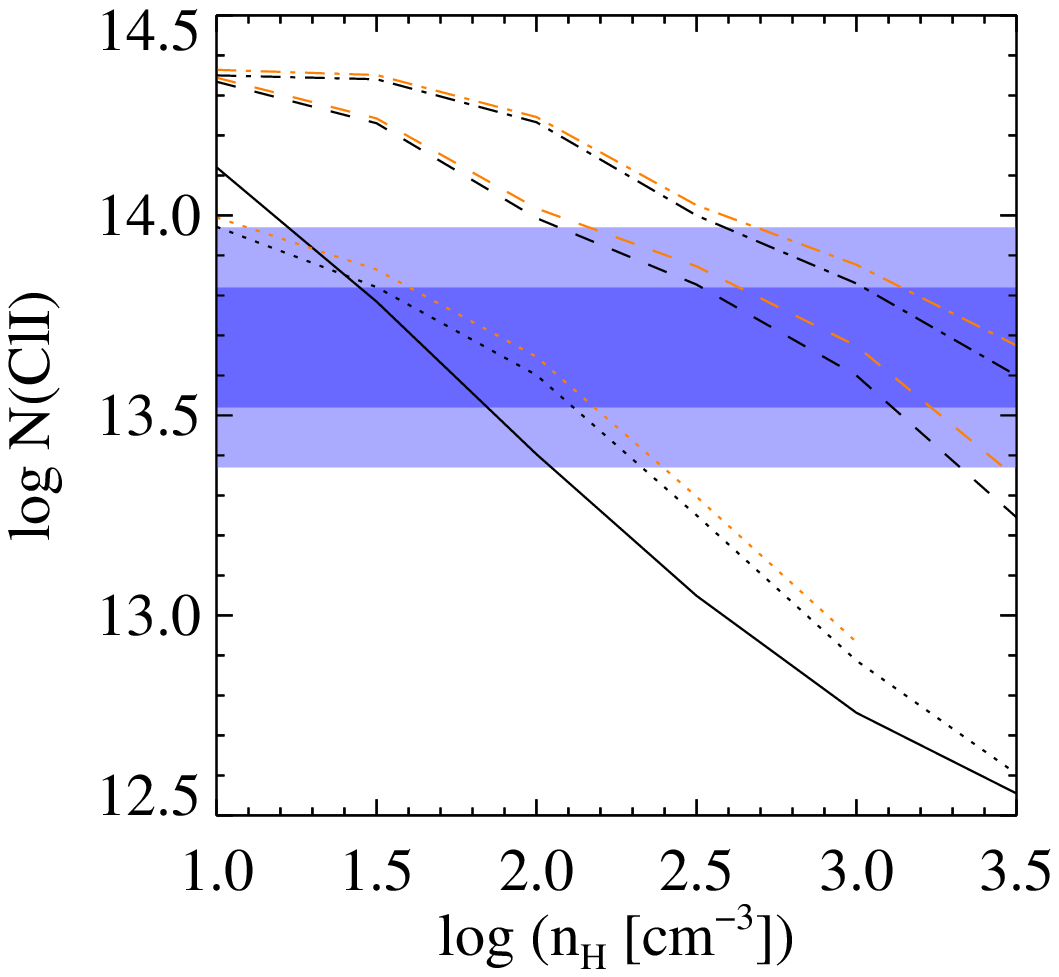} &
\includegraphics[bb=125 280 432 513,clip=,width=0.30\hsize]{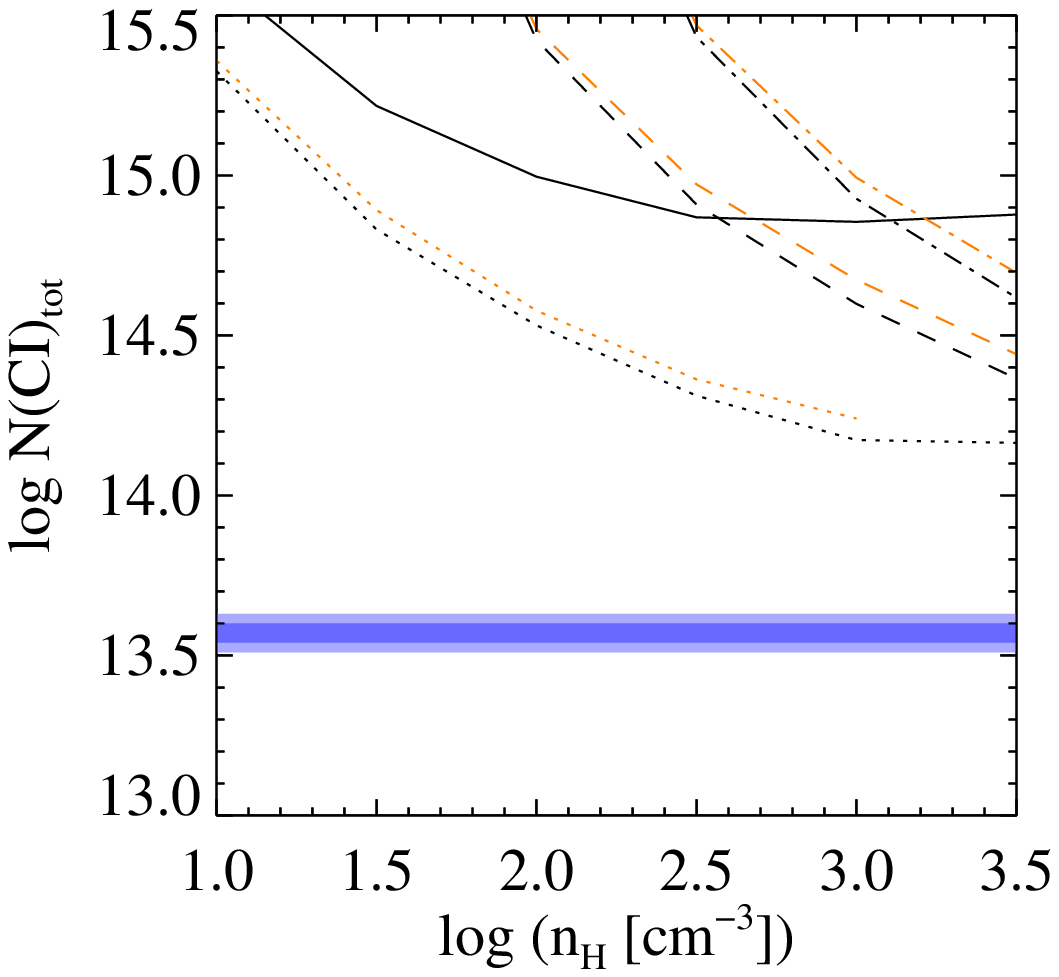} \\
\includegraphics[bb=125 280 432 513,clip=,width=0.30\hsize]{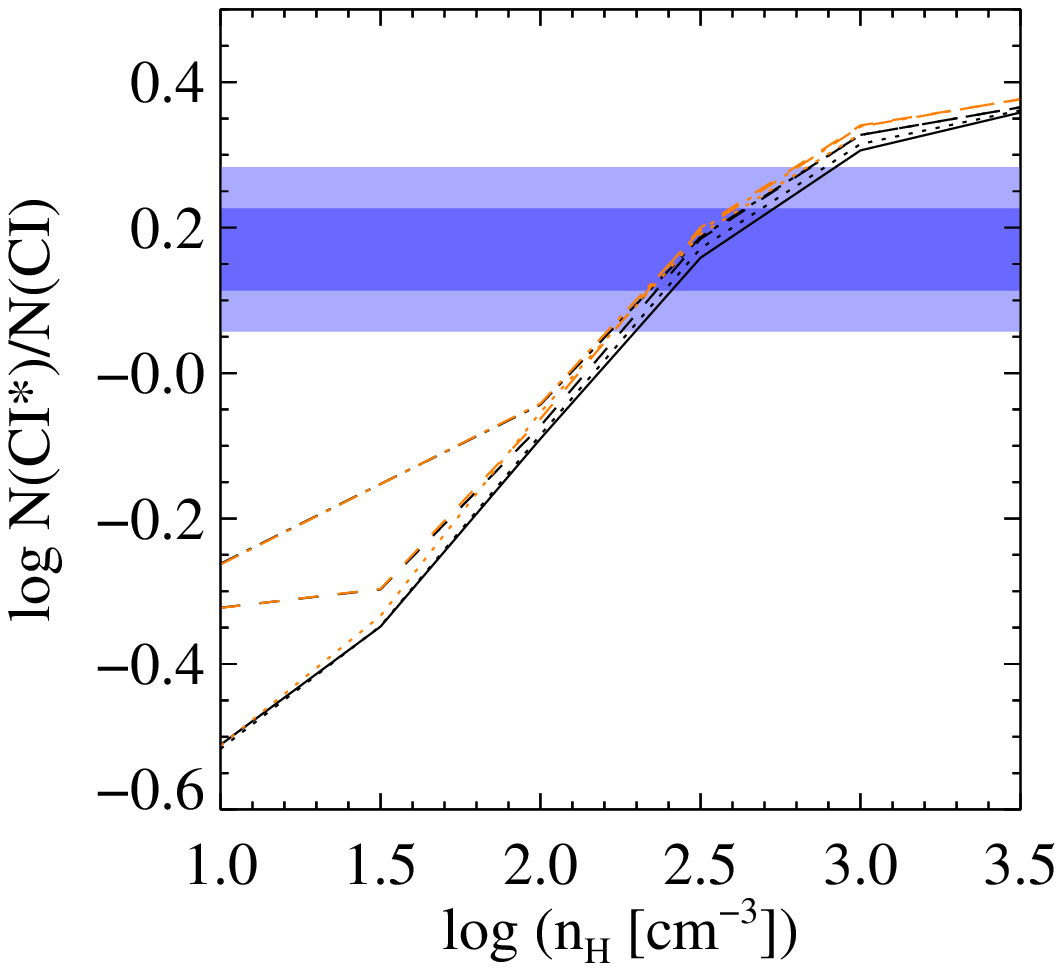} &
\includegraphics[bb=125 280 432 513,clip=,width=0.30\hsize]{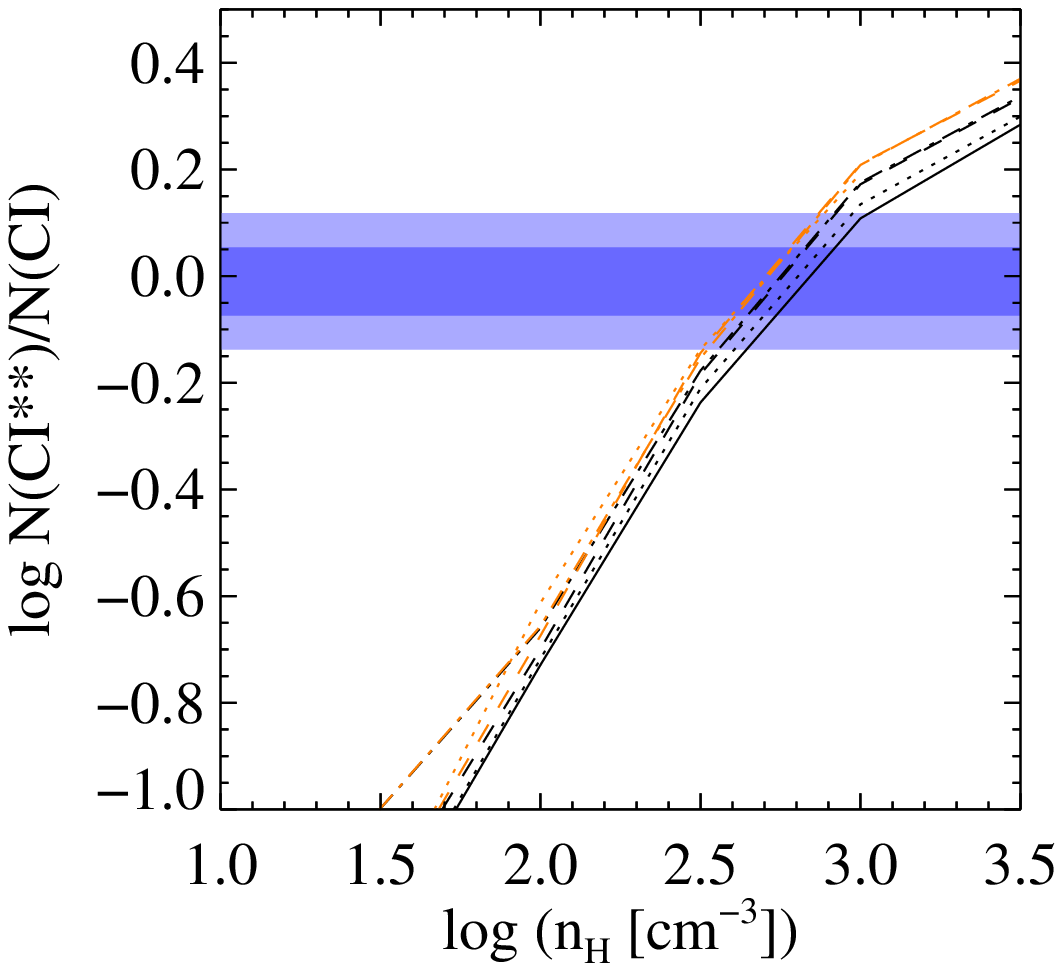} &
\includegraphics[bb=125 280 432 513,clip=,width=0.30\hsize]{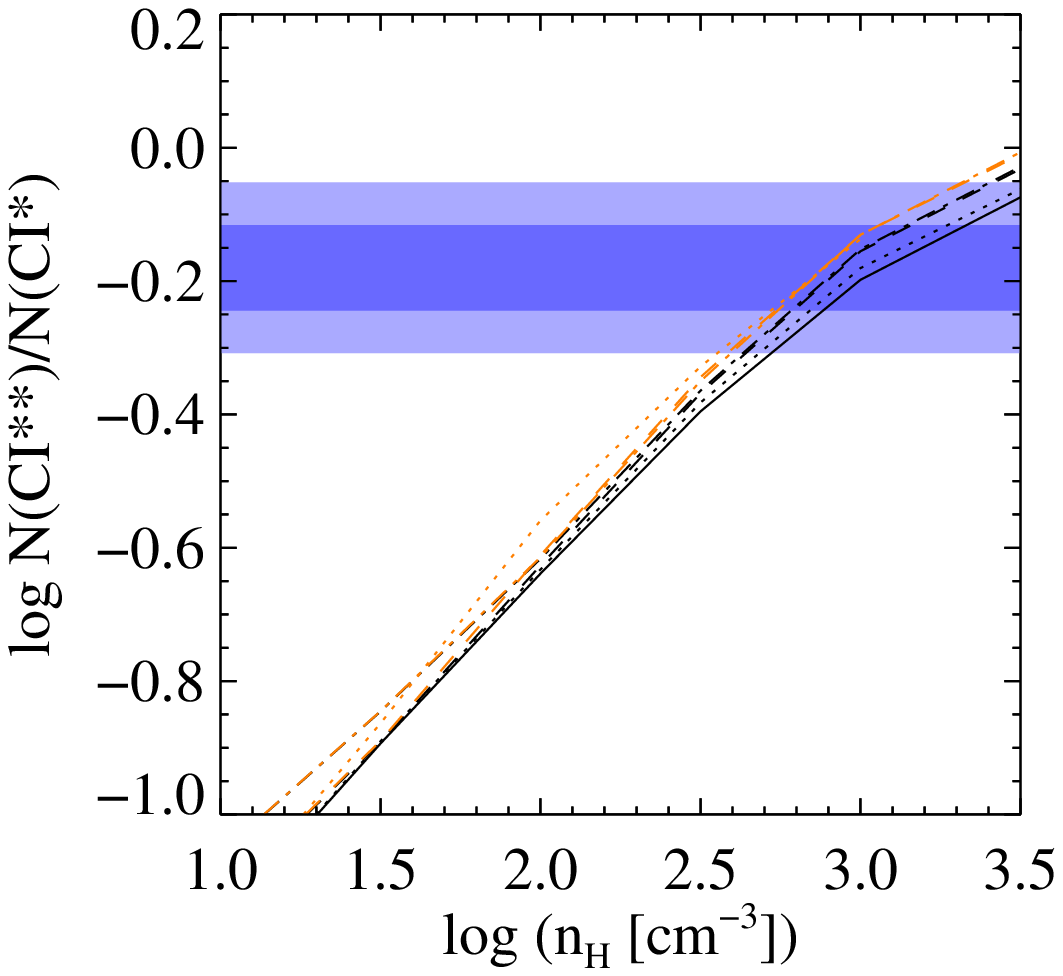} \\
\includegraphics[bb=125 235 432 513,clip=,width=0.30\hsize]{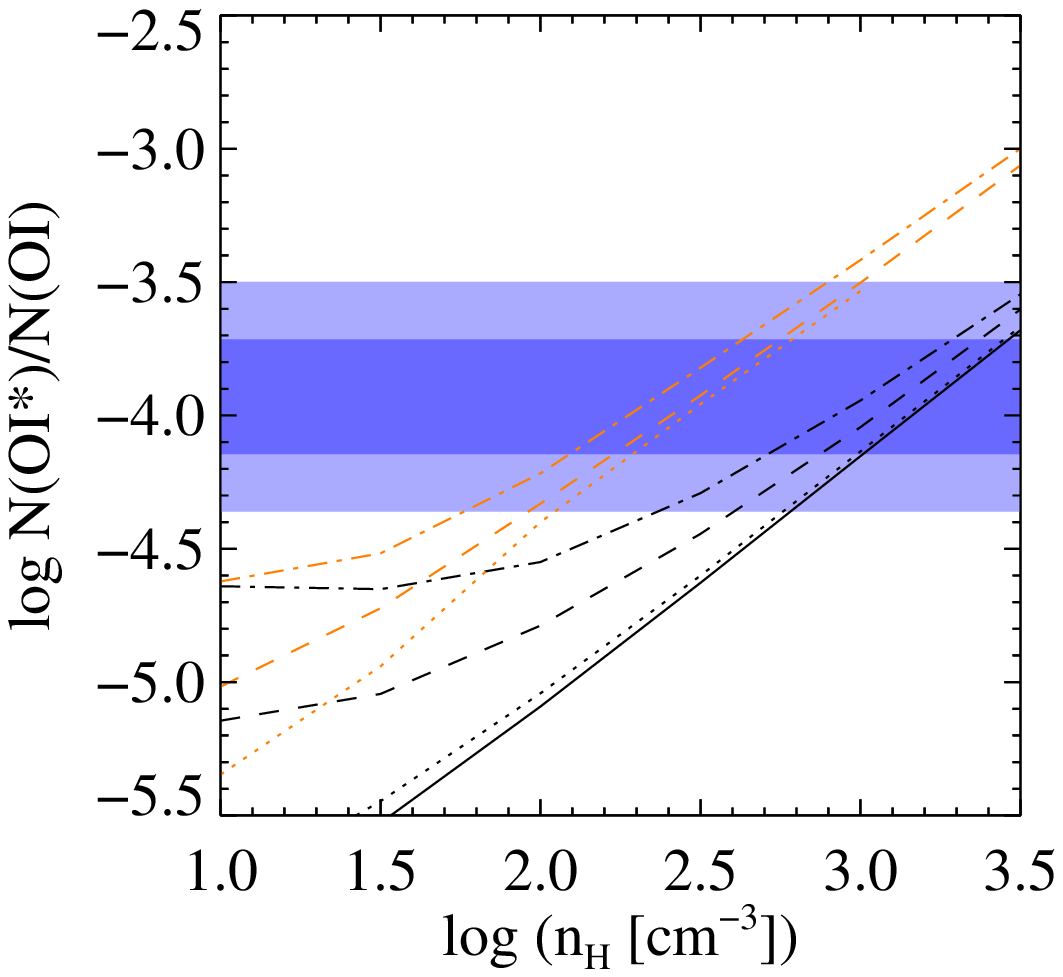} &
\includegraphics[bb=125 235 432 513,clip=,width=0.30\hsize]{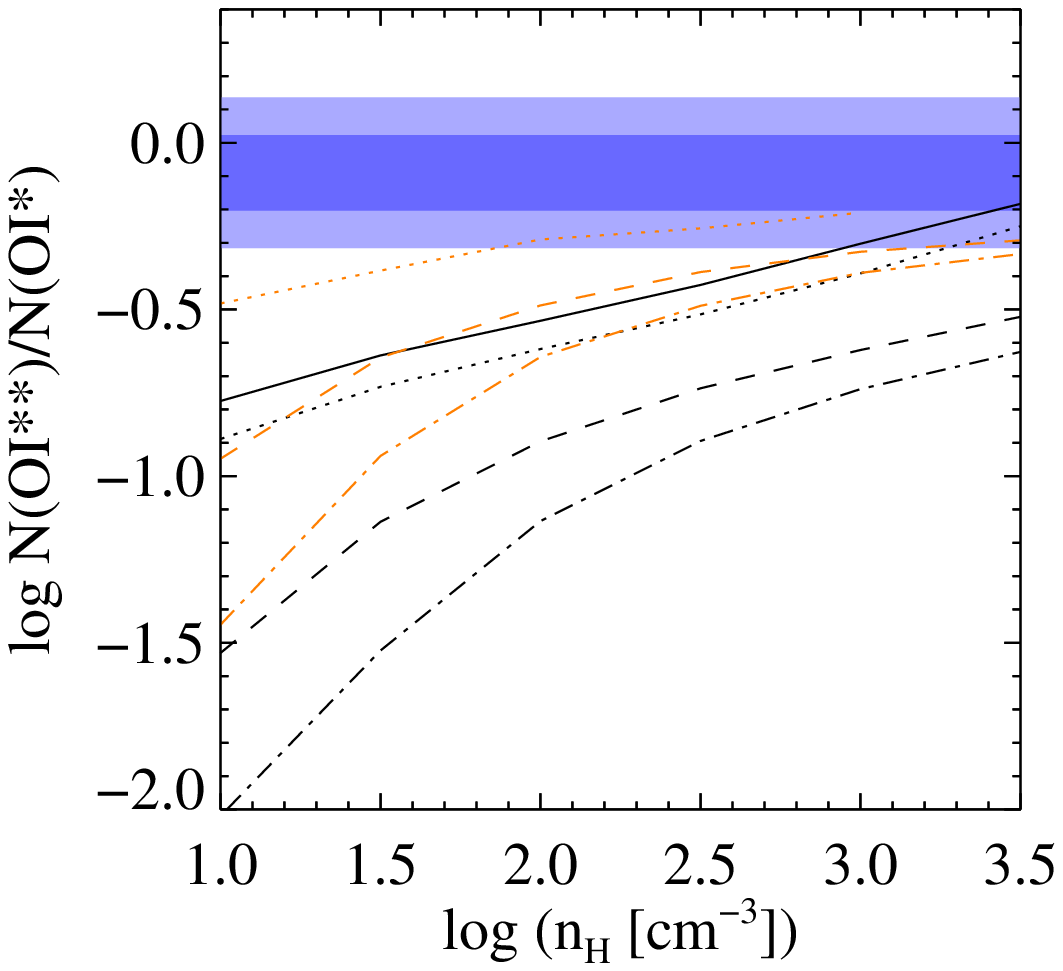} &
\includegraphics[bb=125 235 432 513,clip=,width=0.30\hsize]{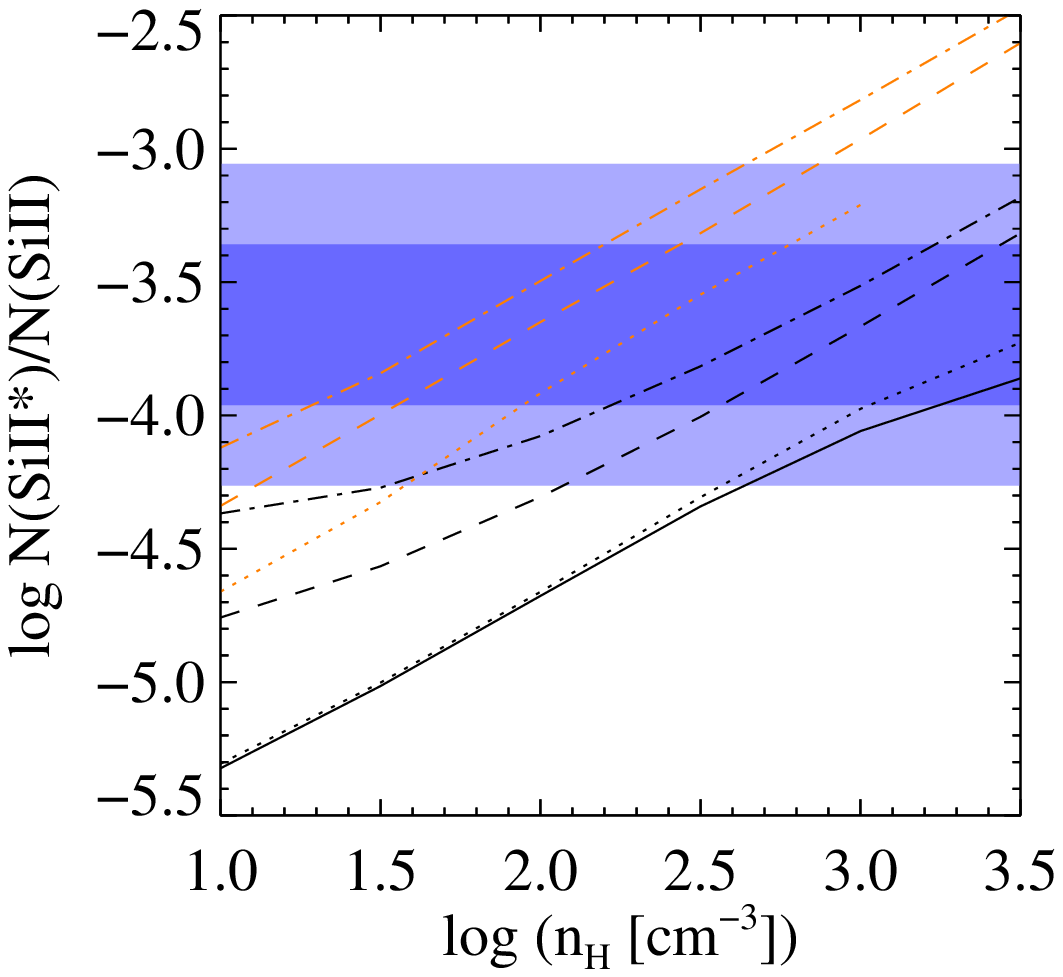} \\
\end{tabular}
\caption{Results of photo-ionisation models. 
For each hydrogen
density the calculation is stopped when the observed $N$(\h2) is reached. 
The blue shaded regions give the 1 and 2$\sigma$ range on the {\sl total} column density (top) 
or column density ratios (bottom) from the observations. The solid, dotted, dashed and dashed-dotted 
lines correspond to an in-situ radiation field of respectively $\chi = 0,1,10,30$ times the local 
Galactic field. The cold-gas-only model ($T=$~80~K, $\log N($H$_2)=20.13$) is represented by black 
lines while the orange lines represent the model with an additional warm component ($T=$~300~K and 
$\log N($H$_2)=19$).
\label{Fig:mod1}}
\end{figure*}

To start with, we consider the case of absorbing gas not associated to any 
star-forming region. In this case we assume the ionising radiation to be the
meta-galactic UV background dominated by QSOs and Galaxies \citep[see for example][]{Haardt12}. 
We assume the metallicity and dust depletion to be identical to the observed values. 
For simplicity we assume the gas temperature to be 80~K
as inferred from $T_{01}$ and did not ask the code to self-consistently
obtain this temperature. The calculations are stopped when the total
\h2\ column density in the model reaches the observed value. We run
the model for a range of hydrogen densities. These ``cold'' models 
correspond to the solid black lines in Fig.~\ref{Fig:mod1}. 
We find that the observed abundance of chlorine in the cold gas can be reproduced 
for $1.4\le{\log~n_{\rm H}}\le 1.8$. However, this density range is inconsistent with 
the \CI\ fine-structure excitation and the density is also too low to explain the 
high excitation of oxygen fine-structure levels. 
Conversely, a higher density model ($\log n_{\rm H} \sim 2.7$), consistent with the 
\CI\ fine-structure excitation under-predicts the abundance of chlorine. This happens 
because H$_2$ is then formed more efficiently and reaches the observed column density 
at a much lower total gas column. Finally, we note that the predicted \CI\ column 
density in the cold phase is about a factor 20 above the observed values.

Therefore, we considered another set of models where we also include
the local radiation field. For this we considered the Galactic UV
background given by \citet{Habing68} scaled by $\chi$. For simplicity, 
we also scale the cosmic ray density by the same amount. We considered 
$\chi = 1, 10$ and 30 (respectively dotted, dashed and dash-dotted curves). 
We also consider a set of model with the addition of a warmer component 
($T\sim300$~K) that contains $\log N($\h2$) \sim 19$, as discussed in 
Sect.~\ref{s:h2diag}. These models (``cold+warm'') are shown as orange curves 
in Fig.~\ref{Fig:mod1}. 

We can see that irrespective of the radiation field and 
the inclusion or not of a warm envelope, the fine-structure excitation of 
C\,{\sc i} indicates a high density as derived above. The inclusion of a 
warm phase makes the different ratios in better agreement with each other, 
although \CI$^{**}$ tends require slightly higher excitation than \CI$^*$. 
As we already consider excitation by cosmic microwave background,
collisions and UV pumping self-consistently in {\sc cloudy}, the additional
pumping could be related to local infra-red radiation field not
explicitly included in our models. However, if we allow for 
1.5$\sigma$ uncertainties in the measured ratios then our models
can consistently reproduce the observations when  2.5$\le {\rm log~n_H}\le$3.0.
For this range in density the observed $N$(\ClI) requires $\chi$ close to or 
slightly below 10. In that case, 
the cold phase should contain between 10\% and 50\% of the total hydrogen 
column density while these values are between 30\% and 80\% %for 
when adding 
the warmer phase.

The addition of the warm phase improves significantly the agreement between 
the predicted and observed ratio of \OI\ fine-structure levels, in particular 
the $\OI^*/\OI$ ratio. In turn, $\SiII^*/\SiII$ becomes now over-predicted 
for the above $\chi$ and $n$ while 
the changes are still not good enough to explain the high \OI$^{**}$/\OI$^*$ 
ratio. In the case of silicon, we remind that we were unable to measure $N(\SiII)$ 
directly due to line saturation, so the disagreement remains reasonable. 
We also note that a lower density in the warm phase ($\log n \sim 2$) than in the 
cold phase would help making the agreement between model and observations better.
Indeed, as noted before from Fig~6 of \citet{Silva02}, it is clear that when T= 150~K (close
to what we use in our models) the very similar value of 10$^{-4}$
found for $N$(\OI$^*$)/$N$(\OI) and $N$(\OI$^{**}$)/$N$(\OI)
cannot be produced by a single density. 
In turn, from the 
bottom panels of the same figure we can see it will be 
easier to find a solution when the gas temperature is 1000\,K. 
Therefore, it is also possible that the fine-structure excitation
of \OI\ originates from an even warmer (i.e $\ge 1000~K$) phase than the one 
probed by \h2\ absorption, presumably with a lower density. 
This phase would not contribute to the observed \ClI\ column density but 
would contain the remaining \HI. 

However, the addition of an even warmer phase without H$_2$ that could 
explain \OI$^*$\ and \OI$^{**}$ will not solve the problem with \CI. Also, 
as can be seen from the corresponding panel in Fig.~\ref{Fig:mod1}, increasing 
further the ionisation rate of carbon is not an option as it also alters the \h2\ equilibrium. 
Reducing the recombination rate can be achieved by reducing the electron 
density that will not alter the \h2\ equilibrium abundance. 
Because \HI\ is self-shielded, the electron density in the \h2\ gas will originate from the 
photo-ionisation of metals, grains and primary and secondary ionisation 
by cosmic rays. Grain recombination also controls the electron density.
One important step towards
understanding this issue is to measure $N$(\CII) accurately. The expected
value suggests that \CII$\lambda$1334 should have detectable damping 
wings. Unfortunately, the S/N of our spectrum is still too low to firmly 
detect the damping wings and constrain $N$(\CII).
Another (simpler) possibility is that the gas phase abundance of carbon is much 
lower than assumed. Assuming 0.4~dex depletion of C onto dust grains as seen 
in the cold ISM, the true abundance of carbon in the DLA should be of the order of 
$[$C/O$]\sim$~-0.7 or even less for the gas phase column density to be consistent with the 
models. This is indeed similar to what is seen in Milky-way halo stars with metallicity
similar to the present DLA \citep[][]{Akerman04}. 
In addition, while [C/H] is difficult to measure in DLAs, available measurements towards very 
low metallicity DLAs do not rule out this possibility \citep{Cooke11,Dutta14}.

\section{Conclusion \label{s:conclusion}}

We have presented high resolution spectroscopic observations of three ESDLAs 
with $\log N(\HI) \ge 21.7$. 
We have firmly detected H$_2$ in two of them, towards \jt\ and \jf. H$_2$ is also 
likely present with a high column density in the third system, towards \jz, 
but additional data is required to firmly establish this.
This recommends a systematic survey for \h2\ in EDLAs as the H$_2$ detection 
rate may be much higher even though the measured overall molecular fractions are not 
as high as those expected in translucent and dense molecular regions. 

We studied in details the system towards \jt, which has the highest \HI\ and H$_2$ 
column densities observed till now in a high-$z$ intervening system ($\log N(\HI)=22.4$) 
and presents several excited fine-structure transitions that help 
constraining the physical conditions. Cobalt and germanium are likely detected in 
our spectrum, although more data would be required to use these species to 
constraint accurately the nucleosynthesis history of this DLA. 
The abundance and excitation of different species shows that H$_2$  is predominantly
originating from a cold dense phase ($T$\,$\sim$\,80\,K, $\log n_{\rm H}$\,$\sim$\,2.5-3) where 
chlorine is found in neutral form due to chemical reactions with H$_2$. 
However, models indicate that the addition of warmer phases is 
required to better explain the high excitation of oxygen and silicon, consistently 
with the wider velocity profile of these species. 
Nevertheless, even considering multiple phases, all models overproduce \CI, if we assume 
a solar abundance of carbon with respect to other elements. This 
may in turn indicate that the [C/O] ratio is actually much lower than solar in this DLA, 
similarly to what has been seen in Milky-Way halo stars.
We conclude that the absorbing gas is likely made of multiple phases having temperature 
in the range 80 to 1000~K irradiated by an local radiation field due to in situ 
star formation. This is also consistent with the picture that extremely strong 
DLAs arise at very small impact parameters from star-forming galaxies \citep{Noterdaeme14}, i.e. likely arise from 
their disc or inner regions.

Finally, direct detection of star-formation activity through the detection of NIR emission 
lines should bring invaluable clues to complete the picture: it should allow us to 
better understand the excitation of the absorbing gas and link its properties with the overall 
properties of the DLA galaxy.

\begin{acknowledgements}
We thank the referee for constructive comments and suggestions,
Gargi Shaw for advises about running Cloudy models and 
Patrick Boiss\'e for useful discussions. We are grateful to Paranal 
Observatory staff for carrying out our observations.
\end{acknowledgements}

\bibliographystyle{aa}
\bibliography{mybib}

\begin{thebibliography}{97}
\expandafter\ifx\csname natexlab\endcsname\relax\def\natexlab#1{#1}\fi

\bibitem[{{Abrahamsson} {et~al.}(2007){Abrahamsson}, {Krems}, \&
  {Dalgarno}}]{Abrahamsson07}
{Abrahamsson}, E., {Krems}, R.~V., \& {Dalgarno}, A. 2007, \apj, 654, 1171

\bibitem[{{Akerman} {et~al.}(2004){Akerman}, {Carigi}, {Nissen}, {Pettini}, \&
  {Asplund}}]{Akerman04}
{Akerman}, C.~J., {Carigi}, L., {Nissen}, P.~E., {Pettini}, M., \& {Asplund},
  M. 2004, \aap, 414, 931

\bibitem[{{Altay} {et~al.}(2013){Altay}, {Theuns}, {Schaye}, {Booth}, \& {Dalla
  Vecchia}}]{Altay13}
{Altay}, G., {Theuns}, T., {Schaye}, J., {Booth}, C.~M., \& {Dalla Vecchia}, C.
  2013, \mnras, 436, 2689

\bibitem[{{Asplund} {et~al.}(2009){Asplund}, {Grevesse}, {Sauval}, \&
  {Scott}}]{Asplund09}
{Asplund}, M., {Grevesse}, N., {Sauval}, A.~J., \& {Scott}, P. 2009, \araa, 47,
  481

\bibitem[{{Balashev} {et~al.}(2015){Balashev}, {Noterdaeme}, {Klimenko},
  {Petitjean}, {Srianand}, {Ledoux}, {Ivanchik}, \&
  {Varshalovich}}]{Balashev15}
{Balashev}, S.~A., {Noterdaeme}, P., {Klimenko}, V.~V., {et~al.} 2015, \aap, in
  press [arXiv:1501.07165]

\bibitem[{{Bird} {et~al.}(2014){Bird}, {Vogelsberger}, {Haehnelt}, {Sijacki},
  {Genel}, {Torrey}, {Springel}, \& {Hernquist}}]{Bird14}
{Bird}, S., {Vogelsberger}, M., {Haehnelt}, M., {et~al.} 2014, ArXiv e-prints

\bibitem[{{Boiss\'e} {et~al.}(1998){Boiss\'e}, {Le Brun}, {Bergeron}, \&
  {Deharveng}}]{Boisse98}
{Boiss\'e}, P., {Le Brun}, V., {Bergeron}, J., \& {Deharveng}, J.-M. 1998,
  \aap, 333, 841

\bibitem[{{Bolatto} {et~al.}(2013){Bolatto}, {Wolfire}, \& {Leroy}}]{Bolatto13}
{Bolatto}, A.~D., {Wolfire}, M., \& {Leroy}, A.~K. 2013, \araa, 51, 207

\bibitem[{{Bowen} {et~al.}(2002){Bowen}, {Pettini}, \& {Blades}}]{Bowen02}
{Bowen}, D.~V., {Pettini}, M., \& {Blades}, J.~C. 2002, \apj, 580, 169

\bibitem[{{Braun}(2012)}]{Braun12}
{Braun}, R. 2012, \apj, 749, 87

\bibitem[{{Burgh} {et~al.}(2007){Burgh}, {France}, \& {McCandliss}}]{Burgh07}
{Burgh}, E.~B., {France}, K., \& {McCandliss}, S.~R. 2007, \apj, 658, 446

\bibitem[{{Chen} {et~al.}(2005){Chen}, {Prochaska}, {Bloom}, \&
  {Thompson}}]{Chen05}
{Chen}, H.-W., {Prochaska}, J.~X., {Bloom}, J.~S., \& {Thompson}, I.~B. 2005,
  \apjl, 634, L25

\bibitem[{{Ciddor}(1996)}]{Ciddor96}
{Ciddor}, P.~E. 1996, \ao, 35, 1566

\bibitem[{{Cooke} {et~al.}(2011){Cooke}, {Pettini}, {Steidel}, {Rudie}, \&
  {Nissen}}]{Cooke11}
{Cooke}, R., {Pettini}, M., {Steidel}, C.~C., {Rudie}, G.~C., \& {Nissen},
  P.~E. 2011, \mnras, 417, 1534

\bibitem[{{Cowan} {et~al.}(2005){Cowan}, {Sneden}, {Beers}, {Lawler},
  {Simmerer}, {Truran}, {Primas}, {Collier}, \& {Burles}}]{Cowan05}
{Cowan}, J.~J., {Sneden}, C., {Beers}, T.~C., {et~al.} 2005, \apj, 627, 238

\bibitem[{{Dawson} {et~al.}(2013){Dawson}, {Schlegel}, {Ahn}, {Anderson},
  {Aubourg}, {Bailey}, {Barkhouser}, {Bautista}, {Beifiori}, {Berlind},
  {Bhardwaj}, {Bizyaev}, {Blake}, {Blanton}, {Blomqvist}, {Bolton}, {Borde},
  {Bovy}, {Brandt}, {Brewington}, {Brinkmann}, {Brown}, {Brownstein}, {Bundy},
  {Busca}, {Carithers}, {Carnero}, {Carr}, {Chen}, {Comparat}, {Connolly},
  {Cope}, {Croft}, {Cuesta}, {da Costa}, {Davenport}, {Delubac}, {de Putter},
  {Dhital}, {Ealet}, {Ebelke}, {Eisenstein}, {Escoffier}, {Fan}, {Filiz Ak},
  {Finley}, {Font-Ribera}, {G{\'e}nova-Santos}, {Gunn}, {Guo}, {Haggard},
  {Hall}, {Hamilton}, {Harris}, {Harris}, {Ho}, {Hogg}, {Holder}, {Honscheid},
  {Huehnerhoff}, {Jordan}, {Jordan}, {Kauffmann}, {Kazin}, {Kirkby}, {Klaene},
  {Kneib}, {Le Goff}, {Lee}, {Long}, {Loomis}, {Lundgren}, {Lupton}, {Maia},
  {Makler}, {Malanushenko}, {Malanushenko}, {Mandelbaum}, {Manera}, {Maraston},
  {Margala}, {Masters}, {McBride}, {McDonald}, {McGreer}, {McMahon}, {Mena},
  {Miralda-Escud{\'e}}, {Montero-Dorta}, {Montesano}, {Muna}, {Myers},
  {Naugle}, {Nichol}, {Noterdaeme}, {Nuza}, {Olmstead}, {Oravetz}, {Oravetz},
  {Owen}, {Padmanabhan}, {Palanque-Delabrouille}, {Pan}, {Parejko},
  {P{\^a}ris}, {Percival}, {P{\'e}rez-Fournon}, {P{\'e}rez-R{\`a}fols},
  {Petitjean}, {Pfaffenberger}, {Pforr}, {Pieri}, {Prada}, {Price-Whelan},
  {Raddick}, {Rebolo}, {Rich}, {Richards}, {Rockosi}, {Roe}, {Ross}, {Ross},
  {Rossi}, {Rubi{\~n}o-Martin}, {Samushia}, {S{\'a}nchez}, {Sayres}, {Schmidt},
  {Schneider}, {Sc{\'o}ccola}, {Seo}, {Shelden}, {Sheldon}, {Shen}, {Shu},
  {Slosar}, {Smee}, {Snedden}, {Stauffer}, {Steele}, {Strauss}, {Streblyanska},
  {Suzuki}, {Swanson}, {Tal}, {Tanaka}, {Thomas}, {Tinker}, {Tojeiro},
  {Tremonti}, {Vargas Maga{\~n}a}, {Verde}, {Viel}, {Wake}, {Watson}, {Weaver},
  {Weinberg}, {Weiner}, {West}, {White}, {Wood-Vasey}, {Yeche}, {Zehavi},
  {Zhao}, \& {Zheng}}]{Dawson13}
{Dawson}, K.~S., {Schlegel}, D.~J., {Ahn}, C.~P., {et~al.} 2013, \aj, 145, 10

\bibitem[{{Dekker} {et~al.}(2000){Dekker}, {D'Odorico}, {Kaufer}, {Delabre}, \&
  {Kotzlowski}}]{Dekker00}
{Dekker}, H., {D'Odorico}, S., {Kaufer}, A., {Delabre}, B., \& {Kotzlowski}, H.
  2000, in Proc. SPIE Vol. 4008, p. 534-545, Optical and IR Telescope
  Instrumentation and Detectors, Masanori Iye; Alan F. Moorwood; Eds., 534--545

\bibitem[{{Dessauges-Zavadsky} {et~al.}(2006){Dessauges-Zavadsky}, {Prochaska},
  {D'Odorico}, {Calura}, \& {Matteucci}}]{Dessauges-Zavadsky06}
{Dessauges-Zavadsky}, M., {Prochaska}, J.~X., {D'Odorico}, S., {Calura}, F., \&
  {Matteucci}, F. 2006, \aap, 445, 93

\bibitem[{{Dutta} {et~al.}(2014){Dutta}, {Srianand}, {Rahmani}, {Petitjean},
  {Noterdaeme}, \& {Ledoux}}]{Dutta14}
{Dutta}, R., {Srianand}, R., {Rahmani}, H., {et~al.} 2014, \mnras, 440, 307

\bibitem[{{Ellison} {et~al.}(2005){Ellison}, {Hall}, \& {Lira}}]{Ellison05}
{Ellison}, S.~L., {Hall}, P.~B., \& {Lira}, P. 2005, \aj, 130, 1345

\bibitem[{{Ellison} {et~al.}(2001){Ellison}, {Ryan}, \&
  {Prochaska}}]{Ellison01b}
{Ellison}, S.~L., {Ryan}, S.~G., \& {Prochaska}, J.~X. 2001, \mnras, 326, 628

\bibitem[{{Fehsenfeld} \& {Ferguson}(1974)}]{Fehsenfeld74}
{Fehsenfeld}, F.~C. \& {Ferguson}, E.~E. 1974, \jcp, 60, 5132

\bibitem[{{Ferland} {et~al.}(1998){Ferland}, {Korista}, {Verner}, {Ferguson},
  {Kingdon}, \& {Verner}}]{Ferland98}
{Ferland}, G.~J., {Korista}, K.~T., {Verner}, D.~A., {et~al.} 1998, \pasp, 110,
  761

\bibitem[{{Fox} {et~al.}(2007){Fox}, {Ledoux}, {Petitjean}, \&
  {Srianand}}]{Fox07a}
{Fox}, A.~J., {Ledoux}, C., {Petitjean}, P., \& {Srianand}, R. 2007, \aap, 473,
  791

\bibitem[{{Fox} {et~al.}(2005){Fox}, {Savage}, \& {Wakker}}]{Fox05}
{Fox}, A.~J., {Savage}, B.~D., \& {Wakker}, B.~P. 2005, \aj, 130, 2418

\bibitem[{{Fynbo} {et~al.}(2006){Fynbo}, {Starling}, {Ledoux}, {Wiersema},
  {Th{\"o}ne}, {Sollerman}, {Jakobsson}, {Hjorth}, {Watson}, {Vreeswijk},
  {M{\o}ller}, {Rol}, {Gorosabel}, {N{\"a}r{\"a}nen}, {Wijers},
  {Bj{\"o}rnsson}, {Castro Cer{\'o}n}, {Curran}, {Hartmann}, {Holland},
  {Jensen}, {Levan}, {Limousin}, {Kouveliotou}, {Nelemans}, {Pedersen},
  {Priddey}, \& {Tanvir}}]{Fynbo06}
{Fynbo}, J.~P.~U., {Starling}, R.~L.~C., {Ledoux}, C., {et~al.} 2006, \aap,
  451, L47

\bibitem[{{Galavis} {et~al.}(1997){Galavis}, {Mendoza}, \&
  {Zeippen}}]{Galavis97}
{Galavis}, M.~E., {Mendoza}, C., \& {Zeippen}, C.~J. 1997, \aaps, 123, 159

\bibitem[{{Gillmon} {et~al.}(2006){Gillmon}, {Shull}, {Tumlinson}, \&
  {Danforth}}]{Gillmon06}
{Gillmon}, K., {Shull}, J.~M., {Tumlinson}, J., \& {Danforth}, C. 2006, \apj,
  636, 891

\bibitem[{{Grillo} \& {Fynbo}(2014)}]{Grillo14}
{Grillo}, C. \& {Fynbo}, J.~P.~U. 2014, \mnras, 439, L100

\bibitem[{{Guimar{\~a}es} {et~al.}(2012){Guimar{\~a}es}, {Noterdaeme},
  {Petitjean}, {Ledoux}, {Srianand}, {L\'opez}, \& {Rahmani}}]{Guimaraes12}
{Guimar{\~a}es}, R., {Noterdaeme}, P., {Petitjean}, P., {et~al.} 2012, \aj,
  143, 147

\bibitem[{{Haardt} \& {Madau}(2012)}]{Haardt12}
{Haardt}, F. \& {Madau}, P. 2012, \apj, 746, 125

\bibitem[{{Habing}(1968)}]{Habing68}
{Habing}, H.~J. 1968, \bain, 19, 421

\bibitem[{{Jenkins} \& {Tripp}(2011)}]{Jenkins11}
{Jenkins}, E.~B. \& {Tripp}, T.~M. 2011, \apj, 734, 65

\bibitem[{{Jorgenson} {et~al.}(2006){Jorgenson}, {Wolfe}, {Prochaska}, {Lu},
  {Howk}, {Cooke}, {Gawiser}, \& {Gelino}}]{Jorgenson06}
{Jorgenson}, R.~A., {Wolfe}, A.~M., {Prochaska}, J.~X., {et~al.} 2006, \apj,
  646, 730

\bibitem[{{Jura}(1974)}]{Jura74a}
{Jura}, M. 1974, \apjl, 190, L33

\bibitem[{{Jura} \& {York}(1978)}]{Jura78}
{Jura}, M. \& {York}, D.~G. 1978, \apj, 219, 861

\bibitem[{{Kanekar} {et~al.}(2011){Kanekar}, {Braun}, \& {Roy}}]{Kanekar11}
{Kanekar}, N., {Braun}, R., \& {Roy}, N. 2011, \apjl, 737, L33

\bibitem[{{Krumholz} {et~al.}(2009){Krumholz}, {Ellison}, {Prochaska}, \&
  {Tumlinson}}]{Krumholz09}
{Krumholz}, M.~R., {Ellison}, S.~L., {Prochaska}, J.~X., \& {Tumlinson}, J.
  2009, \apjl, 701, L12

\bibitem[{{Kulkarni} {et~al.}(2012){Kulkarni}, {Meiring}, {Som}, {P{\'e}roux},
  {York}, {Khare}, \& {Lauroesch}}]{Kulkarni12}
{Kulkarni}, V.~P., {Meiring}, J., {Som}, D., {et~al.} 2012, \apj, 749, 176

\bibitem[{{Le Petit} {et~al.}(2006){Le Petit}, {Nehm{\'e}}, {Le Bourlot}, \&
  {Roueff}}]{LePetit06}
{Le Petit}, F., {Nehm{\'e}}, C., {Le Bourlot}, J., \& {Roueff}, E. 2006, \apjs,
  164, 506

\bibitem[{{Ledoux} {et~al.}(2014){Ledoux}, {Noterdaeme}, {Petitjean}, \&
  {Srianand}}]{Ledoux14}
{Ledoux}, C., {Noterdaeme}, P., {Petitjean}, P., \& {Srianand}, R. 2014, \aap,
  submitted

\bibitem[{{Ledoux} {et~al.}(2006){Ledoux}, {Petitjean}, {Fynbo}, {M{\o}ller},
  \& {Srianand}}]{Ledoux06a}
{Ledoux}, C., {Petitjean}, P., {Fynbo}, J.~P.~U., {M{\o}ller}, P., \&
  {Srianand}, R. 2006, \aap, 457, 71

\bibitem[{{Lehner} {et~al.}(2013){Lehner}, {Howk}, {Tripp}, {Tumlinson},
  {Prochaska}, {O'Meara}, {Thom}, {Werk}, {Fox}, \& {Ribaudo}}]{Lehner13}
{Lehner}, N., {Howk}, J.~C., {Tripp}, T.~M., {et~al.} 2013, \apj, 770, 138

\bibitem[{{Molaro} {et~al.}(2001){Molaro}, {Levshakov}, {D'Odorico},
  {Bonifacio}, \& {Centuri{\'o}n}}]{Molaro01}
{Molaro}, P., {Levshakov}, S.~A., {D'Odorico}, S., {Bonifacio}, P., \&
  {Centuri{\'o}n}, M. 2001, \apj, 549, 90

\bibitem[{{Muzahid} {et~al.}(2014){Muzahid}, {Srianand}, \&
  {Charlton}}]{Muzahid14}
{Muzahid}, S., {Srianand}, R., \& {Charlton}, J. 2014, MNRAS, in press
  [arXiv:1410.3828]

\bibitem[{{Neeleman} {et~al.}(2015){Neeleman}, {Prochaska}, \&
  {Wolfe}}]{Neeleman15}
{Neeleman}, M., {Prochaska}, J.~X., \& {Wolfe}, A.~M. 2015, \apj, 800, 7

\bibitem[{{Noterdaeme} {et~al.}(2012{\natexlab{a}}){Noterdaeme}, {Laursen},
  {Petitjean}, {Vergani}, {Maureira}, {Ledoux}, {Fynbo}, {L{\'o}pez}, \&
  {Srianand}}]{Noterdaeme12a}
{Noterdaeme}, P., {Laursen}, P., {Petitjean}, P., {et~al.} 2012{\natexlab{a}},
  \aap, 540, A63

\bibitem[{{Noterdaeme} {et~al.}(2009{\natexlab{a}}){Noterdaeme}, {Ledoux},
  {Srianand}, {Petitjean}, \& {Lopez}}]{Noterdaeme09co}
{Noterdaeme}, P., {Ledoux}, C., {Srianand}, R., {Petitjean}, P., \& {Lopez}, S.
  2009{\natexlab{a}}, \aap, 503, 765

\bibitem[{{Noterdaeme} {et~al.}(2012{\natexlab{b}}){Noterdaeme}, {Petitjean},
  {Carithers}, {P{\^a}ris}, {Font-Ribera}, {Bailey}, {Aubourg}, {Bizyaev},
  {Ebelke}, {Finley}, {Ge}, {Malanushenko}, {Malanushenko},
  {Miralda-Escud{\'e}}, {Myers}, {Oravetz}, {Pan}, {Pieri}, {Ross},
  {Schneider}, {Simmons}, \& {York}}]{Noterdaeme12c}
{Noterdaeme}, P., {Petitjean}, P., {Carithers}, W.~C., {et~al.}
  2012{\natexlab{b}}, \aap, 547, L1

\bibitem[{{Noterdaeme} {et~al.}(2010){Noterdaeme}, {Petitjean}, {Ledoux},
  {L{\'o}pez}, {Srianand}, \& {Vergani}}]{Noterdaeme10co}
{Noterdaeme}, P., {Petitjean}, P., {Ledoux}, C., {et~al.} 2010, \aap, 523, A80

\bibitem[{{Noterdaeme} {et~al.}(2009{\natexlab{b}}){Noterdaeme}, {Petitjean},
  {Ledoux}, \& {Srianand}}]{Noterdaeme09dla}
{Noterdaeme}, P., {Petitjean}, P., {Ledoux}, C., \& {Srianand}, R.
  2009{\natexlab{b}}, \aap, 505, 1087

\bibitem[{{Noterdaeme} {et~al.}(2014){Noterdaeme}, {Petitjean}, {P{\^a}ris},
  {Cai}, {Finley}, {Ge}, {Pieri}, \& {York}}]{Noterdaeme14}
{Noterdaeme}, P., {Petitjean}, P., {P{\^a}ris}, I., {et~al.} 2014, \aap, 566,
  A24

\bibitem[{{Noterdaeme} {et~al.}(2007){Noterdaeme}, {Petitjean}, {Srianand},
  {Ledoux}, \& {Le Petit}}]{Noterdaeme07}
{Noterdaeme}, P., {Petitjean}, P., {Srianand}, R., {Ledoux}, C., \& {Le Petit},
  F. 2007, \aap, 469, 425

\bibitem[{{Noterdaeme} {et~al.}(2011){Noterdaeme}, {Petitjean}, {Srianand},
  {Ledoux}, \& {L{\'o}pez}}]{Noterdaeme11}
{Noterdaeme}, P., {Petitjean}, P., {Srianand}, R., {Ledoux}, C., \&
  {L{\'o}pez}, S. 2011, \aap, 526, L7+

\bibitem[{{Oppenheimer} \& {Dav{\'e}}(2006)}]{Oppenheimer06}
{Oppenheimer}, B.~D. \& {Dav{\'e}}, R. 2006, \mnras, 373, 1265

\bibitem[{{P{\^a}ris} {et~al.}(2012){P{\^a}ris}, {Petitjean}, {Aubourg},
  {Bailey}, {Ross}, {Myers}, {Strauss}, {Anderson}, {Arnau}, {Bautista},
  {Bizyaev}, {Bolton}, {Bovy}, {Brandt}, {Brewington}, {Browstein}, {Busca},
  {Capellupo}, {Carithers}, {Croft}, {Dawson}, {Delubac}, {Ebelke},
  {Eisenstein}, {Engelke}, {Fan}, {Filiz Ak}, {Finley}, {Font-Ribera}, {Ge},
  {Gibson}, {Hall}, {Hamann}, {Hennawi}, {Ho}, {Hogg}, {Ivezi{\'c}}, {Jiang},
  {Kimball}, {Kirkby}, {Kirkpatrick}, {Lee}, {Le Goff}, {Lundgren}, {MacLeod},
  {Malanushenko}, {Malanushenko}, {Maraston}, {McGreer}, {McMahon},
  {Miralda-Escud{\'e}}, {Muna}, {Noterdaeme}, {Oravetz},
  {Palanque-Delabrouille}, {Pan}, {Perez-Fournon}, {Pieri}, {Richards},
  {Rollinde}, {Sheldon}, {Schlegel}, {Schneider}, {Slosar}, {Shelden}, {Shen},
  {Simmons}, {Snedden}, {Suzuki}, {Tinker}, {Viel}, {Weaver}, {Weinberg},
  {White}, {Wood-Vasey}, \& {Y{\`e}che}}]{Paris12}
{P{\^a}ris}, I., {Petitjean}, P., {Aubourg}, {\'E}., {et~al.} 2012, \aap, 548,
  A66

\bibitem[{{P{\^a}ris} {et~al.}(2014){P{\^a}ris}, {Petitjean}, {Aubourg},
  {Ross}, {Myers}, {Streblyanska}, {Bailey}, {Hall}, {Strauss}, {Anderson},
  {Bizyaev}, {Borde}, {Brinkmann}, {Bovy}, {Brandt}, {Brewington},
  {Brownstein}, {Cook}, {Ebelke}, {Fan}, {Filiz Ak}, {Finley}, {Font-Ribera},
  {Ge}, {Hamann}, {Ho}, {Jiang}, {Kinemuchi}, {Malanushenko}, {Malanushenko},
  {Marchante}, {McGreer}, {McMahon}, {Miralda-Escud{\'e}}, {Muna},
  {Noterdaeme}, {Oravetz}, {Palanque-Delabrouille}, {Pan}, {Perez-Fournon},
  {Pieri}, {Riffel}, {Schlegel}, {Schneider}, {Simmons}, {Viel}, {Weaver},
  {Wood-Vasey}, {Y{\`e}che}, \& {York}}]{Paris14}
{P{\^a}ris}, I., {Petitjean}, P., {Aubourg}, {\'E}., {et~al.} 2014, \aap, 563,
  A54

\bibitem[{{P{\^a}ris} {et~al.}(2011){P{\^a}ris}, {Petitjean}, {Rollinde},
  {Aubourg}, {Busca}, {Charlassier}, {Delubac}, {Hamilton}, {Le Goff},
  {Palanque-Delabrouille}, {Peirani}, {Pichon}, {Rich}, {Vargas-Maga{\~n}a}, \&
  {Y{\`e}che}}]{Paris11}
{P{\^a}ris}, I., {Petitjean}, P., {Rollinde}, E., {et~al.} 2011, \aap, 530, A50

\bibitem[{{P{\'e}roux} {et~al.}(2007){P{\'e}roux}, {Dessauges-Zavadsky},
  {D'Odorico}, {Kim}, \& {McMahon}}]{Peroux07}
{P{\'e}roux}, C., {Dessauges-Zavadsky}, M., {D'Odorico}, S., {Kim}, T.-S., \&
  {McMahon}, R.~G. 2007, \mnras, 382, 177

\bibitem[{{Petitjean} {et~al.}(1992){Petitjean}, {Bergeron}, \&
  {Puget}}]{Petitjean92}
{Petitjean}, P., {Bergeron}, J., \& {Puget}, J.~L. 1992, \aap, 265, 375

\bibitem[{{Petitjean} {et~al.}(2008){Petitjean}, {Ledoux}, \&
  {Srianand}}]{Petitjean08}
{Petitjean}, P., {Ledoux}, C., \& {Srianand}, R. 2008, \aap, 480, 349

\bibitem[{{Pettini} {et~al.}(1997){Pettini}, {Smith}, {King}, \&
  {Hunstead}}]{Pettini97}
{Pettini}, M., {Smith}, L.~J., {King}, D.~L., \& {Hunstead}, R.~W. 1997, \apj,
  486, 665

\bibitem[{{Pettini} {et~al.}(2008){Pettini}, {Zych}, {Murphy}, {Lewis}, \&
  {Steidel}}]{Pettini08}
{Pettini}, M., {Zych}, B.~J., {Murphy}, M.~T., {Lewis}, A., \& {Steidel}, C.~C.
  2008, \mnras, 391, 1499

\bibitem[{{Pontzen} {et~al.}(2008){Pontzen}, {Governato}, {Pettini}, {Booth},
  {Stinson}, {Wadsley}, {Brooks}, {Quinn}, \& {Haehnelt}}]{Pontzen08}
{Pontzen}, A., {Governato}, F., {Pettini}, M., {et~al.} 2008, \mnras, 390, 1349

\bibitem[{{Prochaska} {et~al.}(2003){Prochaska}, {Gawiser}, {Wolfe}, {Cooke},
  \& {Gelino}}]{Prochaska03}
{Prochaska}, J.~X., {Gawiser}, E., {Wolfe}, A.~M., {Cooke}, J., \& {Gelino}, D.
  2003, \apjs, 147, 227

\bibitem[{{Prochaska} {et~al.}(2000){Prochaska}, {Naumov}, {Carney},
  {McWilliam}, \& {Wolfe}}]{Prochaska00}
{Prochaska}, J.~X., {Naumov}, S.~O., {Carney}, B.~W., {McWilliam}, A., \&
  {Wolfe}, A.~M. 2000, \aj, 120, 2513

\bibitem[{{Prochaska} {et~al.}(2011){Prochaska}, {Weiner}, {Chen}, {Mulchaey},
  \& {Cooksey}}]{Prochaska11}
{Prochaska}, J.~X., {Weiner}, B., {Chen}, H.-W., {Mulchaey}, J., \& {Cooksey},
  K. 2011, \apj, 740, 91

\bibitem[{{Prochaska} \& {Wolfe}(2009)}]{Prochaska09a}
{Prochaska}, J.~X. \& {Wolfe}, A.~M. 2009, \apj, 696, 1543

\bibitem[{{Rafelski} {et~al.}(2012){Rafelski}, {Wolfe}, {Prochaska},
  {Neeleman}, \& {Mendez}}]{Rafelski12}
{Rafelski}, M., {Wolfe}, A.~M., {Prochaska}, J.~X., {Neeleman}, M., \&
  {Mendez}, A.~J. 2012, \apj, 755, 89

\bibitem[{{Rao} {et~al.}(2005){Rao}, {Prochaska}, {Howk}, \& {Wolfe}}]{Rao05}
{Rao}, S.~M., {Prochaska}, J.~X., {Howk}, J.~C., \& {Wolfe}, A.~M. 2005, \aj,
  129, 9

\bibitem[{{Roederer}(2012)}]{Roederer12}
{Roederer}, I.~U. 2012, \apj, 756, 36

\bibitem[{{Roy} {et~al.}(2006){Roy}, {Chengalur}, \& {Srianand}}]{Roy06}
{Roy}, N., {Chengalur}, J.~N., \& {Srianand}, R. 2006, \mnras, 365, L1

\bibitem[{{Savage} {et~al.}(1977){Savage}, {Bohlin}, {Drake}, \&
  {Budich}}]{Savage77}
{Savage}, B.~D., {Bohlin}, R.~C., {Drake}, J.~F., \& {Budich}, W. 1977, \apj,
  216, 291

\bibitem[{{Savage} \& {Sembach}(1991)}]{Savage91}
{Savage}, B.~D. \& {Sembach}, K.~R. 1991, \apj, 379, 245

\bibitem[{{Schaye}(2001)}]{Schaye01}
{Schaye}, J. 2001, \apjl, 562, L95

\bibitem[{{Shaw} {et~al.}(2005){Shaw}, {Ferland}, {Abel}, {Stancil}, \& {van
  Hoof}}]{Shaw05}
{Shaw}, G., {Ferland}, G.~J., {Abel}, N.~P., {Stancil}, P.~C., \& {van Hoof},
  P.~A.~M. 2005, \apj, 624, 794

\bibitem[{{Silva} \& {Viegas}(2002)}]{Silva02}
{Silva}, A.~I. \& {Viegas}, S.~M. 2002, \mnras, 329, 135

\bibitem[{{Snow} \& {McCall}(2006)}]{Snow06}
{Snow}, T.~P. \& {McCall}, B.~J. 2006, \araa, 44, 367

\bibitem[{{Sonnentrucker} {et~al.}(2002){Sonnentrucker}, {Friedman}, {Welty},
  {York}, \& {Snow}}]{Sonnentrucker02}
{Sonnentrucker}, P., {Friedman}, S.~D., {Welty}, D.~E., {York}, D.~G., \&
  {Snow}, T.~P. 2002, \apj, 576, 241

\bibitem[{{Srianand} {et~al.}(2008){Srianand}, {Noterdaeme}, {Ledoux}, \&
  {Petitjean}}]{Srianand08}
{Srianand}, R., {Noterdaeme}, P., {Ledoux}, C., \& {Petitjean}, P. 2008, \aap,
  482, L39

\bibitem[{{Srianand} \& {Petitjean}(2000)}]{Srianand00b}
{Srianand}, R. \& {Petitjean}, P. 2000, \aap, 357, 414

\bibitem[{{Srianand} {et~al.}(2005){Srianand}, {Petitjean}, {Ledoux},
  {Ferland}, \& {Shaw}}]{Srianand05}
{Srianand}, R., {Petitjean}, P., {Ledoux}, C., {Ferland}, G., \& {Shaw}, G.
  2005, \mnras, 362, 549

\bibitem[{{Srianand} {et~al.}(2014){Srianand}, {Rahmani}, {Muzahid}, \&
  {Mohan}}]{Srianand14}
{Srianand}, R., {Rahmani}, H., {Muzahid}, S., \& {Mohan}, V. 2014, \mnras, 443,
  3318

\bibitem[{{Tripp} {et~al.}(1998){Tripp}, {Lu}, \& {Savage}}]{Tripp98}
{Tripp}, T.~M., {Lu}, L., \& {Savage}, B.~D. 1998, \apj, 508, 200

\bibitem[{{Tumlinson} {et~al.}(2002){Tumlinson}, {Shull}, {Rachford},
  {Browning}, {Snow}, {Fullerton}, {Jenkins}, {Savage}, {Crowther}, {Moos},
  {Sembach}, {Sonneborn}, \& {York}}]{Tumlinson02}
{Tumlinson}, J., {Shull}, J.~M., {Rachford}, B.~L., {et~al.} 2002, \apj, 566,
  857

\bibitem[{{Vreeswijk} {et~al.}(2004){Vreeswijk}, {Ellison}, {Ledoux}, {Wijers},
  {Fynbo}, {M{\o}ller}, {Henden}, {Hjorth}, {Masi}, {Rol}, {Jensen}, {Tanvir},
  {Levan}, {Castro Cer{\'o}n}, {Gorosabel}, {Castro-Tirado}, {Fruchter},
  {Kouveliotou}, {Burud}, {Rhoads}, {Masetti}, {Palazzi}, {Pian}, {Pedersen},
  {Kaper}, {Gilmore}, {Kilmartin}, {Buckle}, {Seigar}, {Hartmann}, {Lindsay},
  \& {van den Heuvel}}]{Vreeswijk04}
{Vreeswijk}, P.~M., {Ellison}, S.~L., {Ledoux}, C., {et~al.} 2004, \aap, 419,
  927

\bibitem[{{Wagenblast}(1992)}]{Wagenblast92}
{Wagenblast}, R. 1992, \mnras, 259, 155

\bibitem[{{Welty} {et~al.}(2003){Welty}, {Hobbs}, \& {Morton}}]{Welty03}
{Welty}, D.~E., {Hobbs}, L.~M., \& {Morton}, D.~C. 2003, \apjs, 147, 61

\bibitem[{{Welty} {et~al.}(1997){Welty}, {Lauroesch}, {Blades}, {Hobbs}, \&
  {York}}]{Welty97}
{Welty}, D.~E., {Lauroesch}, J.~T., {Blades}, J.~C., {Hobbs}, L.~M., \& {York},
  D.~G. 1997, \apj, 489, 672

\bibitem[{{Welty} {et~al.}(2012){Welty}, {Xue}, \& {Wong}}]{Welty12}
{Welty}, D.~E., {Xue}, R., \& {Wong}, T. 2012, \apj, 745, 173

\bibitem[{{Wolfe} {et~al.}(2005){Wolfe}, {Gawiser}, \& {Prochaska}}]{Wolfe05}
{Wolfe}, A.~M., {Gawiser}, E., \& {Prochaska}, J.~X. 2005, \araa, 43, 861

\bibitem[{{Wolfe} {et~al.}(2004){Wolfe}, {Howk}, {Gawiser}, {Prochaska}, \&
  {Lopez}}]{Wolfe04}
{Wolfe}, A.~M., {Howk}, J.~C., {Gawiser}, E., {Prochaska}, J.~X., \& {Lopez},
  S. 2004, \apj, 615, 625

\bibitem[{{Wolfe} {et~al.}(2003){Wolfe}, {Prochaska}, \& {Gawiser}}]{Wolfe03}
{Wolfe}, A.~M., {Prochaska}, J.~X., \& {Gawiser}, E. 2003, \apj, 593, 215

\bibitem[{{Wolfe} {et~al.}(1986){Wolfe}, {Turnshek}, {Smith}, \&
  {Cohen}}]{Wolfe86}
{Wolfe}, A.~M., {Turnshek}, D.~A., {Smith}, H.~E., \& {Cohen}, R.~D. 1986,
  \apjs, 61, 249

\bibitem[{{York} {et~al.}(2000){York}, {Adelman}, {Anderson}, {Anderson},
  {Annis}, {Bahcall}, {Bakken}, {Barkhouser}, {Bastian}, {Berman}, {Boroski},
  {Bracker}, {Briegel}, {Briggs}, {Brinkmann}, {Brunner}, {Burles}, {Carey},
  {Carr}, {Castander}, {Chen}, {Colestock}, {Connolly}, {Crocker}, {Csabai},
  {Czarapata}, {Davis}, {Doi}, {Dombeck}, {Eisenstein}, {Ellman}, {Elms},
  {Evans}, {Fan}, {Federwitz}, {Fiscelli}, {Friedman}, {Frieman}, {Fukugita},
  {Gillespie}, {Gunn}, {Gurbani}, {de Haas}, {Haldeman}, {Harris}, {Hayes},
  {Heckman}, {Hennessy}, {Hindsley}, {Holm}, {Holmgren}, {Huang}, {Hull},
  {Husby}, {Ichikawa}, {Ichikawa}, {Ivezi{\'c}}, {Kent}, {Kim}, {Kinney},
  {Klaene}, {Kleinman}, {Kleinman}, {Knapp}, {Korienek}, {Kron}, {Kunszt},
  {Lamb}, {Lee}, {Leger}, {Limmongkol}, {Lindenmeyer}, {Long}, {Loomis},
  {Loveday}, {Lucinio}, {Lupton}, {MacKinnon}, {Mannery}, {Mantsch}, {Margon},
  {McGehee}, {McKay}, {Meiksin}, {Merelli}, {Monet}, {Munn}, {Narayanan},
  {Nash}, {Neilsen}, {Neswold}, {Newberg}, {Nichol}, {Nicinski}, {Nonino},
  {Okada}, {Okamura}, {Ostriker}, {Owen}, {Pauls}, {Peoples}, {Peterson},
  {Petravick}, {Pier}, {Pope}, {Pordes}, {Prosapio}, {Rechenmacher}, {Quinn},
  {Richards}, {Richmond}, {Rivetta}, {Rockosi}, {Ruthmansdorfer}, {Sandford},
  {Schlegel}, {Schneider}, {Sekiguchi}, {Sergey}, {Shimasaku}, {Siegmund},
  {Smee}, {Smith}, {Snedden}, {Stone}, {Stoughton}, {Strauss}, {Stubbs},
  {SubbaRao}, {Szalay}, {Szapudi}, {Szokoly}, {Thakar}, {Tremonti}, {Tucker},
  {Uomoto}, {Vanden Berk}, {Vogeley}, {Waddell}, {Wang}, {Watanabe},
  {Weinberg}, {Yanny}, {Yasuda}, \& {SDSS Collaboration}}]{York00}
{York}, D.~G., {Adelman}, J., {Anderson}, Jr., J.~E., {et~al.} 2000, \aj, 120,
  1579

\bibitem[{{Zafar} {et~al.}(2014){Zafar}, {Centuri{\'o}n}, {P{\'e}roux},
  {Molaro}, {D'Odorico}, {Vladilo}, \& {Popping}}]{Zafar14}
{Zafar}, T., {Centuri{\'o}n}, M., {P{\'e}roux}, C., {et~al.} 2014, \mnras, 444,
  744

\bibitem[{{Zwaan} \& {Prochaska}(2006)}]{Zwaan06}
{Zwaan}, M.~A. \& {Prochaska}, J.~X. 2006, \apj, 643, 675

\end{thebibliography}

\end{document}